\documentclass[11pt,a4paper]{article}
\usepackage{jheppub,amsmath,amssymb,slashed,url,bm,textgreek,upgreek}
\usepackage{jheppub}  
\usepackage{amssymb} 
\usepackage{amsmath}
\usepackage{mathtools}
\usepackage{amsfonts}    
\usepackage{dsfont}
\usepackage{pdfpages}
\usepackage{verbatim}
\usepackage{tensor}
\usepackage{mathrsfs}
\usepackage{textgreek} 
\usepackage[mathscr]{euscript}
\usepackage[normalem]{ulem}
\usepackage{tikz}
\usepackage{braket}
\usetikzlibrary{decorations.pathreplacing, decorations.markings,calc,shapes.misc,decorations.pathmorphing,patterns.meta, math}

\newcommand{\beq}{\begin{equation}}
\newcommand{\eeq}{\end{equation}}
 
\newcommand{\bea}{\begin{eqnarray}}
\newcommand{\ea}{\end{eqnarray}}
\newcommand{\barr}{\begin{array}}
\newcommand{\earr}{\end{array}}

\def\d{{\rm d}}
\def\i{{\rm i}}
\def\AdS{{\rm AdS}}

\newcommand\U{\text{U}}
\newcommand\hq{\hat{q}}

\newcommand\SU{\text{SU}}

\title{Supersymmetry breaking in SYK \\
and the black hole spectrum
\vspace{-0.6cm}}

 \author{Matthew Heydeman${}^1$, Yifei Liu${}^2$, and Gustavo J. Turiaci${}^3$}
 \affiliation{${}^1$ Department of Physics \& The Black Hole Initiative, Harvard University, Cambridge, MA, USA}%\emailAdd{author@inst.edu}
 \affiliation{${}^2$ Joseph Henry Laboratories, Princeton University, Princeton, NJ, USA}%\emailAdd{author@inst.edu}
\affiliation{${}^3$ Physics Department, University of Washington, Seattle, WA, USA}%\emailAdd{author@inst.edu}
\abstract{The spectrum and dynamics of near-extremal black holes is strongly modified by quantum effects at low temperatures. When the extremal limit does not preserve any supersymmetry, the density of states goes to zero at extremality and no extremal black holes remain. However, when the extremal limit is supersymmetric, a large microscopic degeneracy survives and there is a gap to the first excited black hole visible from gravity. In this article we study large $N$ quantum mechanical models where supersymmetry is explicitly broken, allowing us to interpolate between these two qualitatively different pictures. We propose and analyze deformations of $\mathcal{N}=2$ SYK models with such a pattern of (super)symmetry breaking which violate the $\U(1)$ $R$-symmetry. These models feature a lifting of the BPS degeneracy and a closing of the spectral gap, and we further show that the large $N$ soft effective action is given by a modification of the $\mathcal{N}=2$ Schwarzian theory in which the $\U(1)_R$ mode becomes massive.}

\begin{document}\maketitle

\section{Introduction}

At the classical level, the thermodynamic behavior of extremal black holes displays strange features. Generally, they have a macroscopic horizon area all the way to extremality, implying they have a large zero-temperature entropy $S_{\rm ext}=A_{\rm ext}/4G_N$. At finite temperature, the approach of their energy to extremality is inconsistent with their description as a statistical quantum system \cite{Preskill:1991tb}. 

These puzzles were resolved by realizing that a specific gravity mode becomes light as the temperature is lowered \cite{Ghosh:2019rcj,Iliesiu:2020qvm}. This implies that its quantum fluctuations are large and the classical description is invalid close enough to extremality. Careful treatment of the reparametrization or ``Schwarzian" mode of the metric is required at the level of the gravitational path integral. The resulting quantum corrected spectrum displays no ground state degeneracy, and an average spacing between near-extremal energy levels exponentially small of order $\exp{\{-S_{\rm ext}\}}$. These features of the spectrum also control the dynamics through the late-time behavior of correlation functions \cite{Mertens:2017mtv,lin2023looking}.

The situation is different in supergravity theories when the extremal limit also preserves some supersymmetry \cite{Heydeman:2020hhw,boruch2022bps}. In this case, a light fermionic mode arises from the gravitino which changes the conclusions in the previous paragraph due to the existence of new super-reparametrizations. The existence of these fermionic modes and their superpartners imply the zero-temperature entropy does translate into a large ground state degeneracy of order $\exp{\{S_{\rm ext}\}}$. One further finds that the gap between these BPS black holes and the first excited state is of order $1/S_{\rm ext}$-- much larger than the naive gap between non-BPS energy levels in a chaotic spectrum which are exponentially suppressed. 

\medskip

The physics of near-extremal black holes is therefore completely different in gravity compared to supergravity (or even within supergravity, depending on whether the extremal solution preserves susy or not), even though all fermionic fields in the solution vanish classically. This raises the question of how the black hole spectrum and dynamics interpolate between these two pictures, when we are in a situation in which supersymmetry is slightly broken by a continuous parameter (this will be made more precise below). How does the gap close, or at least become exponentially reduced? How do the BPS states merge with the excited states? Furthermore, black holes are expected to display a chaotic spectrum, both in gravity \cite{Cotler:2016fpe,Saad:2018bqo,Saad:2019lba,Johnson:2020exp} and supergravity \cite{Stanford:2019vob,Turiaci:2023jfa,Johnson:2023ofr,Johnson:2024tgg,Hernandez-Cuenca:2024icn}. How does the quantum chaotic nature of the spectrum transition between the random matrix ensemble applicable to quantum mechanics and supersymmetric quantum mechanics\footnote{Based on the reasoning in \cite{Johnson:2024tgg} which explains the gap in terms of eigenvalue repulsion from the BPS sector, if one instead considers a random matrix model in which $\mathcal{N}=2$ supersymmetry is weakly broken, it is plausible that microscopic eigenvalue repulsion between the (formerly) BPS states would cause them to penetrate into the gap region.}?

\medskip

It is easy to construct an example where such a supersymmetry breaking parameter is present. For instance, consider the prototypical case of a near-extremal black hole-- the Reissner-Nordstrom solution in $4d$ flat space. In ungauged supergravity with an asymptotically flat vacuum, the extremal black hole with $M=Q$ preserves supersymmetry and the near-BPS spectrum (the modes which become strongly coupled at low temperatures) is described by the $\mathcal{N}=4$ Schwarzian theory \cite{Heydeman:2020hhw}. What happens if we consider such a black hole not in flat space but in gauged supergravity in asymptotically $\AdS_4$? Ignoring any instability, the solution near extremality still has an $\AdS_2$ throat, but supersymmetry is not preserved anymore and the near-extremal dynamics is captured by the $\mathcal{N}=0$ Schwarzian. The supersymmetry breaking parameter in this case is the cosmological constant\footnote{Note that this does not imply that BPS AdS$_4$ black holes do not exist, rather such solutions may be obtained by turning on rotations or magnetic charges\cite{Romans:1991nq,Caldarelli:1998hg}.}.

\medskip

Due to a lack of a solvable gravity model which includes both quantum effects and supersymmetry breaking, the aim of this article is to analyze this question in the simplest possible setup in a holographic quantum system. The spectrum we would like to consider which displays a large BPS degeneracy and a large gap corresponds to that of the $\mathcal{N}=2$ Schwarzian theory, which has, in addition to the Schwarzian mode and its superpartners, a $\U(1)_R$ symmetry mode. Because the $R$-symmetry is abelian, this theory has two parameters corresponding to the freedom in choosing the $R$-charge of the supercharge (denoted conventionally by the odd integer $\hat{q}$)  and the freedom to introduce a $\theta$-angle for a 1d Chern-Simons term. As long as $\theta \neq \pi$ (mod $2\pi$), the solution of this super-Schwarzian model shows the BPS degeneracy is large and the gap is large. In addition to describing the low-energy regime of  $\mathcal{N}=2$ SYK (where $\theta=0$ when the number of fermions $N$ is even)\cite{Fu:2016vas}, it also shares a low-energy phase with supersymmetric AdS black holes \cite{Benini:2015eyy,Cabo-Bizet:2018ehj,boruch2022bps}. For this reason, we view the SYK system as a solvable strongly coupled model which captures some features of the near-horizon dynamics of near-BPS black holes; and indeed similar (though more complicated) models may be obtained directly from the reduction of higher dimensional SCFTs \cite{benini2023quantum,Benini:2024cpf}\footnote{In the context of $\mathcal{N}=4$ SYM in the 1/16th-BPS sector, there has been some progress in finding some features of the mass gap and $\mathcal{N}=2$ Schwarzian physics already in the weakly coupled theory\cite{Chang:2023zqk,Cabo-Bizet:2024gny}. It has further been conjectured that this 1/16th-BPS sector is strongly chaotic\cite{Chen:2024oqv}, and our work can be seen as a toy model for what happens to these BPS states when supersymmetry is weakly broken.}. These models typically have many species of fields and additional global symmetries leading to a more complicated phase structure \cite{Heydeman:2022lse}, and we believe most of the techniques of the present work should generalize to these examples.

\medskip

To analyze the question raised in the previous paragraphs, we will construct an SYK model with the appropriate pattern of supersymmetry breaking. We want the following features to be present:
\begin{enumerate}
    \item The undeformed model should have a large number $O(e^{N})$ of BPS states and a large gap of order $O(1/N)$. A model with this property and a conformal phase is $\mathcal{N}=2$ SYK, as recalled above.
    \item The continuous deformation, when large enough, should remove both the $O(1/N)$ gap and the BPS state degeneracy. This implies that we need the deformation of $\mathcal{N}=2$ SYK to preserve $\mathcal{N}\leq 1$ supersymmetry (see for example table 1 of \cite{Turiaci:2023jfa}). 
    \item In the IR regime, the model should permit a conformal solution for any value of the deformation. This is the analog of demanding that in gravity, the classical $\AdS_2$ throat is present for any value of the background field which breaks susy. In fact, we will study a model with the stronger constraint that the conformal solution is \emph{independent} of the deformation. This can be conveniently implemented by a deformation that affects the UV, but its effect is small in the IR.
    \item We want the supersymmetry-breaking deformation to modify the action for quantum fluctuations around the conformal solution. This means that the Schwarzian effective action will be deformed by terms which lift certain near-zero modes. We will further consider a model where the $\U(1)_R$ symmetry is broken. %This is not independent of requirement 2, since the unbroken $\mathcal{N}=2$ susy immediately implies an unbroken $\U(1)_R$.
\end{enumerate}

These requirements narrow down the type of systems we can consider. We already pointed out we should aim to construct a supersymmetry breaking deformation of $\mathcal{N}=2$ SYK. This is a model of $N$ complex fermions $\psi_i$ with the only non-vanishing anticommutator being $\{ \psi_i, \bar{\psi}_j\} = \delta_{ij}$ and Hamiltonian $H=\{Q,\bar{Q}\}$ with $Q= \i \sum_{i<j<k} C_{ijk} \psi_i \psi_j \psi_k$. The $C_{ijk}$ is a tensor of couplings (which can be disorder averaged for an analytic solution of the model), and interactions are proportional to the square of these couplings. Further, there is a $\U(1)_R$ symmetry $R=\frac{1}{2} \sum_i [\bar{\psi}_i ,\psi_i ]$ and BPS states for any value of the couplings. This model also permits a generalization in which the supercharge contains $\hat{q}$-body fermi interactions; here we specialized to $\hat{q}=3$ but will later consider general and then large $\hat{q}$.

We could consider breaking $\mathcal{N}=2$ down to $\mathcal{N}=0$ by deforming the Hamiltonian with a term which is not the square of a supercharge (introducing the real parameter $\epsilon$):
\beq\label{eq:introcsykdef}
H \to H + \epsilon \sum_{ijkl} J_{ijkl}\, \bar{\psi}_i \bar{\psi}_j \psi_k \psi_l \, \, \, \, \, \textrm{(naive)}
\eeq
The term we have added is essentially the Hamiltonian of complex SYK \cite{Gu:2019jub,Sachdev:1992fk, kitaevTalks, Sachdev:2015efa}, which has a global $\U(1)$ symmetry but not supersymmetry. However, this deformation does not quite work since the exact IR scaling dimension of the fermion according to the $\mathcal{N}=2$ SYK Hamiltonian is $\Delta = 1/6$, see \cite{Fu:2016vas}, while the scaling dimension according to the complex SYK Hamiltonian is $\Delta = 1/4$. This makes the deformation relevant in the IR and the large $N$ classical solution is affected by the deformation. 

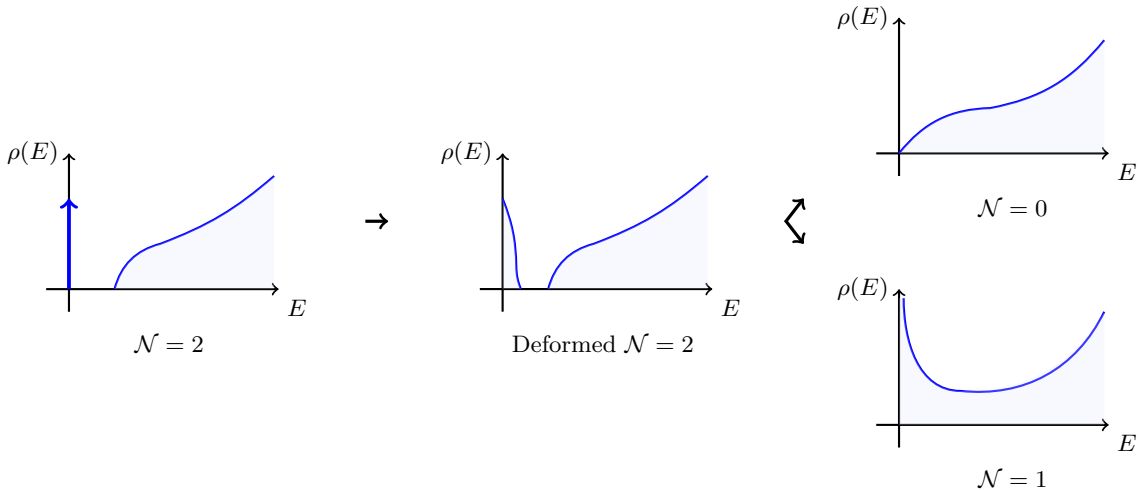
\begin{figure}
    \centering
\begin{tikzpicture}[scale=0.6, baseline={([yshift=0cm]current bounding box.center)}]
\draw[thick,->] (0,-0.5) -- (0,3);
\draw[thick,->] (-0.5,0) -- (4.6,0);
\draw[thick, blue ] (1,0) to[bend left =30] (2,1) to[bend right = 10] (4.5,2.5) ;
%\draw[thick, dashed] (0,1) to[bend left =0] (1,1.1) to[bend right = 10] (4.5,2.6) ;
\node[below] at (5,0) {\footnotesize $E$};
\node at (2.2,0-1.2) {\footnotesize $\mathcal{N}=2$};
\node[left] at (0,3) {\footnotesize $\rho(E)$};
%\node at (2.5,3.5) {\footnotesize Near-BPS};
%\draw[thick] (1,-0.1) -- (1,+0.1);
%\node at (1,-0.5) {\footnotesize $1/\Phi_r$};
%\node[left] at (0,0.5) {\footnotesize $e^{S_0}$};
%\node[left] at (-1.5,1) {$\rho(E) = e^{S_0}\,\delta(E-E_{\text{BPS}}) + \ldots $};
%\node at (-1.8,3) {\small $\langle \rho(E) \rangle_{L\to\infty}$};
\fill[blue!10,nearly transparent]  (1,0) to[bend left =30] (2,1) to[bend right = 10] (4.5,2.5)-- (4.5,0) -- (0,0);
\draw[line width = .5mm, blue, ->] (0,0) -- (0,2) ;
\draw[ ->, very thick ] (6.5,1.5) -- (7,1.5);
\end{tikzpicture}
\hspace{0.3cm}
\begin{tikzpicture}[scale=0.6, baseline={([yshift=0cm]current bounding box.center)}]
\draw[thick,->] (0,-0.5) -- (0,3);
\draw[thick,->] (-0.5,0) -- (4.6,0);
\draw[thick, blue ] (1,0) to[bend left =30] (2,1) to[bend right = 10] (4.5,2.5) ;
%\draw[thick, dashed] (0,1) to[bend left =0] (1,1.1) to[bend right = 10] (4.5,2.6) ;
\node[below] at (5,0) {\footnotesize $E$};
%\node at (2.5,0-1.2) {\footnotesize $J>0$};
\node[left] at (0,3) {\footnotesize $\rho(E)$};
\node at (2.2,0-1.2) {\footnotesize Deformed $\mathcal{N}=2$};
%\node at (2.5,3.5) {\footnotesize Near-BPS};
%\draw[thick] (1,-0.1) -- (1,+0.1);
%\node at (1,-0.5) {\footnotesize $1/\Phi_r$};
%\node[left] at (0,0.5) {\footnotesize $e^{S_0}$};
%\node[left] at (-1.5,1) {$\rho(E) = e^{S_0}\,\delta(E-E_{\text{BPS}}) + \ldots $};
%\node at (-1.8,3) {\small $\langle \rho(E) \rangle_{L\to\infty}$};
\draw[thick, blue ] (0,2) to[bend left =10] (0.3,0.5) to[bend right = 10] (0.4,0) ;
\fill[blue!10,nearly transparent]  (1,0) to[bend left =30] (2,1) to[bend right = 10] (4.5,2.5)-- (4.5,0) -- (0,0);
\fill[blue!10,nearly transparent]  (0,2) to[bend left =10] (0.3,0.5) to[bend right = 10] (0.4,0) -- (0,0) -- (0,2);
%\draw[line width = .5mm, blue, ->] (0,0) -- (0,3) ;
\draw[ ->, very thick ] (6.2,1.5-0.02) -- (6.6,2);
\draw[ ->, very thick ] (6.2,1.5+0.02) -- (6.6,1);
\end{tikzpicture}
\hspace{0.05cm}
\begin{tikzpicture}[scale=0.6, baseline={([yshift=0cm]current bounding box.center)}]
%%%UP
\draw[thick,->] (0,3+-0.5) -- (0,3+3);
\draw[thick,->] (-0.5,3+0) -- (4.6,3+0);
\draw[thick, blue ] (0,3+0) to[bend left =25] (2,3+1) to[bend right = 20] (4.5,3+2.5) ;
\node[below] at (5,3+0) {\footnotesize $E$};
\node[left] at (0,3+3) {\footnotesize $\rho(E)$};
%\draw[thick] (1,3+-0.1) -- (1,3++0.1);
\fill[blue!10,nearly transparent]  (0,3+0) to[bend left =25] (2,3+1) to[bend right = 20] (4.5,3+2.5)-- (4.5,3+0) -- (0,3+0);
\node at (2.5,3-1.2) {\footnotesize $\mathcal{N}=0$};
%%%%DOWN
\draw[thick,->] (0,-3+-0.5) -- (0,-3+3);
\draw[thick,->] (-0.5,-3+0) -- (4.6,-3+0);
\draw[thick, blue ] (0.1,-3+2.8) to[out=-90,in=180] (1.35,-3+0.75) to[bend right = 35] (4.5,-3+2.5) ;
\node[below] at (5,-3+0) {\footnotesize $E$};
\node[left] at (0,-3+3) {\footnotesize $\rho(E)$};
%\draw[thick] (1,-3+-0.1) -- (1,-3++0.1);
\fill[blue!10,nearly transparent]  (0,-3+2.8) -- (0.1,-3+2.8) to[out=-90,in=180] (1.35,-3+0.75) to[bend right = 30] (4.5,-3+2.5)-- (4.5,-3+0) -- (0,-3+0);
\node at (2.5,-3-1.2) {\footnotesize $\mathcal{N}=1$};
\end{tikzpicture}
    \caption{\footnotesize \textbf{Left:} Low-energy spectrum of $\mathcal{N}=2$ SYK  captured by the $\mathcal{N}=2$ Schwarzian theory. \textbf{Middle:} Small deformation turned on. The BPS states will develop a small spread and the gap can also change. \textbf{Right:} Large susy-breaking deformation. Top: deformation that breaks all supercharges, described by $\mathcal{N}=0$ Schwarzian with a $\sqrt{E}$ edge. Bottom: deformation preserves one supercharge, described by the $\mathcal{N}=1$ Schwarzian theory, and $1/\sqrt{E}$ edge. The gap is exponentially small and there are no BPS states.}
    \label{fig:intro}
\end{figure}
\medskip 

To solve the issue mentioned in the previous paragraph one can consider a $\U(1)$ symmetric deformation as in \eqref{eq:introcsykdef} involving $q=2\hat{q}$ fermions. Instead, for reasons that will be clear later, in this paper we propose to break $\mathcal{N}=2$ SYK down to $\mathcal{N}=1$ SYK as a much simpler model that captures similar physics. (In figure \ref{fig:intro} we compare our expectation for a deformation that preserves either one or zero supercharges based on the known Schwarzian spectra.) In both regimes, the fermions have the same scaling dimension in the conformal phase. To do this in the simplest fashion without affecting the large $N$ classical solution in the IR, instead of deforming the interactions, we keep the supercharge $Q$ unmodified in terms of $\psi$ but deform the fermion anticommutation relations. This can be done through the following action:
\beq
\label{eq:introdeformaction}
I = \underbrace{\i \int \d \tau \, \psi^i \partial_\tau \bar{\psi}_i - \int \d \tau \{Q,\bar{Q}\}}_{\mathcal{N}=2~{\rm SYK}} + \underbrace{\epsilon \int \d\tau \Big( \frac{\i}{2} \psi^i \partial_\tau \psi^i - \frac{\i}{2} \bar{\psi}^i \partial_\tau \bar{\psi}^i\Big) }_{\text{deformation}}.
\eeq
In the Hamiltonian description, the information about the first, third and fourth terms is encoded in the algebra, not the Hamiltonian. This seemingly innocent modification of the algebra breaks the $\U(1)$ $R$-symmetry of $\mathcal{N}=2$ SYK and spoils the nilpotency of the supercharge $Q^2 \neq 0 $ when $\epsilon \neq 0$. As we show in section \ref{sectionthree}, the deformation actually preserves one supercharge, the combination $Q+\bar{Q}$, and the Hamiltonian can be written as $H=(Q+\bar{Q})^2$. The existence of a single supersymmetry does not imply the existence of any exact BPS states, and in fact a parametrically small $\epsilon$ lifts the ground state degeneracy in a way we will analyze in this article. However, if one is only interested in the $\U(1)$ breaking aspects of our model and not the supersymmetry, one can consider a version of \eqref{eq:introdeformaction} appropriate for complex SYK, which we analyze at large $q$ in Section \ref{sec:largeqqq}.

\medskip

In the large $N$ limit we can analytically solve the model in terms of the mean field variables-- the two-point functions and the self-energies. This presentation of the model makes it clear that the deformation only affect the so-called UV terms which are neglected in the strict IR limit. This is a crucial feature of our model (other ways to break the ${\rm U}(1)$ symmetry were explored in \cite{Klebanov:2020kck,Baumgartner:2024orz}) and the IR phase is indistinguishable from $\mathcal{N}=2$ SYK and indeed presents the full $\mathcal{N}=2$ super-reparametrizations as symmetries. As we explain later, we verified this explicitly by a numerical solution of the large $N$ Schwinger-Dyson equations. However, it is worth pointing out that the deformation does affect the classical solution by giving a non-zero value to the $\U(1)_R$ breaking correlators such as $\langle \psi^i(\tau) \psi^i(0)\rangle$. The statement is that such classical effects are small in the IR limit. 

\medskip 

The UV terms in the action are crucial in the derivation of the Schwarzian action, since they give rise to the conformal symmetry breaking terms, and we will see that the deformation results in what one might call a ``Higgs-ing" of the $\U(1)_R$ phase mode present in the $\mathcal{N}=2$ Schwarzian theory (as well as one of the two fermionic components). Our result for the low energy large $N$ effective action takes the form:
\beq
I= I_{\mathcal{N}=2~\text{Schwarzian}}+ I_{\rm deformation},
\eeq
with 
\bea
I_{\mathcal{N}=2~\text{Schwarzian}} &=& \frac{N\alpha_S}{J} \int \d\tau \, \Big( - \{ \tan \frac{\pi f}{\beta}, \tau\} + 2 (\partial_\tau a)^2 -4 \partial_\tau \eta \partial_\tau^2 \bar{\eta}+\ldots\Big),\\
I_{\rm deformation}&=& \alpha_{\rm def}NJ \int \d \tau \,\Big( a^2  + \frac{1}{2}(\eta \partial_\tau \eta + \bar{\eta} \partial_\tau \bar{\eta} )-\eta \partial_\tau \bar{\eta} \Big)+\ldots,
\ea
where $f(\tau)$ represents an arbitrary time reparametrization mode, $a(\tau)$ a local $\U(1)$ transformation, and $\eta(\tau)$ a local supersymmetry, see for example \cite{Gu:2019jub}. In the original $\mathcal{N}=2$ Schwarzian theory, the global transformations $a \to a+ \text{const.}$ is a gauge symmetry; this is broken in our deformed theory by giving it an action. The numerical coefficient $\alpha_S$ of the $\mathcal{N}=2$ Schwarzian terms might depend on $\epsilon$ and for $\epsilon=0$ it was computed in \cite{Heydeman:2022lse}. The dots in $I_{\mathcal{N}=2~\text{Schwarzian}}$ represent interactions involving the fermion fields which are uniquely fixed by $\mathcal{N}=2$ supersymmetry. The term $I_{\rm deformation}$ arises from the deformation of our action and $\alpha_{\rm def} \sim O(\epsilon^2)$ at small values of $\epsilon$. We will evaluate the numerical coefficient in the large $\hat{q}$ limit. The dots denote terms that are higher-order in all fields, in particular in the phase mode $a$. The action $I_{\rm deformation}$ preserve one supercharge instead of two (see section \ref{sec:sectionschwarzianwithdefgeneral} for details). If we write the complex fermion $\eta$ in terms of two real fermions $\xi = \eta + \bar{\eta} $  and $\lambda = \eta - \bar{\eta}$, the quadratic action depends only on $\lambda$ and when the effect of the deformation is large, the remaining $\xi$ plays the role of the fermion of the surviving $\mathcal{N}=1$ Schwarzian mode.
\medskip 

Using that $\alpha_{\rm def} \sim O(\epsilon^2)$, the form of the deformation of the Schwarzian action implies that the breaking of $\mathcal{N}=2$ supersymmetry is relevant when $\beta J \gtrsim 1/\epsilon$. Since the gap of $\mathcal{N}=2$ SYK is of order $E_{\rm gap} \sim J/N$, as long as $\epsilon \ll 1/N$ the deformation will mostly affect the BPS states by producing a small spreading in their energy. Indeed we verified this by a simulation of the model and furthermore found the model develops a ground state energy of order $\epsilon^2 e^{-\alpha N}$ with $\alpha$ a numerical coefficient. When $\epsilon \sim 1/N$ the states that were BPS at $\epsilon=0$ mix with the excited states and for $\epsilon \gg 1/N$ the spectrum will take the form dictated by the $\mathcal{N}=1$ Schwarzian, leaving no BPS states and gaps exponentially small in $N$. These features are also reproduced below by a numerical simulation of the model.

\bigskip

The plan of the paper is as follows. In section \ref{sec:N2SYK} we review basic features of $\mathcal{N}=2$ SYK first described in \cite{Fu:2016vas} and further studied in \cite{Berkooz:2020xne,Peng:2020euz,Heydeman:2022lse, lin2023holography,lin2023looking,Turiaci:2023jfa}. In section \ref{sectionthree} we introduce our deformation that breaks $\mathcal{N}=2$ SYK to $\mathcal{N}=1$ SYK. In section \ref{sec:largeqqq} we analyze the model in the large $q$ limit. We also study a $\U(1)$ breaking deformation of complex SYK by modifying the commutation relations, and analyze it in the large $\hat{q}$ limit to illustrate the procedure (the model without supersymmetry is interesting in its own right). As an intermediate step, we also derive the double-scaling limit of the mean-field action for $\mathcal{N}=2$ SYK as a complexified Liouville action. Finally, we conclude with some discussion and future directions in section \ref{sec:discussion}, leaving some technical details for appendices.

\section{Overview of $\mathcal{N}=2$ SYK}\label{sec:N2SYK}

In this section we present a quick overview of the $\mathcal{N}=2$ SYK model. A reader familiar with the model can skip this section, although we will emphasize some aspects that will be important for the new model we introduce in section \ref{sectionthree}.

\subsection{The Hamiltonian formulation}
Much like the complex SYK model \cite{Sachdev:1992fk, georges2001quantum, sachdev2010holographic, Sachdev:2015efa,  Davison:2016ngz, gu2020notes, chowdhury2022sachdev, li2017supersymmetric}%\cite{Sachdev_1993, Georges_2001, Sachdev:2010um,  Sachdev:2015efa, Davison:2016ngz,Gu:2019jub, Chowdhury:2021qpy, Li:2017hdt}
, $\mathcal{N}=2$ SYK consists of $N$ complex fermions which obey the anticommutation relations:
\begin{equation}
\label{eq:oldN=2commutators}
    \{\psi^i, \bar{\psi}_j \} = \delta^i_j \, , \qquad \{\psi^i, \psi^j \} = 0 \, , \qquad \{\bar{\psi}_i, \bar{\psi}_j \} = 0 \, .
\end{equation}
We emphasize here that this is a particular choice we want the fermions to obey which will be consistent with the existence of a global $\U(1)_R$ symmetry; later we will consider other choices in which models with broken $\U(1)_R$ and supersymmetry are most conveniently defined.

A starting point for constructing the Hamiltonian for supersymmetric SYK models involves defining a supercharge $Q$ built from $\hat{q}$-fermion (with $\hat{q}$ odd) interactions\footnote{This theory leads in the IR to a $\mathcal{N}=2$ Schwarzian theory where the $R$-charge of the fermionic reparametrization mode is $\hat{q}$. The Schwarzian theory that arises from black holes in AdS involves fermions with minimal $R$-charge $\hat{q}=1$ \cite{boruch2022bps}. To realize this phase in SYK its necessary to introduce multiple fermions as explained in \cite{Heydeman:2022lse}.}:
\begin{equation}\label{superchargeQ}
    Q= \i^{\frac{\hat{q}-1}{2}} C_{i_1 i_2 \dots i_{\hat{q}}}\psi^{i_1}\psi^{i_2}\ldots \psi^{i_{\hat{q}}},
\end{equation}
where supersymmetry and fermi statistics dictate that the couplings $C_{i_1 i_2 \ldots i_{\hat{q}}}$ are totally antisymmetric. Like all SYK models, we take these couplings to be randomly drawn, and for simplicity we take a Gaussian distribution with variance
\begin{equation}\label{eq:CCJ}
    \langle C_{i_1 i_2 \ldots i_{\hat{q}}} \bar{C}^{i_1 i_2 \ldots i_{\hat{q}}}\rangle \sim J/N^{\hat{q}-1}
\end{equation}
The supercharge, as constructed above, is nilpotent $Q^2=\bar{Q}^2=0$ which follows directly from the anticommutation relation $\{ \psi^i , \psi^j \} =0$. The Hamiltonian can be written as
\begin{equation}
\label{eq:fHamiltonian}
    H= \{ Q, \bar{Q}\} \, .
\end{equation}
Nilpotency of the supercharge is crucial for supersymmetry since it implies that $[Q,H] = [ \bar{Q}, H ] =0$. The derivation is very simple, but since it will be important below we remind the reader
\beq
[Q, H] = [ Q, \{Q, \bar{Q}\} ] = -  [ \bar{Q}, Q^2 ] = 0,~~~~{\rm if}~~~Q^2=0.
\eeq
Thus, nilpotency of the supercharge is an important ingredient in the $\mathcal{N}=2$ algebra\footnote{Importantly, $Q^2=0$ for each memeber of the ensemble. One could investigate theories where the supersymmetry algebra is realized only on average but we will not consider such possibilities.}. For our choice of supercharge \eqref{superchargeQ} the Hamiltonian involves $2(\hat{q}-1)$ body interactions with couplings given by the squares of the random variables $C_{i_1 i_2 \dots i_{\hat{q}}}$. 

%From this point of view, the nonlinear form of the supercharge implies that supersymmetry is realized non-linearly, and we do not need to introduce additional bosons as the superpartners of the fundamental fermions, though supersymmetry still pairs up all states in bose/fermi multiplets. In the path integral formulation discussed below, we will integrate in auxiliary bosons which appear on shell as composites of the fundamental fermions and allow the supersymmetry to be linearly realized.

The $\mathcal{N}=2$ SYK model possesses a $\U(1)_R$ symmetry we denote by $R$ which rotates the phase of the fermions: 
\begin{equation}
    R=\frac{1}{2} \sum_i [\bar{\psi}_i ,\psi_i ]
\end{equation}
This operator is defined such that the quantization of a single fermion produces charges $\pm 1/2$. The algebra implies that $[ R, H]=0$ while $[R , Q ] = \hat{q}\, Q $ and $[ R , \overline{Q} ] = - \hat{q} \, \overline{Q}$ (we will often specialize to the case of $\hat{q}=3$, which results in the standard 4-fermion interaction). $R$ generates an $R$-symmetry because it does not commute with the action of supersymmetry.

Since a motivation for this work is to explain how SYK models with different amounts of supersymmetry are related (as well as the implications for the spectrum of charged near-extremal black holes), we briefly comment on the relationship between complex SYK (CSYK) and $\mathcal{N}=2$ SYK. In terms of the Hamiltonian, we have very schematically
\begin{equation}
    H_{\rm CSYK} \sim \mathcal{J} \bar{\psi}^{\hat{q}-1}\psi^{\hat{q}-1} \, , \qquad     H_{\mathcal{N}=2} \sim C\bar{C} \bar{\psi}^{\hat{q}-1}\psi^{\hat{q}-1} \, ,
\end{equation}
where we defined the complex model to have $2(\hat{q}-1)$-body interactions. From this naive microscopic point of view, the only difference between the models is the statistics of the random couplings via $\mathcal{J}\sim |C|^2$ due to the fact that the supersymmetric Hamiltonian is a square. Despite similarities, the low energy physics is very different. Already from supersymmetry, the spectrum of the latter model is guaranteed to be positive definite in energy, and further the existence of a pair of supercharges implies there are special BPS states which have exactly $E=0$ as a consequence of being in short representations of the $1d$ $\mathcal{N}=2$ Poincare group. We now discuss these features explicitly.

\bigskip

\begin{figure}[t!]
    \centering
    \includegraphics[scale=0.26]{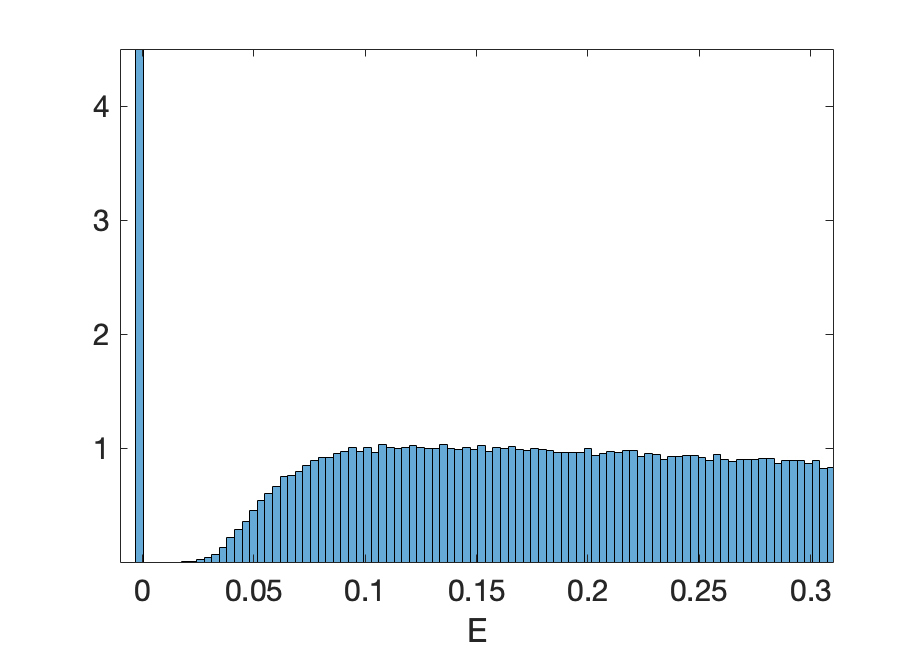}
    \caption{Histogram for the $\mathcal{N}=2$ SYK spectrum averaged over 10000 samples with $N=10$ fermions, as a function of energy in units with $J=1$. The number of states at zero energy is 324.}
    \label{fig:SYK_undef}
\end{figure}

Since the generator of $R$-symmetry commutes with the Hamiltonian, the Hilbert space $\mathcal{H}$ can be decomposed as $\mathcal{H}=\bigoplus_k\mathcal{H}_k$, where $\mathcal{H}_k$ is a subspace that includes quantum states of $R$-charge $k$. In the space $\mathcal{H}_k$, there are two kinds of multiplets: BPS and non-BPS supermultiplets. BPS states are zero-energy states annihilated by the supercharges $Q$ and $\bar{Q}$. Non-BPS states that are raised or lowered to other $R$-charge sectors by the supercharges become components of $(k,k+\hat{q})$ supermultiplets. 
In the exact diagonalization of $\mathcal{N}=2$ SYK model, for generic values of the parameters BPS states appear with a huge degeneracy at $E=0$. For an individual member of the ensemble, non-BPS states have a dense density of states (with energy spacings exponentially small in $N$) starting from a non-zero energy. We show a result from a simulation with $N=10$ in figure \ref{fig:SYK_undef}. The number of ground states is exponential in $N$ (whose justification we review below) and the gap is order $1/N$. The low energy spectrum is described by the $\mathcal{N}=2$ Schwarzian theory 
 \cite{Stanford:2019vob, Mertens:2017mtv, Heydeman:2020hhw,Turiaci:2023jfa} which we also review below.

\subsection*{Counting BPS states with the refined Witten index}

Of interest in this section is what happens exactly at $E=0$, in which there is an exponentially (in $N$) large number of exactly degenerate ground states. As in the discussion of the previous section, this is ultimately due to the unbroken extended $\mathcal{N}=2$ supersymmetry, which guarantees the existence of short multiplets; if $|\chi \rangle$ is a BPS state, then it follows that $    Q|\chi \rangle = \bar{Q}|\chi \rangle = 0$ which implies that $H |\chi \rangle = 0 \, $. What is not obvious from the numerical test alone is that the microscopic degeneracy $d_N(E=0)$ satisfies
\begin{align}
    \log ( d_N(E=0) ) \propto N \, .
\end{align}
The existence of these states is a consequence of supersymmetry, so we attempt an analytic counting by considering a supersymmetric partition function. The naive appropriate quantity to consider is the Witten index \cite{witten1982constraints} which only receives contributions from ground states. The index is defined to be the ordinary thermal partition function with the additional insertion of $(-1)^{\sf F}= e^{-\i \pi R}$, with ${\sf F}$ the fermion number; $\mathcal{I}_N \equiv {\rm Tr}\left[ (-1)^{\sf F} \right] $. The index is invariant under continuous deformations of the theory, including the effect of turning on the random couplings; therefore it can be computed in the free theory.

Actually, as explained in \cite{Fu:2016vas}, for generic odd $\hat{q} \geq 3$, the Witten index of $\mathcal{N}=2$ SYK vanishes due to the cancellation between bosonic and fermionic ground states carrying fractional $\U(1)_R$ charges. In the free theory all fermions decouple, and the individual free fermion index is $\mathcal{I}_1 = (1- 1) = 0$. This is in particular a case where the microscopic index does not match the microscopic degeneracy, which is allowed since the index only lower bounds the degeneracy and there is no argument analogous to \cite{dabholkar2011supersymmetric} to guarantee these quantities will agree.

To remedy this problem following \cite{Fu:2016vas}, we use a discrete $\mathbb{Z}_{\hat{q}}$ global symmetry under which the fermions transform with a $\hat{q}$-th root of unity. We may turn on a discrete chemical potential $r$ for this symmetry without breaking supersymmetry because the $\mathbb{Z}_{\hat{q}} \in U(1)_R$ leaves the supercharge invariant. This allows one to define a non-vanishing refined Witten index, which in this model is given by
\beq\label{eq:wittenindex}
\mathcal{I}_N(r) \equiv {\rm Tr}\left[ (-1)^{\sf F} e^{2 \pi \i r R}\right] =  e^{-\i\pi  N (\frac{r}{\hat{q}}-\frac{1}{2})}(1-e^{\frac{2\pi \i r}{\hat{q}}})^N = \Big(2 \sin \frac{\pi r}{\hat{q}}\Big)^N,
\eeq
This quantity grows with $N$ in the expected way and may be compared with the results from exact diagonalization. 

In the application to $\mathcal{N}=2$ SYK models where we explicitly break supersymmetry, the $\U(1)_R$ symmetry is lost, and thus also the $\mathbb{Z}_{\hat{q}}$ refinement. If we break all supersymmetry, there is nothing protected to compute and no notion of BPS states. If we break only half the supersymmetry, the situation is similar to $\mathcal{N}=1$ SYK where the only refinement possible commuting with supersymmetry is the $(-1)^{\sf F}$ operator itself. The Witten index of this model is again zero due to bose/fermi cancellation in the free fermion partition function. Thus without any additional deformations or global symmetries, we conclude that models which look like $\mathcal{N}=2$ SYK with some supersymmetry broken have vanishing Witten index. 

Strictly speaking, with less supersymmetry it is difficult to compute features of the density of states exactly using the microscopic theory at strong coupling. Thus the situation is similar to higher-dimensional supersymmetric quantum field theories in which some protected quantities may be evaluated exactly, but not the spectrum itself at generic coupling. However, due to their disordered couplings and solvable large $N$ limit, we may learn much more about the $\mathcal{N}=2$ SYK model and its deformations using the path integral approach and its formulation as a mean field theory.

\subsection{The path integral formulation and large $N$ equations}
To find the mean field action and large $N$ solution, we will pass to the Lagrangian formulation and integrate out the disorder. This is most conveniently done in $\mathcal{N}=2$ superspace with coordinates $Z=(\tau, \theta, \bar{\theta})$. Since the steps we outline are standard, we refer the reader to \cite{Fu:2016vas,Heydeman:2022lse, biggs2023supersymmetric} for more details. 

On superspace, supersymmetry transformations are implemented in terms of a complex Grassmann parameter $\eta$ by
\begin{align}
    \tau \rightarrow \tau + \theta \bar{\eta} + \bar{\theta}\eta \, , \qquad
    \theta \rightarrow \theta + \eta \, , \qquad
    \bar{\theta} \rightarrow \bar{\theta} + \bar{\eta} \, , \label{eq:N2RigidSusy}
\end{align}
while the Grassmann coordinates are $\U(1)_R$ charged and transform via a phase $a$.
\begin{align}
    \theta \rightarrow e^{\i a} \theta \, , \qquad
    \bar{\theta} \rightarrow e^{-\i a} \bar{\theta} \, ,
    \label{eq:fThetaCharge}
\end{align}
The action of the model may be written in terms of $N$ chiral fermionic superfields $\Psi^i(\tau, \theta, \bar{\theta})$:
\begin{align}
    \Psi^i(\tau, \theta, \bar{\theta}) = \psi^i(\tau+\theta\bar{\theta})+\sqrt{2}~\theta b^i(\tau) \, ,
\end{align}
where $b^i$ will end up being the auxiliary superpartner bosons. The chirality condition is such that the supercovariant derivative, $D_{\bar{\theta}} \equiv(\partial_{\bar{\theta}}+\theta \partial_\tau)$, annihilates the superfield,
\begin{align}
\label{eq:chiralsuperfield}
D_{\bar{\theta}} \Psi^i(\tau, \theta, \bar{\theta}) = 0 \, .
\end{align}
Since all fields in the problem are complex, we will also introduce anti-chiral superfields defined by conjugation.

The superspace formalism manifests linearized supersymmetry in the Lagrangian density, which we take to be
\begin{equation}
    \mathcal{L}=\frac{1}{2}\!\int \! \d^2\theta \, \bar{\Psi}_i  \Psi^i+\frac{\i^{\frac{{\hat q}-1}{2}}}{\sqrt{2}\hat{q}}\int \! \d\theta \, C_{i_1 i_2 \ldots i_{\hat q}} \Psi^{i_1} \Psi^{i_2} \ldots \Psi^{i_{\hat q}}+\frac{\i^{\frac{{\hat q}-1}{2}}}{\sqrt{2}\hat{q}}\int\!  \d\bar{\theta} \, \bar{C}^{i_1 i_2 \ldots i_{\hat q}} \bar{\Psi}_{i_1} \bar{\Psi}_{i_2} \ldots \bar{\Psi}_{i_{\hat q}}.
    \label{eq:fLagrangian}
\end{equation}
By expanding this expression in components and integrating over the time $\tau$, one obtains the action
\beq\label{eq:sec2action}
I =  \int \d \tau\,  \psi^i \partial_\tau \bar{\psi}_i -  \bar{b}_i b^i + \i^{\frac{\hat{q}-1}{2}} \, C_{i_1 i_2 \dots i_{\hat{q}}} b^{i_1} \psi^{i_2} \dots \psi^{i_{\hat{q}}}  + {\rm h.c.} 
\eeq
Here, we see explicitly the boson $b^i$ and its conjugate $\bar{b}_i$ are auxiliary and may be integrated out. The Hamiltonian \eqref{eq:fHamiltonian} appears by changing formulations and promoting fields to operators with the commutation relations \eqref{eq:oldN=2commutators}. This Lagrangian definition evades the normal ordering ambiguity which is present in evaluating $\{ Q, \bar{Q}\}$.

The mean field action at large $N$ is obtained by integrating in auxiliary bilocal fields which are anti-chiral-chiral superfields. The first such field is $\mathcal{G}_{\Psi\bar{\Psi}}(Z_1,Z_2) = \frac{1}{N} \langle \Psi^i(Z_1)  \bar{\Psi}_i(Z_2) \rangle $ which is interpreted as the superspace 2-point function. This has components
\beq
\label{eq:bilocal2pts}
G_{\psi\bar{\psi}}(\tau_1,\tau_2) \equiv  \frac{1}{N}  \langle  \psi^i(\tau_1)  \bar{\psi}_i(\tau_2) \rangle,~~~~G_{b\bar{b}}(\tau_1,\tau_2) \equiv  \frac{1}{N}  \langle b^i (\tau_1) \bar{b}_i(\tau_2)\rangle.
\eeq
There are analogous fields $G_{\bar{\psi}\psi}$ and $G_{\bar{b}b}$ which are obtained by conjugation. Further, in principle there are other bilinear 2-point functions possible. The first set is
\beq
G_{\bar{\psi}b}(\tau_1,\tau_2) = G_{\psi \bar{b}}(\tau_1,\tau_2) = G_{b \bar{\psi}}(\tau_1,\tau_2) = G_{ \bar{b}\psi}(\tau_1,\tau_2) = 0 \, \quad \textrm{(on-shell)}.
\eeq
Because these are mixed bose/fermi correlators, they have the total statistics of fermions and vanish as classical solutions of the mean field action. They are important however for analyzing the low temperature fluctuations of the model around the classical solution \cite{Heydeman:2022lse}. There is another set of ``diagonal'' 2-point functions one can in principle write down, but here they vanish due to the unbroken $\U(1)_R$ and they do not need to be included in the mean field action:
\beq
G_{\psi \psi}(\tau_1,\tau_2) = G_{ \bar{\psi} \bar{\psi}}(\tau_1,\tau_2) = G_{b b}(\tau_1,\tau_2) = G_{ \bar{b}\bar{b}}(\tau_1,\tau_2) = 0 \, \quad \textrm{($\U(1)$ symmetry)}.
\eeq
Later, the supersymmetry breaking deformation we introduce will also break the $\U(1)_R$ and it will be possible to have a nontrivial solution for these diagonal fields, as in section \ref{section:Low_Energy_Limit}. (In that section we will use $\mathcal{N}=1$ superspace formalism which is similar to the one discussed here but with only one Grassman coordinate $\theta$.) In total, the nonzero bilocal superfield has the expansion
\begin{eqnarray}\label{eqn:defsuperG}
    \mathcal{G}_{\Psi\bar{\Psi}}(Z_1,Z_2)&=&G_{\psi\bar{\psi} }(\tau_1-\theta_1\bar{\theta}_1,\tau_2+\theta_2\bar{\theta}_2)+2 \theta_1\bar{\theta}_2 G_{b\bar{b}}(\tau_1,\tau_2) \, ,
\end{eqnarray}
along with the conjugate field in which barred and unbarred indices are exchanged. The second bilocal to be integrated-in is a chiral-anti-chiral superfield  $\Sigma_{\bar{\Psi}\Psi}(Z_1,Z_2)$, with bosonic components
\begin{eqnarray}
    \Sigma_{\bar{\Psi}\Psi}(Z_1,Z_2)&\supset&\Sigma_{b\bar{b}}(\tau_1+\theta_1\bar{\theta}_1,\tau_2-\theta_2\bar{\theta}_2) +2 \theta_1\bar{\theta}_2 \Sigma_{\psi\bar\psi}(\tau_1,\tau_2)\, ,
\end{eqnarray}
which appears as the Lagrange multiplier enforcing \eqref{eq:bilocal2pts} as well as its fermionic counterparts. The chirality is chosen to be opposite of the 2-point functions. The superfield expansion is analogous to $\mathcal{G}$ above, and contains $\Sigma_{b\bar{b}}(\tau_1,\tau_2)$ as the top component along with $\Sigma_{ \psi\bar{\psi}}(\tau_1,\tau_2)$.

With the above definitions, we will proceed to write down the large $N$ equations for $\mathcal{G}$ and $\Sigma$, which are the Schwinger-Dyson equations for the model. We will not write out all the intermediate steps for the original $\mathcal{N}=2$ model but will instead do so in the next section when we introduce a supersymmetry breaking deformation. Roughly, starting from \eqref{eq:fLagrangian} it is easy to integrate out the couplings assuming a Gaussian distribution. Next, the bilocals are integrated-in using the identity, 
\begin{equation}\label{eq:SSN2}
    1 = \int D \mathcal{G}_{\Psi \bar{\Psi}} D\Sigma_{\bar{\Psi}\Psi}\exp{\left (-N \int \d Z_1 \d \bar{Z}_2 \, \Sigma_{\bar{\Psi}\Psi}(Z_1, Z_2)\left(\mathcal{G}_{\Psi \bar{\Psi}} (Z_1, Z_2) - \frac{1}{N}\Psi^i(Z_1)\bar{\Psi}_i(Z_2)\right) \right )} \, ,
\end{equation}
In this expression and below, we will use the the superspace integrations and delta functions:
\begin{equation}
        \int \d Z \equiv\int \d\tau \d\theta \, , \quad \int \d\bar{Z} \equiv \int \d\tau \d\bar{\theta} \, , \quad \delta(Z_1-Z_2)\equiv (\bar{\theta}_1-\bar{\theta}_2)\delta(\tau_1-\theta_1\bar{\theta}_1-\tau_2+\theta_2\bar{\theta}_2)
\end{equation}
Including the mean fields in the theory deforms the action $I$ by terms which couple them to the fundamental fermions.

The fundamental bosons and fermions couple in the deformed action to the $\Sigma$ fields, and they appear effectively as self energies. The interaction terms may be rewritten in terms of the $G$ variables only, so the total action is quadratic in the fundamental fields, which may then be integrated out. The resulting action only depends on $\mathcal{G}$, $\Sigma$ and becomes weakly coupled as $N$ becomes large; this means we may use the classical equations. Using the procedure we have outlined, one may derive these Schwinger-Dyson equations in superspace:
\bea
\Sigma_{\bar{\Psi}\Psi}(Z_2, Z_3) &=&\frac{1}{2} J~\mathcal{G}_{\bar{\Psi}\Psi}(Z_2, Z_3)^{\hat{q}-1},\nonumber\\
\frac{1}{2}D_{\bar\theta_3} \mathcal{G}_{\Psi\bar{\Psi}}(Z_1, Z_3) &+& \int \d Z_2 \, \mathcal{G}_{\Psi\bar{\Psi}}(Z_1, Z_2) ~\Sigma_{ \bar{\Psi}\Psi}(Z_3, Z_2) = \delta(Z_1-Z_3) \, . \label{eqn:SDinsuperspace} 
\ea

For practical purposes, and also because we plan on deforming this model to break some supersymmetry, we can also write the component version of these equations: 
\begin{align}
    \Sigma_{\psi\bar\psi}(\tau_1,\tau_2)&=J(\hat{q}-1)G_{b\bar{b}}(\tau_1,\tau_2)[G_{\psi\bar{\psi} }(\tau_1,\tau_2)]^{\hat{q}-2} \, , \label{eq:N2conf1} \\
    \Sigma_{b\bar{b}}(\tau_1,\tau_2)&=J[G_{\psi\bar{\psi}}(\tau_1,\tau_2)]^{\hat{q}-1} \, , \label{eq:N2conf2}\\
    G_{\psi\bar{\psi}}^{-1}(i\omega)&= -\i\omega-\Sigma_{\bar\psi\psi} \, ,\label{eq:N2conf3} \\
    G_{b\bar{b}}^{-1}(\i\omega)&=-1-\Sigma_{\bar{b}b} \, . \label{eq:N2conf4}
\end{align}
These equations define the normalization for the coupling variance $J$ implicit in \eqref{eq:CCJ}.

\subsection*{$\mathcal{N}=2$ Super-reparametrizations }
While the full solution of the Schwinger-Dyson equations (\ref{eq:N2conf1}-\ref{eq:N2conf4}) is impossible to find analytically, in the infrared limit the equations are approximately conformal. By infrared, we really mean at characteristic time scales for which $|J\tau|\gg1$ compared to the scale set by $J$. This means we can drop the $-\i \omega$ and ``$-1$'' terms in the equations for $G^{-1}$. Just like in the ordinary SYK model, in this limit the Schwinger-Dyson equations develop a new set of reparametrization symmetries. When the UV $\mathcal{N}=2$ Poincare symmetry is unbroken, we expect in the IR conformal limit to actually have $\mathcal{N}=2$ super-reparametrization symmetry. This is essentially the main argument given in \cite{Fu:2016vas}. To see the super-reparametrization symmetry, we again use the superspace form of the Schwinger-Dyson equations \eqref{eqn:SDinsuperspace} , but drop the UV term: 
\beq
\label{eq:IRFuSD}
\int \d\bar{Z}_2 ~ \mathcal{G}_{\Psi\bar{\Psi}}(Z_1,Z_2)~ \left[\frac{J}{2}\mathcal{G}_{\Psi\bar{\Psi}}(Z_3,Z_2)^{\hat{q}-1}\right] = \delta(Z_1-Z_3).
\eeq

This superspace Schwinger-Dyson equation is invariant under super-reparametrizations; these are transformations of the superspace coordinates $Z=(\tau,\theta,\bar{\theta}) \to Z'=(\tau',\theta',\bar{\theta}')$ which satisfy the additional conditions
\bea
&&D_\theta \bar{\theta}'=0 \qquad D_\theta \tau' = \bar{\theta}' D_\theta \theta',\\
&&D_{\bar{\theta}} \theta'=0 \qquad D_{\bar{\theta}} \tau' = \theta' D_{\bar{\theta}} \bar{\theta}'.
\ea
We can find an explicit parametrization for the set of functional degrees of freedom satisfying the above constraints. For the bosons, we have a time reparametrization $f(\tau)$ and a $\U(1)_R$ phase reparametrization $a(\tau)$. These act as:
\begin{align}
\tau' &= f(\tau) \, , \quad \theta' = e^{\i a(\tau)}\sqrt{\partial_\tau f(\tau)} \theta \, \quad \bar{\theta}' = e^{-\i a(\tau)}\sqrt{\partial_\tau f(\tau)} \bar{\theta} \, .
\end{align}
For the fermionic components, we have a pair of fermionic partner of parametrizations $\eta(\tau)$, $\bar{\eta}(\tau)$ which act as:
\begin{align}
\tau' &= \tau + \theta \bar{\eta}(\tau) + \bar{\theta} \eta(\tau) \, , \quad \theta' = \theta + \eta(\tau + \theta \bar{\theta}) \, \quad \bar{\theta}' = \bar{\theta} + \bar{\eta}(\tau - \theta \bar{\theta}) \label{eqn:fermionreparam}\, .
\end{align}
By composing these, we can obtain general $\mathcal{N}=2$ reparametrizations. The statement that these act as symmetries of the IR theory is the statement that \eqref{eq:IRFuSD} is invariant under the superspace generalization of a time reparametrization of a bilocal field: 
\beq\label{eqn:FususytransfIR}
\mathcal{G}_{\Psi\bar{\Psi}}(Z_1,Z_2) \to (D_{\bar\theta_1} \bar\theta'_1)^{\frac{1}{\hat{q}}}~(D_{\theta_2} \theta'_2)^{\frac{1}{\hat{q}}}~\mathcal{G}_{\Psi\bar{\Psi}}(Z'_1,Z'_2).
\eeq
Using the definitions we have given, one can see this is indeed a symmetry of \eqref{eq:IRFuSD}. This is the first hint that the $\mathcal{N}=2$ SYK model at low temperatures is dual to a supergravity theory. However, this statement must be corrected by two important points. 

The first is that, while the dynamical equations have this symmetry, the conformal solutions to these equations do not. From this point of view, the reparametrization invariance can take us from one set of conformal solutions to another set. This perspective is particularly useful for obtaining the finite temperature correlation functions from the $T=0$ ones. It is also related to the fact that a choice of AdS$_2$ solution of our candidate bulk theory breaks (spontaneously) the full diffeomorphism symmetry to the global conformal group which are the isometries of AdS$_2$. In this case, the supergroup of isometries is $\SU(1,1|1)$, which is the same as the set of isometries of the near horizon region of (for instance) black holes in AdS$_5 \times$ S$^5$, which approach a fibration of AdS$_2 \times$ S$^3 \times$ S$^5$ \cite{Sinha:2006sh}.

The second point is that the reparametrization symmetry and associated conformal solution are valid only in the strict $T=0$ limit. At finite temperatures, we really should include the correction by the UV terms, and once restored, the ($\mathcal{N}=2$) reparametrization invariance is explicitly broken. It is argued in \cite{Fu:2016vas} (though not strictly speaking derived), that the emergent collective mode controlling this explicit breaking is the $\mathcal{N}=2$ super-Schwarzian mode; this has component fields $f(\tau), \, a(\tau), \, \eta(\tau), \, \bar{\eta}(\tau)$ and the effective action is an $\mathcal{N}=2$ supersymmetric generalization of the Schwarzian derivative. This nearly conformal analysis is how one makes contact with super-JT gravity, and ultimately supersymmetric near-BPS black holes.

In the context of the present paper, we want to extend these ideas to SYK models in which we can parametrically break supersymmetry in the action but still have a nearly-conformal limit. We will consider a deformation which breaks the $\mathcal{N}=2$ symmetry to an $\mathcal{N}=1$ subgroup. The naive expectation is that such a deformation would lead to conformal Schwinger-Dyson equations that only have $\mathcal{N}=1$ super-reparametrization invariance with corresponding conformal solutions. This means that the $a(\tau)$ mode and some linear combination of the $\eta(\tau), \, \bar{\eta}(\tau)$ would completely disappear from the theory. Such a model would for many practical purposes look like the ordinary $\mathcal{N}=1$ SYK model. Instead, we will find a deformation of $\mathcal{N}=2$ SYK which has no $\U(1)_R$ symmetry in the UV, yet the conformal equations still satisfy \eqref{eqn:FususytransfIR}. This means in our model discussed in Section \ref{sectionthree}, it is the UV terms that are responsible for the supersymmetry breaking.

\subsection*{Conformal Solution}
Before presenting the conformal solution we would like to comment on the possibility of a background $R$-charge. In \cite{Heydeman:2022lse}, the above Schwinger-Dyson equations (\ref{eq:N2conf1}-\ref{eq:N2conf4}) were studied for a general value of the background charge. From the point of view of the UV Lagrangian, a general background charge corresponds to the inclusion of a chemical potential term $\mu$ for the $\U(1)_R$ symmetry. While both the fundamental fermions and bosons are charged under this symmetry, the bosons are non-dynamical, so the only modification to the Schwinger-Dyson equations is
\begin{align}
    G_{\psi\bar{\psi}}^{-1}(\i\omega)&= -\i\omega + \mu -\Sigma_{\psi\bar\psi} \,  \quad \textrm{(Background $R$-charge.)}
\end{align}
Because an explicit chemical potential corresponds to turning on a background gauge field for $\U(1)_R$, such a background field will generically break supersymmetry as $R$ and $Q$ do not commute. Equivalently, the superspace variables $\theta, \bar{\theta}$ carry explicit $R$-charge so we cannot write the action in superspace. However, in this work we consider a different kind of supersymmetry breaking in which we directly deform the action by terms which are not $\mathcal{N}=2$ supersymmetric. Once this deformation is added, there is no longer any $\U(1)_R$ symmetry and thus no case with background charge to analyze. 

 The ansatz of conformal solutions of $\mathcal{N}=2$ SYK at zero temperature and zero $R$-charge \cite{Fu:2016vas,Heydeman:2022lse} is: 
\bea
G_{\psi \bar{\psi}}(\tau) &=& \frac{g_{\psi\bar\psi}}{|\tau|^{2\Delta}} \, {\rm sgn}(\tau),\hspace{2.5cm}\,G_{b\bar{b}}(\tau) = \frac{g_{b\bar{b}}}{|\tau|^{2 \Delta_b}},\nonumber\\
\Sigma_{\psi \bar\psi}(\tau) &=& \frac{(1-2\Delta)\tan \pi \Delta}{2\pi g_{\psi\bar{\psi}}}\frac{{\rm sgn}(\tau)}{|\tau|^{2-2\Delta}} ,~~~\Sigma_{b\bar{b}} (\tau) = \frac{(2\Delta_b-1)\cot \pi \Delta_b}{2\pi g_{b\bar{b}}} \frac{1}{|\tau|^{2-2\Delta_b}},\nonumber
\ea
To solve (\ref{eq:N2conf1}-\ref{eq:N2conf4}), we use the above ansatz and match the scaling dimensions, which gives
\beq
(\hat{q}-1) \Delta + \Delta_b = 1 \, ,
\eeq
which is related to the fact that we want the interaction term  $C_{i_1\ldots i_{\hat{q}}} b^{i_1} \psi^{i_2}\ldots \psi^{i_{\hat{q}}}$ to be marginal. In seeking relevant deformations of $\mathcal{N}=2$ SYK, the most interesting ones are those in which new interaction terms we add are of the same scaling dimension as this term, so that both types of interactions are nontrivial in the infrared. We will comment more on this point later. 

Further constraints come from the consideration of unbroken supersymmetry. As we have discussed, in the absence of background $R$-charge supersymmetry is preserved. This implies the relation $G_{b\bar{b}}(\tau_1,\tau_2)= - \partial_{\tau_1} G_{\psi \bar{\psi}}(\tau_1-\tau_2)$, so the scaling dimensions can be solved completely,
\beq
\label{eq:susydimconstraint}
\Delta= \frac{1}{2\hat{q}},~~~\Delta_b = \frac{1}{2} + \frac{1}{2\hat{q}} \, .
\eeq
To find the full conformal solution, we may also attempt to match the prefactors $g_{\psi\bar\psi}$, $g_{b\bar{b}}$, leading to,
\beq
(1-2\Delta)\tan  \pi \Delta = (\hat{q}-1)(1-2(\hat{q}-1)\Delta) \tan \pi (\hat{q}-1)\Delta \, ,
\eeq
which is solved by \eqref{eq:susydimconstraint}. Unlike other kinds of SYK models, supersymmetry further implies the prefactors are also related through $g_{b \bar{b}} = 2 \Delta g_{\psi \bar{\psi}}$, leading to the analytical solution
\beq
g_{\psi \bar{\psi}} = \left(\frac{1}{2\pi J} \tan \frac{\pi}{\hat{q}} \right)^{1/\hat{q}},~~~~g_{b \bar{b}} =\frac{1}{\hat{q}} \left(\frac{1}{2\pi J} \tan \frac{\pi}{\hat{q}}  \right)^{1/\hat{q}} \, .
\eeq

This completes our review of the original $\mathcal{N}=2$ SYK model. We have argued that the UV model of interacting fermions has unbroken $\mathcal{N}=2$ supersymmetry generated by a $\hat{q}$-th order in fermions supercharge $Q$ (as well as its conjugate $\bar{Q})$. This implies a conserved $\U(1)_R$ symmetry in which the fundamental fermions have charge $\frac{1}{\hat{q}}$. At large $N$, the model can be reduced to the Schwinger-Dyson equations for the bilocal 2-point functions and self energies. In the conformal limit, these equations enjoy a large supersymmetric extension of time reparametrization symmetry, and they permit an exact analytical solution in which the bilinears take the form of conformal 2-point functions with large anomalous dimensions of the form $\Delta = \frac{1}{2 \hat{q}}$. Going away from the conformal limit breaks the super-reparametrization symmetry, and the reparametrization modes $f(\tau), \, a(\tau), \, \eta(\tau), \, \bar{\eta}(\tau)$ acquire an effective action determined by symmetry to be the $\mathcal{N}=2$ super-Schwarzian derivative. For later reference the precise form of the action is
\bea
I &=& -\frac{\alpha_S N}{J} \int \d \tau \d \theta \d \bar{\theta}\,\Big\{ \frac{\partial_\tau \bar{D} \bar{\theta}'}{\bar{D} \bar{\theta}'} -\frac{\partial_\tau D \theta'}{D\theta'} - \frac{2\partial_\tau \theta' \partial_\tau \bar{\theta}'}{(\bar{D} \bar{\theta}')(D\theta')}\Big\}\\
&=&-\frac{\alpha_S N}{J} \int \d\tau \Big( \{ \tan \frac{\pi f(\tau)}{\beta},\tau\} - 2(\partial_\tau a)^2 + {\rm fermions} \Big). \label{eq:N2EffectiveAction}
\ea
where $\alpha_S$ is a numerical constant. For $\hat{q}=3$ the coefficient was computed in \cite{Heydeman:2022lse} which obtained $\alpha_S \approx 0.00842$. The value in the large $\hat{q}$ limit was obtained in \cite{Peng:2020euz}.

In Section \ref{sectionthree}, we will repeat the above analysis for a new version of supersymmetric SYK in which the UV model is only invariant under the single combination of supercharges $Q+\bar{Q}$, with no conserved $\U(1)$ charge $R$. Surprisingly, the infrared limit of this model is identical to our above discussion, and we analyze the consequences of this, including a new near-conformal window and the breaking of supersymmetry by the UV terms.

\section{A deformation of the $\mathcal{N}=2$ SYK model} \label{sectionthree}

In this section we introduce a new deformation of the $\mathcal{N}=2$ SYK model which breaks one of the two supercharges and reduces it to a quantum mechanics theory with $\mathcal{N}=1$ supersymmetry. The advantage of the specific one introduced here is that it will simplify the IR limit of the theory considerably. We will first describe this deformation, then we will analyze the modified Schwinger-Dyson equations, the conformal limit, and present some numerical results.

\subsection{Breaking both supersymmetry and $\U(1)_R$}\label{sec:susybreakdef}

The deformation we want to implement can be described in multiple ways. Let us begin by writing down the action for the $\mathcal{N}=2$ SYK model reviewed in the previous section, in terms of $N$ complex fermion fields $\psi^i(\tau)$, with $i=1,\ldots, N$. It is given by
\beq\label{eq:Itotal}
I = I_{\rm UV} + I_{\rm IR},
\eeq
where we defined the UV/IR pieces of the action, for obvious reasons, as 
\beq\label{eq:IUVIR}
I_{\rm UV} = \frac{1}{2} \int \d \tau\, \left[ \psi^i \partial_\tau \bar{\psi}_i -  \bar{b}_i b^i \right]+ {\rm h.c.},~~~I_{\rm IR} =  \int \d \tau\,  \i \sum_{j<k} C_{ijk} b^i \psi^j \psi^k  + {\rm h.c.} 
\eeq
We look at the fermions first. In a Hamiltonian formulation, the UV part of the action $I_{\rm UV}$ is first order in time-derivatives of the fermions and therefore encodes the anticommutation relations between the operators $\psi^i$ acting on the Hilbert space, i.e.  $\{ \psi^i , \bar{\psi}_j \} = \delta^i_j$ and $\{ \psi^i, \psi^j\} = \{ \bar{\psi}_i, \bar{\psi}_j \} = 0$. The bosonic field $b^i$ is non-propagating and can immediately be integrated out, resulting in a Hamiltonian for the fermions given by
\beq\label{eq:Qs3}
H=(Q+ \overline{Q})^2,~~~~Q = \i \sum_{i<j<k}C_{ijk} \psi^i \psi^j \psi^k.
\eeq
The supercharge, as constructed above, is nilpotent; $Q^2=\overline{Q}^2=0$, which follows directly from the anticommutation relation $\{ \psi^i , \psi^j \} =0$. 
As we explain in the previous section, nilpotency of the supercharge $Q^2$ implies supersymmetry $[Q,H]=[\bar{Q},H]=0$ using $H=\{Q,\bar{Q}\}$. 
The theory also has an $R$-symmetry generated by $R=\frac{1}{2} \sum_i [\overline{\psi}_i ,\psi^i ]$ and the fermion algebra implies that $[ R, H]=0$ while $[R , Q ] = \hq\, Q $ and $[ R , \overline{Q} ] = - \hq \, \overline{Q}$ (for the specific choice in \eqref{eq:Qs3}, $\hq=3$).

An obvious way to break the supersymmetry is to modify the Hamiltonian, deforming the component $I_{\rm IR}$ of the action. This type of deformations do not have a clear gravity interpretation since they can affect the conformal phase at low energies. Instead, we will keep the same Hamiltonian, but modify the way the fermionic operators act on the Hilbert space. In other words, we will deform $I_{\rm UV}$. The new $I_{\rm IR}$ is unchanged and the UV part becomes
\beq\label{eq:def}
I_{\rm UV} \to I_{\rm UV} +   \epsilon \int \d \tau \frac{1}{2} \left[ \i\psi^i \partial_\tau \psi^i - \i b^i b^i +{\rm h.c.}\right].
\eeq
Here, $\epsilon$ is an overall free coefficient which controls the strength of the deformation. In the Hamiltonian formulation, the supercharge still takes the form \eqref{eq:Qs3} but now the fermionic operators satisfy the defining relations
\begin{equation}\label{eq:PsiCRe}
\begin{split}
    \{\psi^i,\bar\psi_j\}=\frac{1}{1-\epsilon^2}\delta^i_j,~~
    \{\psi^i,\psi^j\}=\frac{\i\epsilon}{1-\epsilon^2}\delta^i_j,~~
    \{\bar\psi_i,\bar\psi_j\}=\frac{-\i\epsilon}{1-\epsilon^2}\delta^i_j.
\end{split}
\end{equation}
These expressions show that the limit $\epsilon \to \pm 1$ is singular, and this happens for an elementary reason. While the deformed UV action \eqref{eq:def} looks well-behaved for any $\epsilon$, it is easy to see that when $\epsilon\to \pm 1$ there is both a fermionic and a bosonic zero-mode\footnote{For example it is easy to see that the UV action is independent of fluctuations that satisfy $b^i = \pm \overline{b}_i$ where the sign depends on the sign of $\epsilon$. Similarly, the fermion zero-mode corresponds to fluctuations satisfying $\psi^i = \pm \overline{\psi}_i$. These are the zero-modes mentioned in the text.}. Since the purpose of this article is to study small deformations that break supersymmetry, we will restrict to the range  $-1< \epsilon < 1$.  

Naively, one could think that since after the deformation one can still construct the operators $Q$, $\overline{Q}$ and quantize the theory with a Hamiltonian $H= (Q+\overline{Q})^2$,  $\mathcal{N}=2$ supersymmetry is unbroken. Nevertheless, this conclusion is too fast, since the non-vanishing anticommutator $\{ \psi^i, \psi^j \} \neq 0$ implies that now $Q^2 \neq 0$ and $\overline{Q}^2 \neq 0$. Therefore it is no longer true that the two supercharges commute with the Hamiltonian. Moreover, it is clear from \eqref{eq:def} that the deformation breaks the ${\rm U}(1)$ $R$-symmetry, since the fermionic algebra now implies $[R,H]\neq 0$.

The discussion above might suggest that supersymmetry is completely broken once the deformation is turned on. This is again too fast. Since the relation $H=  (Q+\overline{Q})^2$ is still valid, it is evident that
\beq
\label{eq:deformedQ}
\mathcal{Q} \equiv Q + \overline{Q} ,~~~\Rightarrow~~~[\mathcal{Q}, H] =0.
\eeq
Therefore, when the deformation is turned on, the $\mathcal{N}=2$ supersymmetry is broken down to $\mathcal{N}=1$ supersymmetry. 

We introduced our deformation in the UV term, but this is really a basis-dependent statement. Since the $\psi$ fermions satisfy an unconventional anticommutation relation, under our deformation, it is possible to introduce a new basis that puts them in the canonical form. Define the fermions $\chi^i , \overline{\chi}^i$ such that
\beq\label{eq:psivseta}
\psi^i =  \frac{\cos \frac{\tilde{\epsilon}}{2}}{\cos \tilde{\epsilon}}\, \chi^i +\i \frac{\sin \frac{\tilde{\epsilon}}{2}}{\cos \tilde{\epsilon}}\, \overline{\chi}^i,~~~~~ \overline{\psi}{}^i =  \frac{\cos \frac{\tilde{\epsilon}}{2}}{\cos \tilde{\epsilon}}\, \overline{\chi}^i -\i \frac{\sin \frac{\tilde{\epsilon}}{2}}{\cos \tilde{\epsilon}}\, \chi^i ,
\eeq
where we defined 
\beq
\epsilon = \sin (\tilde{\epsilon}),~~~-\pi/2<\tilde{\epsilon}<\pi/2.
\eeq
This is a useful parametrization since it naturally restricts to $|\epsilon|<1$. One can check that the relations \eqref{eq:PsiCRe} imply the canonical anticommutators $\{ \chi^i, \overline{\chi}^j\} = \delta^{ij}$ and $\{ \chi^i , \chi^j \} = \{ \overline{\chi}^i , \overline{\chi}^j\} = 0$ for the $\chi$ fermions. It is evident that as $\epsilon \to 0$, or equivalently $\tilde{\epsilon} \to 0$, we get $\psi^i \to \chi^i$ and the model reduces back to $\mathcal{N}=2$ SYK. In the $\chi$-variables, the breaking of supersymmetry becomes more transparent. In this presentation, the supercharge \eqref{eq:Qs3} should be written in terms of $\chi$ and $\overline{\chi}$, and the right hand side will involve both of them, implying again that $Q^2 \neq 0 $ for generic values of $C_{ijk}$. We cannot  pick a basis where, simultaneously, $Q$ is holomorphic in the fermion fields and fermions have canonical commutation relations.

As reviewed in section \ref{sec:N2SYK}, one can define a refinement of the Witten index of $\mathcal{N}=2$ SYK that guarantees the presence of a large number of ground states growing exponentially with $N$ at large $N$. The refinement exploits the $\mathbb{Z}_{\hq} \subset {\rm U}(1)$ subgroup of the $R$-symmetry that commutes with the supercharge. Since the deformation introduced here breaks the ${\rm U}(1)$ $R$-symmetry, it also breaks the $\mathbb{Z}_{\hq}$ symmetry. In the absence of a refinement, the Witten index of $N$ complex fermions exactly vanishes. Therefore, the remaining supersymmetry no longer protects the zero-energy ground states, and we will later verify this in a simulation of the model.

\subsection*{Superspace formulation}
Since our deformation breaks $\mathcal{N}=2$ supersymmetry down to $\mathcal{N}=1$, there should be a superspace formulation of our theory. In order to do this, introduce the $\mathcal{N}=1$ super-line parametrized by $(\tau, \theta)$. We can define a set of $N$ complex superfields $\Psi^i(\tau, \theta)$ which are expanded as 
\beq
\Psi^i(\tau,\theta) = \psi^i(\tau) + \theta \, b^i(\tau),~~~\overline{\Psi}_i(\tau,\theta) = \bar{\psi}_i (\tau) + \theta \, \bar{b}_i(\tau).
\eeq
Notice that complex conjugation acts on $\psi$ and $b$ but does not act on $\theta$ since we are using $\mathcal{N}=1$ superspace. Therefore, even though we work with complex superfields, this is \textbf{not} the same as $\mathcal{N}=2$ superspace presented in section \ref{sec:N2SYK} since there is only one odd coordinate $\theta$\footnote{It might be possible to think of this formulation as a fermionic dimensional reduction from the $\mathcal{N}=2$ superline $(\tau, \theta, \overline{\theta})$ down to the $\mathcal{N}=1$ superline $(\tau,\theta)$ upon restricting $\theta = \overline{\theta}$. We did not find this to be a useful perspective when evaluating the action and therefore will not use this language.}. Using this formulation we can write the $\mathcal{N}=2$ SYK action as
\beq
I =  \int \d \tau \int \d \theta\,\left[-\frac{1}{2}\Psi^i D_\theta \overline{\Psi}_i + \frac{\i}{3} C_{ijk} \Psi^i \Psi^j \Psi^k + {\rm h.c.} \right],
\eeq
where $D_\theta = \partial_\theta + \theta \partial_\tau$. After performing the $\theta$ integral we reproduce the action written in \eqref{eq:Itotal} and \eqref{eq:IUVIR}. The superspace action can be put in a more standard form by introducing the real and imaginary parts of the superfield $\Psi$ as real $\mathcal{N}=1$ superfields, but working in the complex formulation will be more practical.

In the $\mathcal{N}=1$ superspace formulation, the deformation introduced in the beginning of this section can be written as 
\beq\label{eq:defsupers}
I = \int \d \tau \int \d \theta\,\left[-\frac{1}{2}\Psi^i D_\theta \overline{\Psi}_i+ \frac{\i}{3} C_{ijk} \Psi^i \Psi^j \Psi^k - \frac{\i}{2} \epsilon \Psi^i D_\theta \Psi^i \right]+ {\rm h.c.}.
\eeq
We can verify that this deformation is the same as the one introduced earlier in this section. Indeed, we can check by explicit integration over the fermionic coordinate $\theta$ that
\beq
-\i  \int \d \theta \, \Psi^i D_\theta \Psi^i + {\rm h.c.} = \i \, \psi^i \partial_\tau \psi^i -\i \, b^i b^i + {\rm h.c.}
\eeq
Inserting this in the action \eqref{eq:defsupers} we see that it reproduces the deformation \eqref{eq:def}. 

\bigskip

The analysis above shows once again that the deformation \eqref{eq:def} does preserve $\mathcal{N}=1$ supersymmetry. It is instructive to elaborate on this more by describing more precisely the symmetry transformations. The original $\mathcal{N}=2$ SYK supersymmetry transformation (which our model has when $\epsilon=0$) is given by 
\begin{equation}\label{transformation}
\begin{split}
\delta_Q\bar\psi_i=\varepsilon_1\bar{b}_i,\quad\delta_Q b^i=\varepsilon_1\partial_\tau\psi^i,\quad\delta_Q\psi^i=\delta_Q\bar{b}_i=0\\
\delta_{\bar{Q}}\psi^i=\varepsilon_2b^i,\quad\delta_{\bar{Q}}\bar{b}_i=\varepsilon_2\partial_\tau\bar\psi_i,\quad\delta_{\bar{Q}}\bar\psi_i=\delta_{\bar{Q}} b^i=0
\end{split}
\end{equation}
These transformations are a symmetry of the action when $\epsilon=0$ but not when the deformation is turned on. By the discussion presented above in the Hamiltonian formulation, we expect $\mathcal{Q} = Q + \overline{Q}$ to be a conserved supercharge. Therefore $\delta_\mathcal{Q} = \delta_Q + \delta_{\overline{Q}}$ should be a symmetry. Explicitly, the transformation is 
\begin{equation}\label{susyq=1}
\begin{split}
\delta_\mathcal{Q}\bar\psi^i=\varepsilon\bar{b}^i,\quad\delta_\mathcal{Q} b^i=\varepsilon\partial_\tau\psi^i,\quad\delta_\mathcal{Q}\psi^i=\varepsilon b^i,\quad\delta_\mathcal{Q}\bar{b}^i=\varepsilon\partial_\tau\bar\psi^i
\end{split}
\end{equation}
To find the supercharge, it is useful to promote the supersymmetry to a local transformation, meaning that the parameter $\varepsilon(\tau)$ depends on time as in the Noether trick. The change in the action for such a local transformation is given by
\begin{equation}
\begin{split}
    \delta I
    =&\int \d \tau \, \partial_\tau\left(-\frac{1}{2}\sum_{i}\varepsilon(\tau)(\psi^i\bar{b}_i+\bar{\psi}_ib^i)-\i \epsilon\sum_i \varepsilon(\tau)(\psi^ib^i-\bar\psi_i\bar{b}_i)\right)\\
    &+\int \d \tau \,\varepsilon(\tau)\partial_\tau\sum_{i}\sum_{j<k}\left(\frac{\i}{3}C_{ijk}\psi^i\psi^j\psi^k+\frac{\i}{3}\bar{C}^{ijk}\bar\psi_i\bar{\psi}_j\bar{\psi}_k\right)
\end{split}
\end{equation}
The first line is a total derivative and therefore can be ignored. The second line is also a total derivative for global supersymmetry transformations satisfying $\partial_\tau \varepsilon =0$. This shows that indeed $\delta_{\mathcal{Q}}$ is a symmetry of the theory independent of the deformation parameter. Finally, from the second line we can read of the  conserved fermionic charge
\begin{equation}
\begin{split}
    \mathcal{Q}=\sum_{i<j<k}(\i C_{ijk}\psi^i\psi^j\psi^k+\i\bar{C}^{ijk}\bar\psi_i\bar{\psi}_j\bar{\psi}_k).
\end{split}
\end{equation}
This preserved supercharge coincides with $Q+\overline{Q}$, with $Q$ given in equation \eqref{eq:Qs3}, and therefore the analysis is consistent with the Hamiltonian considerations in the beginning of this section.

As a consistency check of the formalism, we verify whether the anticommutator $\{\mathcal{Q},\psi^i\}$ is equal to the boson $b^i$ since it has to be consistent with the supersymmetry transformation $\delta_{\mathcal{Q}}\psi^i=b^i$. We begin analyzing the action. The equations of motion that come from the scalar fields are given by
\begin{equation}
\begin{split}
    b^i=&\i\sum_{ j<k}\bar{C}^{ijk}\bar{\psi}_j\bar{\psi}_k+\i\epsilon\bar{b}_i=\frac{1}{1-\epsilon^2}\sum_{ j<k}(\i\bar{C}^{ijk}\bar{\psi}_j\bar{\psi}_k-\epsilon C_{ijk}\psi^j\psi^k), \\
    \bar{b}_i=&\i\sum_{j<k}C_{ijk}\psi^j\psi^k-\i\epsilon b^i=\frac{1}{1-\epsilon^2}\sum_{j<k}(\i C_{ijk}\psi^j\psi^k+\epsilon\bar{C}^{ijk}\bar{\psi}_j\bar{\psi}_k).
\end{split}
\end{equation}
The first equality in the first and second line is what directly come from the equations of motion for the scalars. The second equality in the the first and second line arise from the explicit solution of the equations of motion. When deriving the anticommutator $\{\mathcal{Q},\psi^i\}$, one needs to be careful with the anticommutations because the $\psi$ fermions are not canonical when the deformation is turned on. The result is 
\begin{equation}
\begin{split}
    \{\mathcal{Q},\psi^l\}
    =&\frac{1}{1-\epsilon^2}\sum_{i<j}(\i\bar{C}^{ijl}\bar\psi_i\bar{\psi}_j-\epsilon C_{ijl}\psi^i\psi^j)
\end{split}
\end{equation}
Thus, the relation $\{\mathcal{Q},\psi^i\}=b^i$ holds, along with its complex conjugate counterpart.  This is a consistency check that the $\mathcal{N}=1$ supersymmetry is still preserved by the model.  

\begin{comment}
\bigskip

To conclude the presentation of our model we would like to make the following observation, even though it will not be very important for what comes next. We can apply the change of variables in \eqref{eq:psivseta} at the level of the complex $\mathcal{N}=1$ superfields introduced in this section, namely
\beq
\Psi^i = \frac{\cos \frac{\tilde{\epsilon}}{2}}{\cos \tilde{\epsilon}} \,\upeta^i +\i  \frac{\sin\frac{\tilde{\epsilon}}{2}}{\cos \tilde{\epsilon}}\,\overline{\upeta}^i,
\eeq
where we remind the reader that $\epsilon = \sin(\tilde{\epsilon})$. $\upeta^i$ denotes a complex superfield with lowest component given by $\eta^i (\tau)$. In terms of the superfield $\upeta$ the action becomes 
\bea
I &=&\frac{1}{2} \int \d \tau \int \d \theta\, \Big\{ \,\upeta^i D_\theta \overline{ \upeta}{}^i   \nonumber\\
&&\hspace{-0.5cm} +  \frac{\i}{3} \frac{\cos^3 \frac{\tilde{\epsilon}}{2}}{\cos^3 \tilde{\epsilon}} C_{ijk} \Big(\upeta^i + \i \tan \frac{\tilde{\epsilon}}{2} \,\overline{\upeta}^i\Big)
\Big(\upeta^j + \i \tan \frac{\tilde{\epsilon}}{2} \,\overline{\upeta}^j\Big)
\Big(\upeta^k + \i \tan \frac{\tilde{\epsilon}}{2} \, \overline{\upeta}^k\Big) \Big\}+{\rm h.c.}\label{eq:defetass}
\ea
We see that our deformation can be put either purely in the kinetic term, as in \eqref{eq:defsupers}, or purely in the interactions in \eqref{eq:defetass}. One of the main points is that the discussion of the IR conformal phase simplifies considerably if we put the deformation in the kinetic term instead. 

\end{comment}

\subsection{Exact diagonalization}\label{sec:ED}
To gain an intuition for how the deformation affects the large $N$ spectrum of the theory, it is instructive to directly diagonalize the Hamiltonian with randomly chosen $C_{ijk}$ numerically. Fermions satisfying the canonical commutation relations are easy to represent in a Hilbert space made of $N$ qubits as $\psi_i = Z_1 \otimes \ldots \otimes Z_{i-1} \otimes (X_i+\i Y_i)/2$ where $\{X_i,Y_i,Z_i\}$ are the Pauli matrices acting on the $i$-th qubit. To carry out the exact diagonalization of the model, it is therefore easier to work in the  $\chi$-basis since these fermions can be represented in the Hilbert space in the standard fashion. The deformation is then translated to a modification of the Hamiltonian which we write as $H=\mathcal{Q}^2$ with an $\epsilon$-dependent supercharge $\mathcal{Q}$. 

\begin{figure}[t!]
    \centering
    \includegraphics[width=\textwidth]{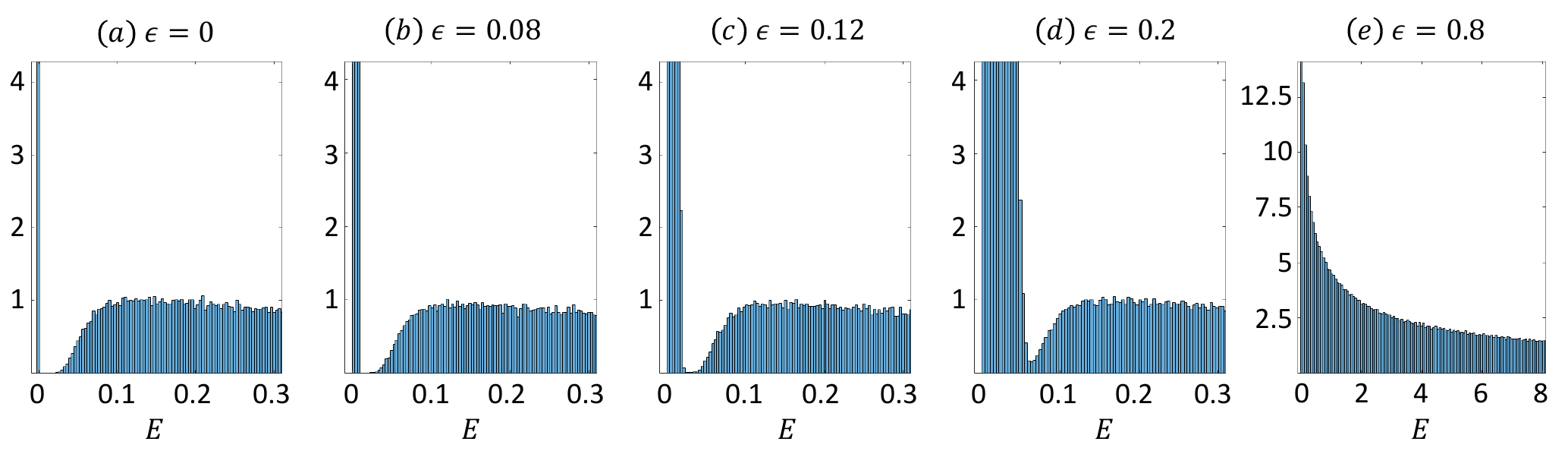}
    \caption{\footnotesize  The histogram of eigenvalues of the Hamiltonian with different values of $\epsilon$ and fixed $J=2$, $N=10$. Each histogram includes 2000 repetitions of the Hamiltonian with different random couplings. The y-axis is the number of states in each bin divided by the number of repetitions, i.e., the average number of states. In (a), the number of states at $E=0$ is 324. Similarly, in the rest of the figures, the number of states near $E=0$ exceeds the y-axis shown in the figure, and the total number of states below the gap stays at 324 before the gap closes. }\label{spectrum_SUSY_with_deformation}
\end{figure}
The result for a spectrum after averaging over disorder is shown in figure \ref{spectrum_SUSY_with_deformation}. Each spectrum includes energy eigenvalues in 2000 repetitions of exact diagonalization, each with renewed random couplings. When $\epsilon$ is zero, we reproduce the spectrum for the $\mathcal{N}=2$ SYK model. The degeneracy of the BPS states at $E=0$ is protected by $\mathcal{N}=2$ supersymmetry. The spectrum has a gap between the degenerate BPS states and the non-BPS continuum (note we are working with even $N$ so there is no time reversal anomaly). The spectrum is similar to the situation in supergravity corresponding to the spectrum of near-BPS black holes, except that the density of states in the continuum stops increasing as energy becomes bigger. The reason is that the $\mathcal{N}=2$ supersymmetric SYK model has emergent conformal symmetry in the IR limit, which makes the Schwarzian/gravity description a good approximation at low energies, but this conformal phase is not a good description at higher energies. 

When the deformation is turned on, half of the $\mathcal{N}=2$ supersymmetry is broken, and the BPS states spread out. This process is captured as $\epsilon$ increases by a small amount, where the BPS states spread towards the continuum until the gap closes around $\epsilon=0.15$. This implies that for small enough $\epsilon$ we can still distinguish between the states that would-be-BPS at $\epsilon=0$ and the continuum of non-BPS states. We will see later that the maximal value of $\epsilon$ where the BPS states can be isolated scales as $\epsilon \sim O(1)/N$. When $\epsilon$ becomes bigger, e.g., $\epsilon=0.8$, the spectrum is more smoothed over and becomes the typical spectrum for $\mathcal{N}=1$. Additionally, we can count the number of states below the gap before it closes when turning on the deformation. The number stays the same, although these nearly zero-energy states gain small amounts of energy with increasing $\epsilon$.

\begin{figure}[t!]
    \centering
\includegraphics[width=0.45\textwidth]{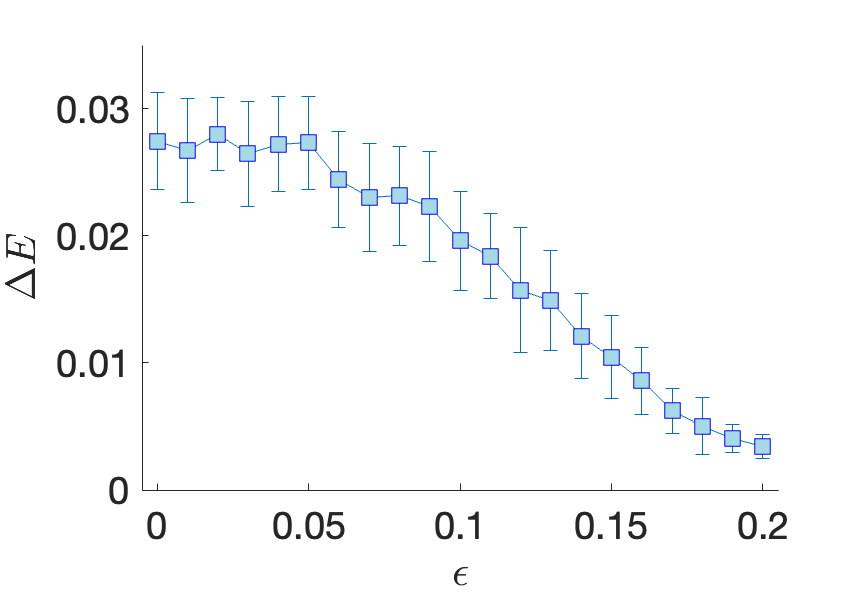}
    \caption{\footnotesize  Gap size $\Delta E$ decreases with $\epsilon$. At each $\epsilon$, we present the average and the standard deviation of gap sizes among 50 spectra. Each spectrum included 100 repetitions of exact diagonalization of different Hamiltonians. The fermion number is $N=10$. }\label{gap_statistics}
\end{figure}

Figure \ref{spectrum_SUSY_with_deformation} shows the gap size decreases with increasing $\epsilon$. To give a quantitative illustration, we collect the gap size from multiple spectra at each $\epsilon$ and calculate the average and standard deviation of gap sizes, shown in figure \ref{gap_statistics}. The size of the Hamiltonian grows exponentially with the number of fermion as $2^N\times 2^N$, so in practice $N$ must be confined to relatively small values in the numerical calculation. This low value of $N$ causes the energy eigenvalues in the continuum to be relatively sparse. To compensate for this, we include many repetitions over the disorder in one spectrum so that the energy eigenvalues in the continuum become dense enough, and the average energy difference between the nearest energy eigenvalues is much smaller than the gap size. This is the premise for distinguishing the gap in a spectrum. On top of that, we collect multiple spectra at each $\epsilon$ to extract the statistical properties of the gap size among them, since every Hamiltonian is generated with random couplings. From figure \ref{gap_statistics}, we see that the gap starts from around 0.027 and decreases to around 0.003 at $\epsilon\approx0.19$, around which the gap closes.
\begin{figure}[h]
    \centering
    \includegraphics[width=0.45\textwidth]{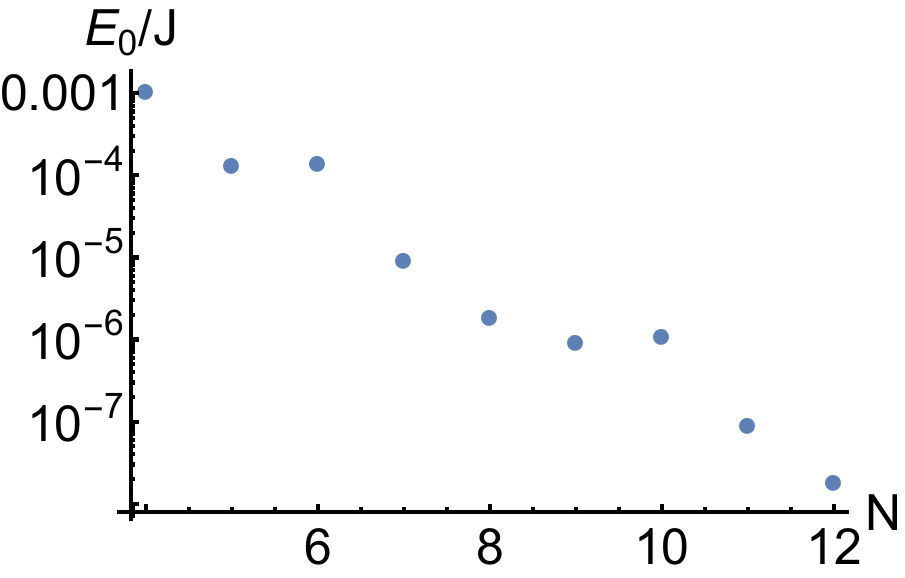}
    ~~~
    \includegraphics[width=0.45\textwidth]{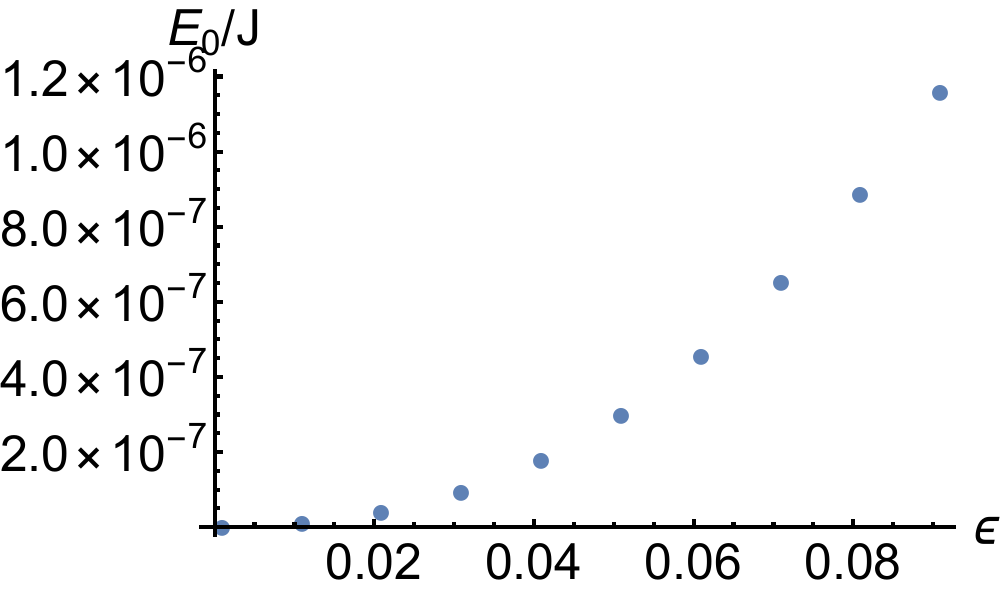}
    \caption{\footnotesize \emph{Left:} Plot of $E_0/J$ as a function of $N$, with $\epsilon=0.1$. The vertical axis is logarithmic to illustrate that the behavior is well approximated by $E_0 \sim e^{- \alpha N}$ with $\alpha\sim 1.8$. Other values of $\epsilon$ were consistent with this scaling. \emph{Right:} Plot of $E_0/J$ as a function of $\epsilon$ for $N=10$. Other values of $N$ show similar behavior. The behavior is well approximated by $E_0 \sim \epsilon^2$.}\label{fig:GSE}
\end{figure}

We have seen in the simulation that as the deformation is turned on, the ground states of $\mathcal{N}=2$ SYK spread out. In the undeformed theory these states are protected, but not after one supercharge is broken. It is reasonable then to ask whether, at finite $\epsilon$, supersymmetry is spontaneously broken or not. In other words, does the true ground state of the deformed model has zero energy or not? We show evidence in figure \ref{fig:GSE}  from exact diagonalization that the ground state energy is non-zero for the deformed theory and its value is exponentially small in $N$. The later feature is similar to the result for $\mathcal{N}=1$ SYK found in \cite{Fu:2016vas}. To leading order in a small deformations we find $E_0/J \sim \epsilon^2 e^{-\alpha N}$, with $\alpha \sim 1.8$. The coefficient $\alpha$ we found is similar to the value for $\mathcal{N}=1$ SYK, although this might be a coincidence.

\subsection{Mean field action and large $N$ equations}
In this section we average over couplings and derive the mean field action for the deformed $\mathcal{N}=2$ SYK model. In the large $N$ limit this theory becomes classical and the equations of motion are the Schwinger-Dyson equations. The derivation has become quite standard by now, and for this reason we point out only the main steps. 

The first step consists in integrating out the couplings with a Gaussian weight. This leads to an action that is bi-local in time and can be conveniently written in superspace as
\beq
I = I_1 + I_2,
\eeq
where
\bea
I_1 &=&-\frac{1}{2}\int \d \tau \int \d \theta\,\left[\Psi^i D_\theta \overline{\Psi}_i  +  \i \epsilon \Psi^i D_\theta \Psi^i \right]+ {\rm h.c.}\nonumber\\
I_2&=&-\frac{1}{9}\int \d \tau_1 \d \theta_1 \int \d \tau_2 \d \theta_2 \, \overline{\Psi}_i \overline{\Psi}_j \overline{\Psi}_k(\tau_1,\theta_1) \, \langle \overline{C}_{ijk} C^{i'j'k'}\rangle\, \Psi^{i'} \Psi^{j'}  \Psi^{k'}(\tau_2,\theta_2) .  
\ea
We can insert now the variance of the couplings which is proportional to $J$ and also includes a sum over multiple contractions due to the antisymmetry property of $C_{ijk}$. Written in components, the result for $I_2$ becomes
\beq\label{eq:intac}
I_2 = JN\int \d\tau_1\int \d\tau_2\Big(\frac{1}{N}\sum_i\bar{b}_i(\tau_1)b^i(\tau_2)\Big)\Big(\frac{1}{N}\sum_i\bar{\psi}_i(\tau_1)\psi^i(\tau_2)\Big)^2 + \ldots
\eeq
The dots denote terms which are subleading in the large $N$ limit. 

The next step is to introduce both the two-point functions and the self-energies. We denote the two-point functions as
\bea
    G_{\psi\bar\psi}(\tau_1,\tau_2)&=&\frac{1}{N}\sum_i\psi^i(\tau_1)\bar\psi_i(\tau_2),~~
   G_{b\bar{b}}(\tau_1,\tau_2)=\frac{1}{N}\sum_ib^i(\tau_1)\bar{b}_i(\tau_2),\label{eq:Gdef1}\\
    G_{\psi\psi}(\tau_1,\tau_2)&=&\frac{1}{N}\sum_i\psi^i(\tau_1)\psi_i(\tau_2),~~~ G_{bb}(\tau_1,\tau_2)=\frac{1}{N}\sum_i b^i(\tau_1)b_i(\tau_2).\label{eq:Gdef2}
\ea
These are all the independent two-point functions which have a non-vanishing expectation value in the large $N$ limit. First, correlators involving a boson and a fermion vanish since the two-point function such as $G_{b\psi}$ would be a fermionic field in the mean field action. Second, other non-vanishing correlators such as $G_{\bar{\psi}\bar{\psi}}$ are determined by $G_{\psi\psi}$ by complex conjugation and similarly for $G_{\psi \bar{\psi}}$. In the case of $\mathcal{N}=2$ SYK we can further drop correlators with a non-vanishing $R$-charge such as $G_{\psi\psi}$ or $G_{bb}$. In our case, when the deformation is turned on, the ${\rm U}(1)$ $R$-symmetry is broken and there is no reason for those correlators to vanish. We can summarize these correlators in two bi-local superfields
\bea
\mathcal{G}_{\Psi\overline{\Psi}}(\uptau_1,\uptau_2) &=& \frac{1}{N} \sum_i \Psi^i (\uptau_1) \overline{\Psi}{}^i (\uptau_2) = G_{\psi \bar{\psi}}(\tau_1,\tau_2) + \theta_2 \theta_1 G_{b \bar{b}} (\tau_1,\tau_2) + \ldots,\\
\mathcal{G}_{\Psi\Psi}(\uptau_1,\uptau_2) &=& \frac{1}{N} \sum_i \Psi^i (\uptau_1) \Psi^i (\uptau_2)=G_{\psi \psi}(\tau_1,\tau_2) + \theta_2 \theta_1 G_{b b} (\tau_1,\tau_2) + \ldots,
\ea
where we define the superline coordinate $\uptau_i = (\tau_i,\theta_i)$ for $i=1,2$. The dots denote fermionic two-point functions which as we mentioned above we can ignore to leading order in the large $N$ limit. 

Having introduced the bilocal fields $G$, we next introduce the self-energies which appear as Lagrange multipliers imposing the relations \eqref{eq:Gdef1}  and \eqref{eq:Gdef2} between the fields $G$ and the fundamental fermionic/bosonic degrees of freedom. As superfields the self-energies are
\bea
\Sigma_{\Psi \bar{\Psi}}(\uptau_1, \uptau_2) &=& \Sigma_{b\bar{b}} (\tau_1,\tau_2) + \theta_2\theta_1 \Sigma_{\psi \bar{\psi}}(\tau_1,\tau_2) + \ldots,\\
\Sigma_{\Psi \Psi}(\uptau_1, \uptau_2) &=& \Sigma_{bb} (\tau_1,\tau_2) + \theta_2\theta_1 \Sigma_{\psi \psi}(\tau_1,\tau_2) + \ldots.
\ea
In the large $N$ limit we can neglect the fermionic components such as $\Sigma_{b \psi}$ for example. When introducing them as Lagrange multipliers, they add a term in the action given by
\bea
I &\to& I + N \int \d \uptau_1 \int \d \uptau_2 \,\Sigma_{\Psi \bar{\Psi}}(\uptau_1,\uptau_2) \Big\{\mathcal{G}_{\Psi \bar{\Psi}}(\uptau_1,\uptau_2) - \sum_i \frac{1}{N} \Psi^i(\uptau_1) \bar{\Psi}^i(\uptau_2)\Big\}\nonumber\\
&&+ N \int \d \uptau_1 \int \d \uptau_2\, \Sigma_{\Psi \Psi}(\uptau_1,\uptau_2) \Big\{\mathcal{G}_{\Psi \Psi}(\uptau_1,\uptau_2) -\sum_i \frac{1}{N} \Psi^i(\uptau_1) \Psi(\uptau_2)^i\Big\},
\ea
where the measure of integration is $\d \uptau \equiv \d \tau \d \theta$. (The first line can be compared with its counterpart in $\mathcal{N}=2$ superspace, equation \eqref{eq:SSN2}.) We now have an action that depends on the mean fields $\mathcal{G}$ and $\Sigma$ and also on the fundamental degrees of freedom $\Psi$. As explained e.g. in \cite{Kitaev:2017awl} we can replace $\Psi$ for $\mathcal{G}$ in the interaction term \eqref{eq:intac} which results in an action quadratic in the fundamental fields $\Psi$. The last step is then to integrate out the fields $\Psi$, leaving us with the final form of the mean field action involving the bilocal fields alone. 

To motivate the form of the final answer, let us analyze in some detail the form of the fermion and boson kinetic terms. In superspace language it has the form
\beq
I \supset \int \d \uptau_1 \int \d \uptau_2\sum_i \begin{bmatrix}\Psi^i&\bar\Psi_i\end{bmatrix}
\begin{bmatrix}-\frac{\i\epsilon}{2}\upsigma-\Sigma_{\Psi\Psi}&-\frac{1}{2}\upsigma-\frac{1}{2}\Sigma_{\Psi\bar\Psi}\\-\frac{1}{2}\upsigma-\frac{1}{2}\Sigma_{\Psi\bar\Psi}&\frac{\i\epsilon}{2}\upsigma-\Sigma_{\bar\Psi\bar\Psi}\end{bmatrix}
\begin{bmatrix}\Psi^i\\\bar\Psi_i\end{bmatrix}, \label{eq:quadkern}
\eeq
where we introduce the differential operator $\upsigma(\uptau_1,\uptau_2) = \delta(\uptau_1-\uptau_2) D_{\theta_2}$ and the delta function is, explicitly, $\delta(\uptau_1 - \uptau_2) = \delta (\tau_1-\tau_2) (\theta_1-\theta_2)$. The coordinate dependence, suppressed above to avoid cluttering, is as follows: The row superfields on the left are evaluated at $\uptau_1$ and the column superfields on the right at $\uptau_2$. The self energies in the quadratic kernel are evaluated at $(\uptau_1,\uptau_2)$. After integrating out the fields $\Psi$, what remains is the Berezinian of the quadratic kernel in \eqref{eq:quadkern}.

Since the previous description of the kinetic part of the mean field action might not be particularly transparent, we will present it in components as well. Expanding the superfields that appear in \eqref{eq:quadkern} and integrating over the fermionic coordinates we get
\begin{equation}
\begin{split}
    I
    \supset& \int \d \tau_1 \int \d \tau_2 \\
    &\sum_i\Big\{ \begin{bmatrix}\psi^i&\bar\psi_i\end{bmatrix}(\tau_1)\begin{bmatrix}-\frac{\i\epsilon}{2}\sigma_f-\Sigma_{\psi\psi}&-\frac{1}{2}\sigma_f-\frac{1}{2}\Sigma_{\psi\bar\psi}\\-\frac{1}{2}\sigma_f-\frac{1}{2}\Sigma_{\psi\bar\psi}&\frac{\i\epsilon}{2}\sigma_f-\Sigma_{\bar\psi\bar\psi}\end{bmatrix}(\tau_1,\tau_2)\begin{bmatrix}\psi^i\\\bar\psi_i\end{bmatrix}(\tau_2)\\
&+\begin{bmatrix}b^i&\bar{b}_i\end{bmatrix}(\tau_1)\begin{bmatrix}-\frac{\i\epsilon}{2}\sigma_b-\Sigma_{bb}&-\frac{1}{2}\sigma_b-\frac{1}{2}\Sigma_{b\bar{b}}\\-\frac{1}{2}\sigma_b-\frac{1}{2}\Sigma_{b\bar{b}}&\frac{\i\epsilon}{2}\sigma_b-\Sigma_{\bar{b}\bar{b}}\end{bmatrix}(\tau_1,\tau_2)\begin{bmatrix}b^i\\\bar{b}_i\end{bmatrix}(\tau_2) \Big\}\\
\end{split}
\end{equation}
where we defined $\sigma_f(\tau_1,\tau_2 ) = \delta'(\tau_1-\tau_2)$ and $\sigma_b(\tau_1,\tau_2)  = \delta(\tau_1-\tau_2)$. Finally, to make the discussion in this section as explicit as possible, we show the final form of the mean field action written in terms of components and keeping only the terms that are relevant in the large $N$ limit. We further generalize from $\hq =3$ to arbitrary $\hq$:
\begin{equation}\label{Eqn:effective_action}
\begin{split}
    \frac{I_\text{eff}}{N}=&-\log \text{Pf}\Big\{\Big(-\frac{\i\epsilon}{2}\sigma_f-\Sigma_{\psi\psi}\Big)\Big(\frac{\i\epsilon}{2}\sigma_f-\Sigma_{\bar\psi\bar\psi}\Big)-\frac{1}{4}\left(-\sigma_f-\Sigma_{\psi\bar\psi}\right)^2\Big\}\\
    &+\frac{1}{2}\log \det\Big\{\Big(-\frac{\i\epsilon}{2}\sigma_b-\Sigma_{bb}\Big)\Big(\frac{\i\epsilon}{2}\sigma_b-\Sigma_{\bar{b}\bar{b}}\Big)-\frac{1}{4}\Big(-\sigma_b-\Sigma_{b\bar{b}}\Big)^2\Big\}\\
    &+\int \d\tau_1\int \d\tau_2\Big\{G_{\psi\bar\psi}(\tau_1,\tau_2)\Sigma_{\psi\bar\psi}(\tau_1,\tau_2)+G_{\psi\psi}(\tau_1,\tau_2)\Sigma_{\psi\psi}(\tau_1,\tau_2).\\
    &+G_{\bar\psi\bar\psi}(\tau_1,\tau_2)\Sigma_{\bar\psi\bar\psi}(\tau_1,\tau_2)+G_{b\bar{b}}(\tau_1,\tau_2)\Sigma_{b\bar{b}}(\tau_1,\tau_2)+G_{bb}(\tau_1,\tau_2)\Sigma_{bb}(\tau_1,\tau_2)\\
    &+G_{\bar{b}\bar{b}}(\tau_1,\tau_2)\Sigma_{\bar{b}\bar{b}}(\tau_1,\tau_2)-JG_{b\bar{b}}(\tau_1,\tau_2)G_{\psi\bar\psi}(\tau_1,\tau_2)^{\hq-1}\Big\}.\\
\end{split}
\end{equation}
The last step we need is to derive the large $N$ classical equations of motion that arise from varying $G$ and $\Sigma$ fields in the action $I_{\rm eff}$ above. These are the Schwinger-Dyson equations of the model which could as well have been derived by diagrammatic methods. The result is the following-- upon varying the correlators we get the equations  
\bea
    \Sigma_{\psi\psi}(\tau_1,\tau_2)&=&\Sigma_{\bar\psi\bar\psi}(\tau_1,\tau_2)=\Sigma_{bb}(\tau_1,\tau_2)=\Sigma_{\bar{b}\bar{b}}(\tau_1,\tau_2)=0\label{SDequations1}\\
    \Sigma_{\psi\bar\psi}(\tau_1,\tau_2)&=&J(\hq-1)G_{b\bar{b}}(\tau_1,\tau_2)G_{\psi\bar\psi}(\tau_1,\tau_2)^{\hq-2},\\
    \Sigma_{b\bar{b}}(\tau_1,\tau_2)&=&JG_{\psi\bar\psi}^{\hq-1}(\tau_1,\tau_2),
\ea
which we wrote for general values of $\hq$. Upon varying the self-energies we get the following equations which are conveniently written in Fourier space 
\bea
    G_{\psi\bar\psi}^{-1}(\i\omega)&=&-\i\omega-\Sigma_{\psi\bar\psi}+\frac{\epsilon^2\omega^2}{-\i\omega-\Sigma_{\psi\bar\psi}}\\
    G_{b\bar{b}}^{-1}(\i\omega)&=&-1-\Sigma_{b\bar{b}}-\frac{\epsilon^2}{-1-\Sigma_{b\bar{b}}}\\
    -G_{\bar\psi\bar\psi}^{-1}(\i\omega)&=&G_{\psi\psi}^{-1}(\i\omega)=\epsilon\omega+\frac{\left(-\i\omega-\Sigma_{\psi\bar\psi}\right)^2}{\epsilon\omega}\\
    -G_{\bar{b}\bar{b}}^{-1}(\i\omega)&=&G_{bb}^{-1}(\i\omega)=-\i\epsilon-\frac{(-1-\Sigma_{b\bar{b}})^2}{\i\epsilon}\label{SDequations7}
\ea
In the $\epsilon\to 0$ limit these equations reproduce \eqref{eq:N2conf1} to \eqref{eq:N2conf4}. In the next section we will solve these equations for any $\epsilon$, first in the IR limit of late times or low frequencies, and then numerically for all times/frequencies.

Before moving on, we will perform two checks on these equations. Firstly, we can verify the interaction terms are correct by taking the $\epsilon \to 0 $ limit. In this limit the equations become those of the $\mathcal{N}=2$ SYK model \cite{Fu:2016vas}. In particular we see that as $\epsilon$ goes to zero, the correlators that carry $R$-charge such as $G_{\psi \psi}$ and $G_{bb}$ vanish. Secondly, we can verify the other terms are correct by taking the free fermion limit $J\to0$ for arbitrary $\epsilon$. In this limit all self-energies vanish. The equations for the correlators become
\bea
G_{\psi\bar\psi}(\i \omega) &=& \frac{\i}{\omega} \frac{1}{1-\epsilon^2},~~~~G_{b\bar{b}}(\i\omega) = - \frac{1}{1-\epsilon^2},\\
G_{\psi\psi}(\i\omega) &=& \frac{\i}{\omega} \frac{\i\epsilon}{1-\epsilon^2},~~~~G_{bb}(\i\omega) =- \frac{\i \epsilon}{1-\epsilon^2}.
\ea
The correlators derived from these expressions coincide with the free ones arising from a theory with action \eqref{eq:def}. Our conventions for the Fourier transforms are implicit in this, since for $\epsilon=0$ we should reproduce the free fermion correlators.

\subsection{Low energy limit}\label{section:Low_Energy_Limit}
In this section we want to find an approximate solution of the Schwinger-Dyson equations in the IR limit, meaning the regime of late times or low frequency in Fourier space. Assuming the solution is time-translation invariant implies that all correlators depend only on $\tau= \tau_1-\tau_2$. In terms of $\tau$, the IR regime corresponds to $J|\tau| \gg1$ and $J |\beta-\tau|\gg 1$. The main result of this section, which we will justify, is that the IR limit of the mean fields is precisely the same as $\mathcal{N}=2$ SYK and independent of the deformation parameter $\epsilon$. Namely, the only non-zero fields are
\bea
G_{\psi \bar{\psi}}(\tau) &=& \frac{g_{\psi\bar\psi}}{(\frac{\beta}{\pi} \sin \frac{\pi \tau}{\beta})^{2\Delta}} ,~~~G_{b\bar{b}} = \frac{g_{b\bar{b}}}{(\frac{\beta}{\pi} \sin \frac{\pi \tau}{\beta})^{2\Delta_b}},\\
\Sigma_{\psi \bar\psi}(\tau) &=& \frac{(1-2\Delta)\tan \pi \Delta}{2\pi g_{\psi\bar{\psi}}}\frac{1}{(\frac{\beta}{\pi} \sin \frac{\pi \tau}{\beta})^{2-2\Delta}} ,\\
\Sigma_{b\bar{b}} (\tau) &=& \frac{(2\Delta_b-1)\cot \pi \Delta_b}{2\pi g_{bb}} \frac{1}{(\frac{\beta}{\pi} \sin \frac{\pi \tau}{\beta})^{2-2\Delta_b}},
\ea
where the fermionic and bosonic scaling dimensions are
\beq
\Delta = \frac{1}{2\hq} \,,~~~\Delta_b = \frac{1}{2} + \frac{1}{2\hq}\,,
\eeq
and the prefactors are given by
\beq
g_{\psi\bar\psi} = \Big( \frac{\tan \frac{\pi}{2\hq}}{2\pi J}\Big)^{1/\hq},~~~g_{b\bar{b}} = \frac{1}{\hq} \Big( \frac{\tan \frac{\pi }{2\hq} }{2\pi J}\Big)^{1/\hq}.
\eeq
These expressions are valid for $0<\tau<\beta$, and other values can be obtained by the (anti)-commutation relations of fundamental fields, e.g. $G_{\psi\bar\psi}(-\tau) = - G_{\psi \bar\psi } (\tau)$. At zero temperatures these correlators simplify a bit, giving
\bea
G_{\psi \bar{\psi}}(\tau) &=& \frac{g_{\psi\bar\psi}}{|\tau|^{2\Delta}} \, {\rm sgn}(\tau),\hspace{2.5cm}\,G_{b\bar{b}}(\tau) = \frac{g_{b\bar{b}}}{|\tau|^{2 \Delta_b}},\nonumber\\
\Sigma_{\psi \bar\psi}(\tau) &=& \frac{(1-2\Delta)\tan \pi \Delta}{2\pi g_{\psi\bar{\psi}}}\frac{{\rm sgn}(\tau)}{|\tau|^{2-2\Delta}} ,~~~\Sigma_{b\bar{b}} (\tau) = \frac{(2\Delta_b-1)\cot \pi \Delta_b}{2\pi g_{bb}} \frac{1}{|\tau|^{2-2\Delta_b}},\nonumber
\ea
this time written for any $\tau$, positive or negative. This implies that in Fourier space $G_{\psi\bar\psi} \propto \omega^{2\Delta-1}=\omega^{-\frac{\hq-1}{\hq}}$ while $G_{b\bar{b}} \propto \omega^{2\Delta_b-1}=\omega^{\frac{1}{\hq}}$. For the self-energies $\Sigma_{\psi\bar{\psi}}\propto \omega^{\frac{\hq-1}{\hq}}$ and $\Sigma_{b\bar{b}} \propto \omega^{-1/\hq}$. We will use these scaling relations repeatedly below.

At low temperatures we expect the free energy to be given by
\beq
-\beta F = - \beta E_0 + S_0 + \frac{2\pi^2 \alpha_S N/J}{\beta},
\eeq
The Schwarzian coupling for $\epsilon=0$ was computed in \cite{Heydeman:2022lse}, given by $\alpha_S \approx 0.00842$ for $\hat{q}=3$. $S_0$ is the zero-temperature entropy which can be determined purely from the IR solution and therefore we expect it to be $\epsilon$-independent. We therefore have $S_0 = N \log ( 2 \cos \frac{\pi}{2\hq} )$. Of course, as we have seen in section \ref{sec:ED}, this zero-temperature entropy (meaning one first take $N\to\infty$ and then $T\to0$) does not translate into a large ground state degeneracy since the $\mathcal{N}=1$ supersymmetry present at $\epsilon\neq 0$ is no longer sufficient to protect the $\mathcal{N}=2$ BPS states.

We now explain why this solution is correct at low frequencies. In order to do this we begin our analysis in the zero-temperature case since the equations simplify. In this case we keep the leading order terms in the small $\omega$ expansion. The Schwinger-Dyson equations that determine $G_{\psi\bar\psi}$ and $G_{b\bar{b}}$ are 
\begin{equation}\label{SD_equations_1}
\begin{split}
    &\Sigma_{\psi\psi}(\tau_1,\tau_2)=\Sigma_{\bar\psi\bar\psi}(\tau_1,\tau_2)=\Sigma_{bb}(\tau_1,\tau_2)=\Sigma_{\bar{b}\bar{b}}(\tau_1,\tau_2)=0\\
    &\Sigma_{\psi\bar\psi}(\tau_1,\tau_2)=J(\hq-1)G_{b\bar{b}}(\tau_1,\tau_2)G_{\psi\bar\psi}(\tau_1,\tau_2)^{\hq-2},\ \Sigma_{b\bar{b}}(\tau_1,\tau_2)=JG_{\psi\bar\psi}^{\hq-1}(\tau_1,\tau_2)\\
    &G_{\psi\bar\psi}^{-1}(\i\omega)=-\Sigma_{\psi\bar\psi}, \
    G_{b\bar{b}}^{-1}(\i\omega)=-\Sigma_{b\bar{b}}.
\end{split}
\end{equation}
They are the same IR equations as in the original $\mathcal{N}=2$ SYK model, so all properties like reparametrization invariance and conformal symmetry of solutions should follow. In particular it is the reparametrization invariance that allows us to go from zero temperature to finite temperature. The IR limit of the Schwinger-Dyson equations for $G_{\psi\psi}$ and $G_{bb}$ is more subtle.
To start with, we can keep the leading order term in those two equations,
\begin{equation}\label{SD_equations_2}
\begin{split}
    G_{\psi\psi}(\i\omega)=\frac{\epsilon\omega}{\Sigma_{\psi\bar\psi}^2},~~~ \ G_{bb}(\i\omega)=-\frac{\i \epsilon}{\Sigma_{b\bar{b}}^2}.
\end{split}
\end{equation}
Since the charged mean fields do not appear anywhere else in the Schwinger-Dyson equations, the $\mathcal{N}=2$ SYK uncharged mean field solution automatically solves the rest of the equations. The only question that remains is whether then the charged fields as dictacted by \eqref{SD_equations_2} are small or not in the IR limit. If we were working at small deformations we would be able to neglect the right hand side of these equations since they are proportional to $\epsilon$. We emphasize that the result in this section goes beyond the observation of the previous sentence. Instead we propose that the undeformed conformal solution is also the IR solution of our model regardless of $\epsilon$. 

Before addressing whether the charged fields are indeed subleading, the following exercise is instructive. Assume we indeed include \eqref{SD_equations_2} as part of our IR Schwinger-Dyson equations. Do these equations present reparametrization invariance? This is necessary in order to talk about a conformal phase in SYK models. Besides this, without reparametrization invariance the finite temperature solution would not be possible to construct from the zero temperature one. We now show that a term such as \eqref{SD_equations_2} breaks reparametrization invariance. To do this we introduce $\widetilde{\Sigma}_{\psi\psi}$ and $\widetilde{\Sigma}_{bb}$ defined by
\beq\label{eq:deftildsigm}
G_{\psi \psi} * \widetilde{\Sigma}_{\psi\psi} = -1,~~~G_{bb} * \widetilde{\Sigma}_{bb} = -1,
\eeq
meaning the inverse of $G_{\psi\psi}(\tau_1,\tau_2)$ and $G_{bb}(\tau_1,\tau_2)$ interpreted as two-index operators. In general, under the reparameterization, any mean field $X(\tau_1,\tau_2)$ (which could be either a self energy or correlator) transforms as
\begin{align}
    &X(\tau_1,\tau_2)\rightarrow [f'(\tau_1)f'(\tau_2)]^{\Delta_X} X(f(\tau_1),f(\tau_2)),\label{reperametrization_Sigma}
\end{align}
with $\Delta_X$ being its scaling dimension which can be determined at zero temperature. In temporal space, we can combine equations \eqref{eq:deftildsigm} with \eqref{SD_equations_2} to obtain
\begin{equation}
    \widetilde\Sigma_{\psi\psi}(\tau,\tau')=\frac{\i}{\epsilon}\int \d\tau_1 \d\tau_2 \, \Sigma_{\psi\bar\psi}(\tau,\tau_1)\Sigma_{\psi\bar\psi}(\tau_1,\tau_2)\text{sgn}(\tau_2-\tau')\label{eq:wtsss}
\end{equation}
Assuming reparametrization invariance, the left hand side would transform as in equation \eqref{reperametrization_Sigma}, while the right hand side transforms instead as 
\begin{align}
    &\frac{\i}{\epsilon}[f'(\tau)]^{1-2\Delta}\int \d f(\tau_1) \d f(\tau_2) [f'(\tau_1)]^{1-4\Delta}[f'(\tau_2)]^{-2\Delta}\Sigma_{\psi\bar\psi}(f(\tau),f(\tau_1))\\
    &\quad\Sigma_{\psi\bar\psi}(f(\tau_1),f(\tau_2))\text{sgn}(f(\tau_2)-f(\tau'))\notag
\end{align}
Since the extra factors with the power $1-4\Delta$ and $-2\Delta$ do not vanish, the right side does not match the reparametrization of the left hand side of \eqref{eq:wtsss}. The same issue arises for the equation that determines $\widetilde{\Sigma}_{bb}$. 

We see that the Schwinger-Dyson equations for $G_{\psi\psi}$ and $G_{bb}$ in equation \eqref{SD_equations_2} do not have the reparameterization invariance characteristic of a conformal phase. Fortunately, we now show that the correlators $G_{\psi\psi}$ and $G_{bb}$ derived from them are subleading compared to their neutral counterparts, in the IR limit. This justifies ignoring the right hand side of \eqref{SD_equations_2} in the low frequency regime. We will carry out the analysis at zero-temperature first. To leading order in the small $\omega$ limit, equation \eqref{SD_equations_2} implies that
\beq
G_{\psi\psi} (\i \omega) \sim \frac{1}{J} \, \left|\frac{\omega}{J}\right|^{\frac{2-\hq}{\hq}},~~~G_{bb}( \i \omega) \sim \left|\frac{\omega}{J}\right|^{\frac{2}{\hq}}.
\eeq
We now ignore numerical prefactors on $\hq$ and $\epsilon$ and emphasize the dependence on $J$ and $\omega$. The IR limit corresponds to $\left|\frac{\omega}{J}\right| \ll 1$. Comparing the charged mean fields against their uncharged counterpart we see that they are subleading in this limit $G_{\psi \psi} /G_{\psi \bar\psi} \sim \left|\frac{\omega}{J}\right|^{\frac{1}{\hq}}\to 0$ and $G_{bb}/G_{b \bar{b}} \sim \left|\frac{\omega}{J}\right|^{\frac{1}{\hq}}\to 0$. Since reparametrization invariance is lost, this argument does not guarantee they are subleading at finite temperatures. This will be verified from a numerical solution of the Schwinger-Dyson equations in the next section. This argument can fail if we first take the large $\hat{q}$ limit before taking the IR limit. In this case we will see in section \ref{sec:largeqqq} that the $\U(1)$ breaking correlators can be non-negligible in the IR.

\subsection{Large $N$ numerical solution} 
\begin{figure}[t!]
    \centering
    \includegraphics[width=\textwidth]{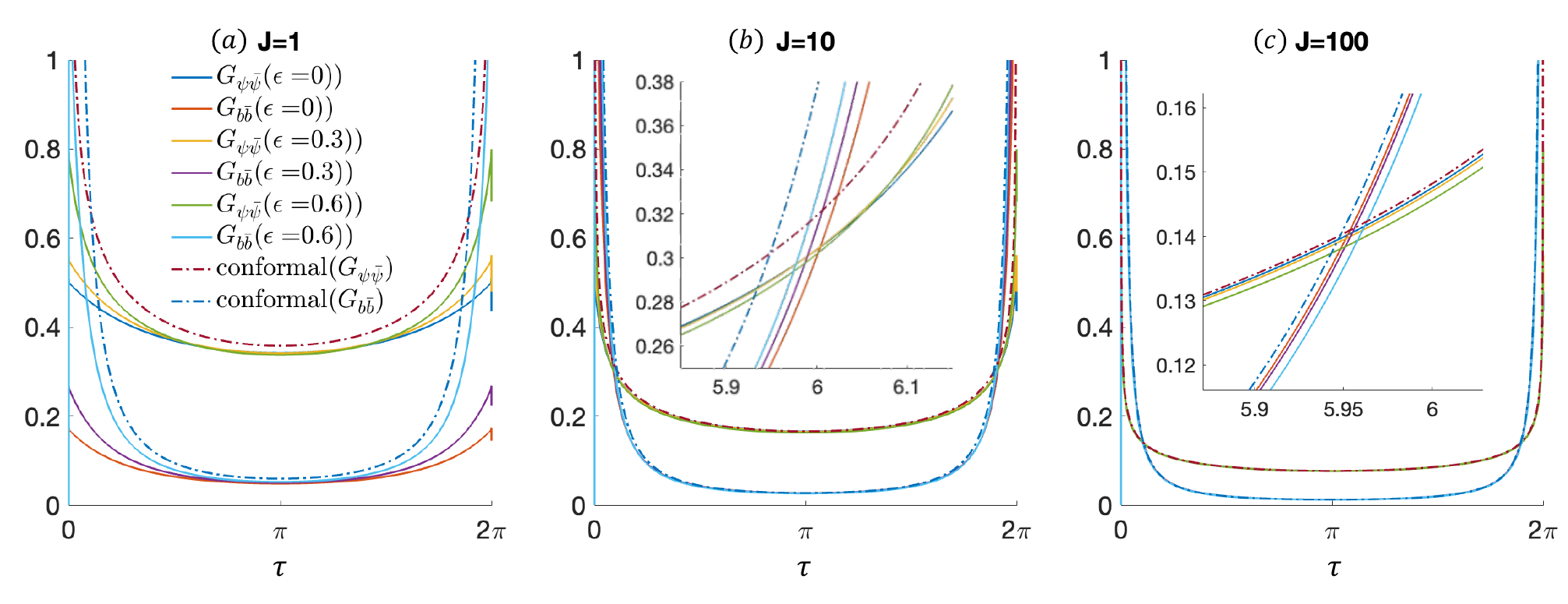}
    \caption{Comparison of numerical solutions of two-point functions $G_{\psi\bar\psi}$ and $G_{b\bar{b}}$ at finite temperature with 18 time steps and the conformal solutions with different $J$'s. 
    The legend in (a) applies to all three plots. The solid lines are the numerical solutions, and the dashed lines are the conformal solutions. In (b) and (c), we zoom in on the crossing area of $G_{\psi\bar\psi}$ and $G_{b\bar{b}}$ to how convergent they are. }
    \label{numerical_G}
\end{figure}

We now verify the claims in the previous section by solving the full Schwinger-Dyson equations for $G_{\psi\bar\psi}$, $G_{b\bar{b}}$, $G_{\psi\psi}$, and $G_{bb}$ numerically with non-zero $\epsilon$, finite temperature, and for all times. For simplicity we work out the case $\hq=3$ and fix units such that $\beta=2\pi$. As shown in figure \ref{numerical_G}, the two-point functions approach the analytical solutions in the conformal limit when $J$ increases.
\begin{figure}[t!]
    \centering
    \includegraphics[width=\textwidth]{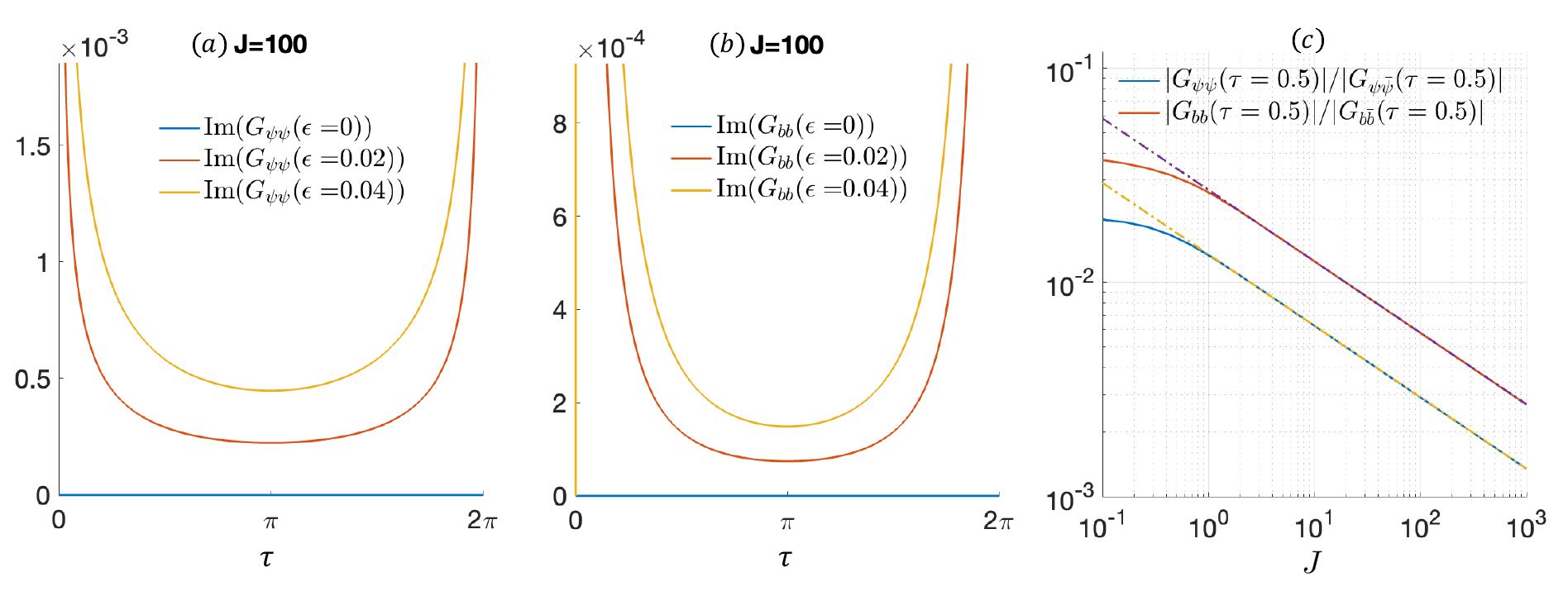}
    \caption{Plots (a) and (b) are numerical results of the two-point functions $G_{\psi\psi}$ and $G_{bb}$ at finite temperature with $J=100$, 18  time steps, and different $\epsilon$'s. (c) shows their relative magnitudes compared to their off-diagonal counterparts $G_{\psi\bar\psi}$ and $G_{b\bar{b}}$ when increasing $J$ with $\epsilon=0.02$. 
    }
    \label{numerical_Gbb}
\end{figure}

All two-point functions have a similar shape appropriate for a conformal phase, and the magnitude of the functions can be approximated by the absolute value at $\tau=\pi$. We read the numerical values of $G_{\psi\psi}$ and $G_{\psi\bar\psi}$ at $\tau=\pi$ and calculate their relative magnitude and showed it in the figure \ref{numerical_Gbb}(c).  The relative magnitudes follow a power law of $J^{-1/3}$. As $J$ increases, $G_{\psi\psi}$ is more suppressed compared to $G_{\psi\bar\psi}$.  These results also apply to the two-point functions for bosons. The numerical result is another justification that we can ignore the leading-order equations for $G_{\psi\psi}$ and $G_{bb}$ in the conformal limit, where $\tau J$ becomes large.

\newpage
\subsection{The super-reparametrization action for the broken $\U(1)$ mode}
\label{sec:sectionschwarzianwithdefgeneral}
Up to this point, we have largely analyzed the low energy consequences of our deformed $\mathcal{N}=2$ SYK model from the point of view of the conformal solutions $G_{\psi \bar{\psi}}$, $G_{b \bar{b}}$, etc to the Schwinger-Dyson equations. A key observation was the fact that the model may be written in a way \eqref{eq:def} such that we have only deformed the UV terms; i.e. we have added new kinetic terms proportional to the parameter $\varepsilon$ which vanish when we study the conformal limit of the Schwinger-Dyson equations in which the inverse temperature $\beta \rightarrow \infty$.

As is well known in SYK, the interesting dynamical behavior of the model is found at finite $\beta$ in which we perturb away from the conformal limit. The relevant low energy perturbations $\delta G_{\psi \bar{\psi}}$, $\delta G_{b \bar{b}}$, $\dots$ around the conformal solution constitute the soft modes of the theory associated to the breaking of (super)-conformal symmetry. These low energy degrees of freedom are reparametrization modes which have a dynamical action given by the $\mathcal{N}=2$ supersymmetric extension of the Schwarzian derivative. We postpone the details of this calculation for our deformed model until Section \ref{sec:largeqqq} (in that section we work in the large $\hat{q}$ limit, but we believe the resulting effective action works more generally at finite $\hat{q}$). In this subsection, we will review the soft mode effective action for the undeformed $\mathcal{N}=2$ SYK and explain in general terms the effect of the deformation. 

The form of the supersymmetry breaking modification we study has two essential features we emphasize in this section. The first is that it explicitly breaks the $\U(1)_R$ symmetry in the UV in terms of the fundamental fermions. This means that, although the conformal limit with $\mathcal{N}=2$ super reparametrization invariance remains unmodified, the full theory away from the conformal limit cannot have an exact $\U(1)_R$. We will argue and later explicitly show that from the point of view of the IR effective action, the mode corresponding to $\U(1)_R$ reparametrizations essentially becomes massive. While all the $\mathcal{N}=2$ reparametrizations remain degrees of freedom in the effective theory, their action is modified. In addition, away from the conformal limit, global $\U(1)_R$ transformations no longer correspond to a zero mode, and it is no longer removed from the path integral, in the sense that we no longer divide by global ${\rm SU}(1,1|1)$ transformations in defining the measure of the $\mathcal{N}=2$ super-Schwarzian fields. The UV breaking of an $\mathcal{N}=2$ SYK model to an $\mathcal{N}=1$ SYK model manifests itself in the fact that the IR theory possesses only one global supercharge. The original fields of the $\mathcal{N}=2$ Schwarzian decompose into global $\mathcal{N}=1$ multiplets, where an $\mathcal{N}=1$ Schwarzian action remains massive, while the $\U(1)_R$ and one of the fermion modes becomes massive. One can roughly argue that some of the ``gauged" symmetries of the original model are broken, and these modes become massive dynamical fields.

Related to the comments above, the second essential feature of our model is that the modification we propose only effects the UV terms. In additional to preserving $\mathcal{N}=2$ reparametrization invariance in the conformal limit, it also allows us to argue that the finite temperature effective action remains local. Recall that the mean field action for the bilinears is a bi-local action, and it is somewhat nontrivial that the near-zero mode action is the purely local (super) Schwarzian derivative. When we break some symmetries with our UV deformation, it is no longer guaranteed that the resulting IR effective action remains local. Because some of the naive conformal $\mathcal{N}=2$ reparametrizations do not respect these UV boundary conditions, an argument similar to those made about ``fictitious'' reparametrizations \cite{Fu:2016vas} which become massive via local terms also seems to apply in our case. We will review this argument below.

The above discussion leads us to suggest that the infrared effective action for the soft modes of our deformed model consists of the $\mathcal{N}=2$ Super-Schwarzian action deformed by a local term which lifts the zero mode of the $\U(1)_R$ reparametrization which we call $a(\tau)$. For small fluctuations, the simplest such term that can be added is proportional to $a(\tau)^2$, and in Section \ref{sec:largeqqq} we will confirm this explicitly and compute the coefficient as a function of $\epsilon$ at large $\hat{q}$. More globally however, it is important to note that the field $a(\tau)$ originates as a phase mode of a local $\U(1)$ transformation, therefore it is a compact boson\footnote{In gravity, this arises because $a(\tau)$ arises as the boundary mode of a $\U(1)$ gauge field on the AdS$_2$ disk.}. Thus, $a(\tau)^2$ is not invariant under compact shifts $a(\tau) \sim a(\tau) + 2 \pi \hat{q}$, and we expect the nonlinear completion of the effective action to have this periodicity. In Appendix \ref{app:U1effective}, we verify that the correct nonlinear action includes $\left(\cos(2 a)-1 \right)$, but we do not determine the full nonlinear supersymmetric completion of this theory.

Anticipating the effective action contains $a(\tau)^2$, we can use the symmetries of the problem to obtain the other terms of the quadratic expansion of the action. As we have emphasized, the deformation no longer preserves global $\mathcal{N}=2$ supersymmetry (this was broken already at the level of the fundamental fermion interactions). However, the remaining global symmetry is still $\mathcal{N}=1$ supersymmetry, and in the rest of this section we will use this symmetry to find the supersymmetric completion of the quadratic action which contains $a(\tau)^2$. 

Before adding the deformation, recall that in $\mathcal{N}=2$ superspace, the global supersymmetry transformations of \eqref{eq:N2RigidSusy} are:
\begin{align}
\label{eq:N2globalsusy}
\tau &\rightarrow \tau + \bar{\theta} \varepsilon \, , \\
\theta &\rightarrow \theta + \varepsilon \, , \\
\bar{\theta} &\rightarrow \bar{\theta} \, ,
\end{align}
for the chiral transformation, as well as the antichiral
\begin{align}
\tau &\rightarrow \tau + \theta \bar{\varepsilon} \, , \\
\theta &\rightarrow \theta  \, , \\
\bar{\theta} &\rightarrow \bar{\theta} + \bar{\varepsilon} \, ,
\end{align}
In the conformal limit, these global supersymmetries enlarge to local super-reparametrizations, and the action of these as well as their bosonic counterparts on correlators was given in Eq.~\eqref{eqn:FususytransfIR}. Away from the conformal limit, the reparametrizations are no longer exact symmetries and instead become dynamical modes which themselves enjoy the global $\mathcal{N}=2$ supersymmetry. We have already discussed on choice of parametrization for the super-reparametrization modes in \eqref{eqn:fermionreparam}. To see how the $\mathcal{N}=2$ reparametrization modes transform under global SUSY, we use the parametrization of \cite{Fu:2016vas}, 
\begin{align}
\label{eq:reparamsuperfields}
    \theta' & = \rho(\tau + \theta \bar{\theta}) \left ( \theta + \eta (\tau + \theta \bar{\theta})\right ) \, , \\
    \bar{\theta}' & = \bar{\rho}(\tau - \theta \bar{\theta}) \left ( \bar{\theta} + \bar{\eta} (\tau - \theta \bar{\theta})\right ) \, , \\
    \tau' &= f(\tau) + \theta \bar{g}(\tau) + \bar{\theta} g(\tau) + \theta \bar{\theta}h(\tau) \, .
\end{align}
The natural super-line elements transform in a simple way under the chiral and anti-chiral transformations:
\begin{align}
    \delta_\varepsilon (\tau + \theta \bar{\theta}) = 0 \, , \, \, \, \delta_{\bar{\varepsilon}} (\tau + \theta \bar{\theta}) = 2 \theta \bar{\varepsilon} \, , \, \, \, \,     \delta_\varepsilon (\tau - \theta \bar{\theta}) = 2 \bar{\theta} \varepsilon  \, , \, \, \, \delta_{\bar{\varepsilon}} (\tau + \theta \bar{\theta}) = 0 \, ,
\end{align}
For the chiral transformations, we see that the superspace component fields behave as
\begin{align}
    \theta' &\rightarrow \rho (\eta + \varepsilon) + \theta \rho + \theta \bar{\theta} (\rho' \eta + \rho \eta' + \rho' \varepsilon) \, , \\
    \bar{\theta}' &\rightarrow \bar{\rho} \bar{\eta} + \bar{\theta}( \bar{\rho} + 2\varepsilon(\bar{\rho} \bar{\eta}'+ \bar{\rho}'\bar{\eta})) - \theta \bar{\theta} (\bar{\rho}' \bar{\eta} + \bar{\rho} \bar{\eta}' ) \, .
\end{align}
From this we read off the global supersymmetry transformations of the Schwarzian component fields:
\begin{align}
\label{eq:linearizedsusyfields}
    \delta_\varepsilon \eta = \varepsilon \, , \, \, \, \delta_\varepsilon \bar{\eta} = 2\varepsilon\bar{\eta}\bar{\eta}' \, , \, \, \, \delta_\varepsilon \rho = 0 \, , \, \, \, \delta_\varepsilon \bar{\rho} = 2\varepsilon(\bar{\rho}\bar{\eta}' + \bar{\rho}'\bar{\eta})
\end{align}
A similar calculation for the anti-chiral transformation gives
\begin{align}
    \delta_{\bar \varepsilon} \eta = 2\bar{\varepsilon}\eta \eta' \, , \, \, \, \delta_{\bar \varepsilon}  \bar{\eta} = \bar{\varepsilon} \, , \, \, \, \delta_{\bar \varepsilon}  \rho = 2\bar{\varepsilon}(\rho \eta' + \rho'\eta) \, , \, \, \, \delta_{\bar \varepsilon}  \bar{\rho} = 0
\end{align}

The remaining discussion is simplified if we are only interested in the Schwarzian action to quadratic order, which is all we will need if we are interested in the supersymmetric completion of an $a(\tau)^2$ term. We place the theory at $T=0$ on the real line (rather than the thermal circle) and seek the linearized supersymmetry transformations and quadratic action. In what follows, we use the explicit parametrization Eq.~(8.23-8.27) of \cite{Peng:2020euz}. We will first consider infinitessimal reparametrizations on the line given by
\begin{align}
    f(\tau) = \tau + \varphi(\tau) \, .
\end{align}
Now, naively we would like to construct the $\mathcal{N}=2$ Super-Schwarzian action to quadratic order in the fields which is invariant under supersymmetry transformations which are linear in the fields. However, when we view the finite reparametrizations Eq.~\eqref{eq:reparamsuperfields} as superfields, the nonlinearity of these superfields means the correct supersymmetry transformations are not just the linearization of Eq.~\eqref{eq:linearizedsusyfields}. Instead, we must linearize the superfields first, then act with the corresponding chiral and anti-chiral supertranslations. Using the parametrization where we set $i a(\tau)$ to be the axion-like phase of the field $\rho(\tau)$, we obtain
\begin{align}
    \rho(\tau) &= 1 + \i a(\tau) + \frac{\varphi'(\tau)}{2} \, , \quad \qquad \bar{\rho}(\tau) = 1 - \i a(\tau) + \frac{\varphi'(\tau)}{2} \, , \\
    g(\tau) &= \eta(\tau) \, , \quad \qquad \bar{g}(\tau) = \bar{\eta}(\tau) \, , \quad  \qquad  h(\tau) = \mathcal{O}(\eta^2) \, .
\end{align}
Therefore, the linearized superfields are
\begin{align}
\label{eq:reparamsuperfieldsLIN}
    \theta' & = \eta + \theta (1 + \i a + \frac{\varphi'}{2}) + \theta \bar{\theta} \eta' \, , \\
    \bar{\theta}' & = \bar{\eta} + \bar{\theta} (1 - \i a + \frac{\varphi'}{2}) - \theta \bar{\theta} \bar{\eta}' \, , \\
    \tau' &= \tau + \varphi + \theta \bar{\eta} + \bar{\theta}\eta \, .
\end{align}
Note the unfortunate clash of notation; on the left we have indicated the coordinates after a reparametrization with a prime, whereas on the right the prime denotes the time derivative $\partial_\tau$. Further, although the $\theta \bar{\theta}$ term of $\tau'$ has dropped out, the fact that the $\eta$, $\bar{\eta}$ fields transform inhomogeneously as in Eq.~\eqref{eq:linearizedsusyfields} means that in checking the variation, the term quadratic in $\eta$'s will generate a necessarily linear term. In the end the nonlinear parts will drop out.

It is now a simple exercise to act with the chiral and anti chiral supertranslations, Eq.~\eqref{eq:N2globalsusy}. From this, one may deduce the linearized action of supersymmetry on the component fields. As before, the chiral and anti-chiral transformations are generated by constant Grassmann $\varepsilon$ and  $\bar{\varepsilon}$, respectively, not to be confused with the infinitessimal reparametrization $\varphi(\tau)$:
\begin{align}
\label{eq:N2componentsusyLIN}
    \delta_{\varepsilon} \varphi &= \varepsilon \bar{\eta} \, , \quad \delta_{\varepsilon} \eta = \varepsilon (1+ \i a + \frac{\varphi'}{2}) \, , \quad \delta_{\varepsilon} \bar{\eta} = 0 \, , \quad \delta_{\varepsilon} a = \i \varepsilon \bar{\eta}' \, , \\
    \delta_{\bar \varepsilon} \varphi &= \bar{\varepsilon} \eta \, , \quad \delta_{\bar \varepsilon} \eta = 0 \, , \quad \delta_{\bar \varepsilon} \bar{\eta} = \bar{\varepsilon} (1 - \i a + \frac{\varphi'}{2})  \, , \quad \delta_{\bar \varepsilon} a = -\i \bar{\varepsilon} \eta' \, .
\end{align}

We now recall \cite{Fu:2016vas,Stanford:2017thb,Peng:2020euz} the quadratic expansion of the $\mathcal{N}=2$ super-Schwarzian action on the infinite line with $f(\tau) = \tau + \varphi(\tau)$:\footnote{Note that in SYK models on the thermal circle, one usually recovers the quadratic expansion of the Schwarzian theory with $f(\tau) = \tan\left(\frac{\tau + \varphi(\tau)}{2} \right)$.}
\begin{align}
\label{eq:N2quadraticaction}
    I_{\mathcal{N}=2 \textrm{ quad}} =  \int \! \d\tau \, \, \frac12(\varphi'')^2 +2 (a')^2 - 4 \eta' \bar{\eta}'' \, .
\end{align}
This action is invariant under the supersymmetry transformations Eq.~\eqref{eq:N2componentsusyLIN}. Up to the overall prefactor, it is the linearization of the effective action \eqref{eq:N2EffectiveAction}.

After adding the deformation in the UV, the microscopic fermion Lagrangian is only invariant under the $\mathcal{N}=1$ supersymmetry transformation generated by $Q+\bar{Q}$. The remaining supersymmetry generated by $Q-\bar{Q}$ and the $\U(1)_R$ symmetry are broken. This pattern of preserved supersymmetry is the infrared manifestation of that found in the UV fermion description, Eq.~\eqref{eq:deformedQ}. We now proceed to write down quadratic action which describes the unbroken $\mathcal{N}=1$ Schwarzian theory coupled to the broken modes, and this action retains an unbroken global $\mathcal{N}=1$ supersymmetry.

Starting with Eq.~\eqref{eq:N2quadraticaction}, we write the real and imaginary components of the fermions as
\begin{align}
    \xi \equiv \eta + \bar{\eta} \, , \qquad  \lambda \equiv \eta - \bar{\eta} \, .
\end{align}
We also consider linear combinations of the transformations generated by $\varepsilon$ and $\bar{\varepsilon}$,
\begin{equation}
    \delta_{\varepsilon_1} \equiv \delta_{\varepsilon}+\delta_{\bar{\varepsilon}=\varepsilon}\, , \qquad \delta_{\varepsilon_2} \equiv \delta_{\varepsilon}+\delta_{\bar{\varepsilon}=-\varepsilon} \, .
\end{equation}
In these variables the supersymmetries Eq.~\eqref{eq:N2componentsusyLIN} act as
\begin{align}
    \delta_{\varepsilon_1}\varphi &= \varepsilon_1 \xi \, , \qquad \delta_{\varepsilon_1}\xi = 2\varepsilon_1 +\varepsilon_1 \varphi' \, , \qquad \delta_{\varepsilon_1}\lambda = 2 \i \varepsilon_1 a \, , \qquad \delta_{\varepsilon_1}a = -\i \varepsilon_1 \lambda' \, , \\
    \delta_{\varepsilon_2}\varphi &= -\varepsilon_2 \lambda \, , \qquad \delta_{\varepsilon_2}\xi = 2\i \varepsilon_2 a \, , \qquad \delta_{\varepsilon_2}\lambda = 2 \varepsilon_2 + \varepsilon_2 \varphi' \, , \qquad \delta_{\varepsilon_2}a =\i \varepsilon_2 \xi' \, ,
\end{align}
and the action is
\begin{align}
    I_{\mathcal{N}=2 \textrm{ quad}} =  \int \! \d\tau \, \, \frac12(\varphi'')^2 +2 (a')^2 - \xi'\xi'' + \lambda' \lambda'' \, .
\end{align}

Now, consider deforming this action by the bosonic term $\delta_\alpha S = \frac{\alpha}{2}a^2$, where $\alpha(\epsilon)$ is some function of the deformation parameter $\epsilon$; we compute this coefficient for our specific model at large $\hat{q}$ in the next section. Without any reference to SYK, this is just a free parameter, but in the context of our deformed SYK, the total factor is naturally proportional to the number of fermions; earlier we called $\alpha \equiv \alpha_{\rm def}NJ$ to distinguish it from the Schwarzian coupling, which is set to $1$ in this subsection. The variation of this term under the $Q+\bar{Q}$ transformation is
\begin{eqnarray}
    \delta_{\varepsilon_1}(\frac{\alpha(\epsilon)}{2}a^2) = -\i \varepsilon_1 \alpha(\epsilon)  a \lambda' \, . 
\end{eqnarray}
Cancelling this variation requires our deformed action to be
\begin{align}
\label{eq:quadraticdeformedSch}
    I_{\mathcal{N}=2 \textrm{ quad}}(w) =  \int \! \d\tau \, \, \frac12(\varphi'')^2 - \xi'\xi'' +2 (a')^2 + \frac{\alpha (\epsilon)}{2}a^2  + \lambda' \lambda'' + \frac{\alpha(\epsilon)}{4}\lambda \lambda' \, .
\end{align}
This new action is invariant under $\delta_{\varepsilon_1}$, but not $\delta_{\varepsilon_2}$. We could attempt to add new fermionic terms to make the action invariant under both transformations, but this would require a modification of the $\varphi$ part of the action, so we would no longer necessarily be describing the quadratic expansion of the Schwarzian theory. In the infrared, or when $\alpha_{\rm def}NJ$ turns out to be large, we essentially have a higher derivative $\mathcal{N}=1$ scalar multiplet $(\varphi', \xi')$ decoupled from a $\mathcal{N}=1$ fermi multiplet $(\lambda, a)$ in which the scalar $a$ is auxiliary. Because we do not derive the full nonlinear completion of this action, we cannot make further statements about the interaction terms between these multiplets in the full theory, but see Appendix \ref{app:U1effective} for some first steps in this direction.

We conclude this section with a heuristic picture of why one should expect the infrared effective action with the deformation $\epsilon \neq 0$ to be purely local with non-derivative terms for $a(\tau)$. At first sight, the proposed quadratic Eq.~\eqref{eq:quadraticdeformedSch} is a minimal answer compatible with the expected pattern of symmetry breaking. However, given that the full mean field action Eq.~\eqref{Eqn:effective_action} is bilocal, it is not guaranteed that the relevant fluctuation modes necessarily give a local answer. 

The Schwarzian theory describes the effective action of nearly zero modes. In the context of \eqref{Eqn:effective_action}, dropping the UV terms implies reparametrizations of the form Eq.~\eqref{eqn:FususytransfIR} generated by $f(\tau), a(\tau)$, etc are exact zero modes of the problem. As in the usual SYK model\cite{Maldacena:2016hyu}, the breaking of reparametrization symmetry is a UV effect. Performing a small and slowly varying reparametrization around the conformal solution leads to a small change in the action at long wavelengths. This is in contrast to a situation presented in the appendix of \cite{Maldacena:2016upp}, where a nonlocal effective action may (for instance) be obtained when the model contains an operator with $1<\Delta<\frac32$ which dominates over the Schwarzian mode. 

Because our model with $\epsilon \neq 0$ is identical in the conformal limit to the original $\mathcal{N}=2$ SYK model, it has the same set of infrared reparameterization modes and does not have operators which would spoil the locality\footnote{For example, if we were to break supersymmetry by adding a complex SYK Hamiltonian with $q=2\hat{q}$ it would be harder to argue for locality.}. However, the new UV terms which depend on $\epsilon$ are not compatible with the short distance $\U(1)_R$ and $Q-\bar{Q}$ symmetries; they are simply absent in the microscopic theory, even though these symmetries manifest themselves as the $(a(\tau), \lambda(\tau))$ fields in Eq.~\eqref{eq:quadraticdeformedSch}. For example, performing a $\U(1)$ phase reparametrization gives:
\begin{equation}
\begin{split}
    G_{\psi\bar\psi}(\tau_1,\tau_2)\rightarrow e^{\frac{\i}{\hat{q}}(a(\tau_1)-a(\tau_2))}G_{\psi\bar\psi}(\tau_1,\tau_2)\\
    G_{\psi\psi}(\tau_1,\tau_2)\rightarrow e^{\frac{\i}{\hat{q}}(a(\tau_1)+a(\tau_2))}G_{\psi\psi}(\tau_1,\tau_2)\\
    G_{\bar\psi\bar\psi}(\tau_1,\tau_2)\rightarrow e^{-\frac{\i}{\hat{q}}(a(\tau_1)+a(\tau_2))}G_{\bar\psi\bar\psi}(\tau_1,\tau_2)\\
\end{split}
\end{equation}
along with analogous transformations for the bosonic bilinears which have charge $\frac{1-\hq}{\hq}$. At short distances, $\tau_2 \rightarrow \tau_1$ and we have
\begin{align}
    G^{UV}_{\psi\bar\psi}(\tau)&\rightarrow G^{UV}_{\psi\bar\psi}(\tau) \, \\
    G^{UV}_{\psi\psi}(\tau)&\rightarrow e^{\frac{2 \i a(\tau)}{\hat{q}}}G^{UV}_{\psi\psi}(\tau) \, \\
    G^{UV}_{\bar\psi\bar\psi}(\tau)&\rightarrow e^{-\frac{2 \i a(\tau)}{\hat{q}}}G^{UV}_{\bar\psi\bar\psi}(\tau) \, .
\end{align}
These same transformations also act on the $\Sigma$ fields and change the action \eqref{Eqn:effective_action} due to the presence of the extra $\epsilon$ terms in the determinants. The presence of the additional phases for the charged bilinears are ultimately responsible for the non-derivative interactions involving $a(\tau)$.

The fact that a would-be near zero mode from the IR description dramatically changes the UV correlator suggests that this mode is non-propagating, or at least massive at the relevant energy scale. In fact, a similar set of modes is already present in $\mathcal{N}=1,2$ SYK models\cite{Fu:2016vas,Heydeman:2022lse}. In that context, it is important to note that supersymmetric SYK models in the conformal limit typically have additional spurious reparametrization symmetries in addition to those governed by the super-Schwarzian. To see this, one can note that Eq.~\eqref{eq:IRFuSD} is actually invariant under a larger set of symmetries than Eq.~\eqref{eqn:FususytransfIR}; these are written in superspace in \cite{Fu:2016vas} and include:
\begin{align}
    G_{\psi\bar\psi}(\tau_1, \tau_2)&\rightarrow \Omega(\tau_1) \Omega(\tau_2) G_{\psi\bar\psi}(\tau_1, \tau_2) \, , \\
    G_{b\bar b}(\tau_1, \tau_2)&\rightarrow \Omega(\tau_1)^{1-\hat{q}} \Omega(\tau_2)^{1-\hat{q}} G_{b\bar b}(\tau_1, \tau_2) \, .
\end{align}
These spurious scaling symmetries (and their superpartners) again affect the form of the UV correlators. It was argued in \cite{Fu:2016vas} that they lead to local non-derivative terms in the effective action which suppress fluctuations of these modes; schematically they enter in the form $I_\Omega = \int \d\tau  (\Omega(\tau) - 1 )^2$.

The $\mathcal{N}=2$ SYK model with $\epsilon \neq 0$ also has a set of spurious reparametrizations which essentially drop out of the problem. The new feature is that given the above discussion, we can now offer an interpretation of the UV $\epsilon$ terms-- they have the effect of making the $(a(\tau), \lambda(\tau))$ act as spurious modes\footnote{As we have already discussed, one important difference between $a(\tau)$ and $\Omega(\tau)$ is that $a(\tau)$ is a phase and is thus compact. Additionally, $\Omega(\tau)$ acts on all UV correlators, whereas $a(\tau)$ only effects the charged bilinears.}. In the next section, we will undertake a direct verification that the model with $\epsilon \neq 0$ is described by an action of this kind.

\section{$\U(1)$ breaking in complex SYK and $\mathcal{N}=2$ SYK: Large $q$ analysis}\label{sec:largeqqq}

The purpose of this section is to show, in the large $q$ limit, that the low energy action of our deformed $\mathcal{N}=2$ SYK model takes the form argued for in section \ref{sec:sectionschwarzianwithdefgeneral} based on the symmetries of the problem.

\subsection{Complex SYK: $\U(1)$ breaking deformation}
We begin our analysis by studying an analogous problem in complex SYK, which we will refer to as CSYK. This theory consists on $N$ complex fermionic fields $\psi^i$ with action 
\beq
I = I_{\rm UV} + I_{\rm IR},
\eeq
where
\beq
I_{\rm UV} = \frac{1}{2} \int \d\tau \, \psi^i \partial_\tau \bar{\psi}_i + {\rm h.c.},~~I_{\rm IR} = \i \int \d \tau \, \sum_{i,j,k,l} J_{ijkl} \bar{\psi}^i \bar{\psi}^j \psi^k \psi^l.
\eeq
In the Hamiltonian formulation $\psi^i$ are operators acting on a $2^{N}$ dimensional Hilbert space with anti-commutation relations $\{ \psi^i, \psi^j\}=\{ \bar{\psi}^i, \bar{\psi}^j\}=0$ and $\{ \psi^i, \bar{\psi}^j\}=\delta^{i}_j$. The theory has a $\U(1)$ symmetry generated by $J=\frac{1}{2} \sum_i [\bar{\psi}_i, \psi^i]$ such that the quantization of a fermion leads to charges $\pm 1/2$. The model again has random Gaussian couplings $J_{ijkl}$ which have no relation to the $\U(1)$ symmetry generator.

Our aim is to construct a model with the same conformal solution as CSYK but with the $\U(1)$ symmetry broken. This can be done in a way completely analogous to what we did in section \ref{sec:susybreakdef}. We modify the UV action in the following fashion
\beq
I_{\rm UV} \to I_{\rm UV} + \frac{1}{2} \epsilon \int \d\tau \big\{\i \, \psi^i \partial_t \psi^i \,+{\rm h.c.}\big\}
\eeq
In the Hamiltonian description this leads to deformed canonical relations as in \eqref{eq:PsiCRe}. The new UV action breaks the $\U(1)$ symmetry acting as $\psi \to e^{\i \alpha} \psi$. We can introduce fermions $\eta^i$, linearly related to $\psi$ and $\bar{\psi}$ as in \eqref{eq:psivseta}, such that the UV action is invariant under $\eta \to e^{\i \alpha} \eta$. While this transformation is a symmetry of $I_{\rm UV}$, its action is not compatible with the $\U(1)$ symmetry transformation of the IR piece $I_{\rm IR}$. Since we cannot find a $\U(1)$ action preserving $I_{\rm UV}$ and $I_{\rm IR}$ simultaneously the symmetry is explicitly broken when $\epsilon \neq 0$.

\subsubsection*{Large $q$ SYK at $\epsilon=0$}
We first review the large-$q$ analysis of CSYK without the $\U(1)$-breaking deformation. In this case we only need to keep track of the $\psi-\bar{\psi}$ two-point function and the associated self-energy. In the large $q$ limit we impose the following ansatz \cite{Maldacena:2016hyu}
\beq
G_{\psi \bar{\psi}}(\tau_1,\tau_2) = \frac{{\rm sgn}(\tau_{12})}{2}\left[ 1 + \frac{1}{q} g(\tau_1,\tau_2)\right],~~~\Sigma_{\psi\bar{\psi}}(\tau_1,\tau_2) = \frac{\sigma(\tau_1,\tau_2)}{q}. 
\eeq
where $g(\tau_1,\tau_2)$ and $\sigma(\tau_1,\tau_2)$ are finite as $q\to\infty$. Since the self-energy is small, order $1/q$, we can Taylor-expand the free-fermion term in the mean field action to quadratic order in $\sigma(\tau_1,\tau_2)$. Since the rest of the action is linear in the self-energy, we end up with a Gaussian action for $\sigma(\tau_1,\tau_2)$ and therefore the self-energy can easily be integrated out. The resulting action for the two-point function $g(\tau_1,\tau_2)$ is 
\beq
I =\frac{N}{2q^2} \int \d \tau_1 \d \tau_2 \, \left\{ \frac{1}{4}\partial_{\tau_1} g(\tau_1,\tau_2) \, \partial_{\tau_2} \, \overline{g}(\tau_1,\tau_2) - \mathcal{J}^2 \, e^{\frac{1}{2}g(\tau_1,\tau_2)}e^{\frac{1}{2}\overline{g}(\tau_1,\tau_2)}\right\}.
\eeq
This action takes the form of a 2d Liouville theory for the combination $g+\bar{g}$, decoupled (ignoring boundary conditions) from the other combination $g-\bar{g}$. For simplicity we will work in an ensemble with vanishing background $\U(1)$ chemical potential. The equation of motion for the two-point function derived from this action and its solution $g(\tau_1,\tau_2)=g_0(\tau_1,\tau_2)$, are given by
\beq
-\partial_{\tau_1}\partial_{\tau_2} g(\tau_1,\tau_2) = 2\mathcal{J}^2 e^{g(\tau_1,\tau_2)},~~~e^{g_{0}(\tau_1,\tau_2)} = \left(\frac{\cos \frac{\pi v}{2}}{\cos [\pi v (\frac{1}{2}-\frac{|\tau_{12}|}{\beta})]}\right)^2,
\eeq
where
\beq
\beta \mathcal{J} = \frac{\pi v}{\cos \frac{\pi v}{2}}.
\eeq
Working at $\mu=0$ leads to the same solution as real SYK, as expected. 

We can obtain the expression of the free energy to leading order in large $N$ by evaluating the on-shell action. This calculation can be found in \cite{Maldacena:2016hyu} or \cite{Davison:2016ngz}. The Schwarzian action describes instead a subset of fluctuations around the classical solution, parametrized by 
\beq
g(\tau_1,\tau_2)  \to g_{0}(\tau_1,\tau_2)  + \delta g(\tau_1,\tau_2).
\eeq
We can divide the fluctuations into those that satisfy $\delta g(\tau_1,\tau_2) = \delta \bar{g}(\tau_1,\tau_2)$ (antisymmetric) and $\delta g(\tau_1,\tau_2) =- \delta \bar{g}(\tau_1,\tau_2)$ (symmetric). We will be interested in modes that are relevant at low temperatures.

\medskip

We begin by analyzing the antisymmetric sector. The quadratic action becomes 
\beq
I=I_{\rm classical}- \frac{N}{4q^2} \int \d \tau_1 \d \tau_2 \,\, \delta g (\tau_1,\tau_2)\, \Big\{ \frac{1}{2}\partial_{\tau_1}\partial_{\tau_2} + \mathcal{J}^2 e^{g_0(\tau_1,\tau_2)} \Big\} \delta g(\tau_1,\tau_2) .
\eeq
Following \cite{Maldacena:2016hyu} we can expand the quantum fluctuations $\delta g$ in a complete basis of functions that diagonalize the kernel appearing in the quadratic action. More precisely, we should find a complete set of functions of two times $\delta g_\lambda$ satisfying 
\beq
 \Big\{ \frac{1}{2} \partial_{\tau_1}\partial_{\tau_2} + \mathcal{J}^2 e^{g_{0}(\tau_1,\tau_2)} \Big\} \delta g_\lambda(\tau_1,\tau_2)= \lambda \mathcal{J}^2 \,e^{g_0(\tau_1,\tau_2)}\, \delta g_\lambda(\tau_1,\tau_2)
\eeq
Such a mode has an action proportional to $\lambda$ if the complete set of functions is chosen to be orthogonal
\bea
 \int \d \tau_1 \d \tau_2 \, \mathcal{J}^2e^{g_0(\tau_1,\tau_2)}\delta g_{\lambda'}(\tau_1,\tau_2) \, \delta g_{\lambda} (\tau_1,\tau_2)=\delta_{\lambda'\lambda}
\ea
The values of $\lambda$ are discrete and arise from imposing the boundary condition that $\delta g(\tau_1,\tau_1)=0$. It is useful to expand the fluctuation modes as
\beq
\delta g (\tau_1,\tau_2) = e^{-\i n y}\, \psi_n(x),~~~x=\frac{2\pi \tau_{12}}{\beta},~ y=\frac{\pi(\tau_1+\tau_2)}{\beta}.
\eeq
and the eigenvalue equation becomes
\beq
\left[ n^2 + 4 \partial_x^2 - \frac{2 (1-\lambda) v^2}{\sin^2 \frac{v x+ (1-v) \pi}{2}}\right] \psi_n (x) = 0
\eeq
This is precisely the same eigenvalue equation involved in diagonalizing the ladder kernel in the fermion four-point function. The scaling dimension $h$ of operators propagating in the four-point function is related to the action eigenvalue $\lambda$ by $h(h-1) =2(1-\lambda)$. This is not a coincidence and it is explained in section 4 of \cite{Maldacena:2016hyu}. The problem of finding both the complete set of modes and their eigenvalues was solved in the antisymmetric sector in \cite{Maldacena:2016hyu} (which includes the Schwarzian mode).

\medskip

What property identifies the relevant low-energy modes? Most quantum fluctuations have an action which remains finite at zero temperatures $\lambda \neq 0 $ as $\beta \mathcal{J} \to \infty$. These modes are too ``heavy'' to affect the low temperature regime. The special feature of the Schwarzian and phase mode is that their action vanishes as the temperature is decreased
\beq
\lambda = \frac{\lambda_0}{\beta\mathcal{J}} + \ldots, 
\eeq
where the dots denote terms further subleading in $1/\beta \mathcal{J}$. The coefficient $\lambda_0$ encodes the quadratic action of the low energy modes. Since $h(h-1)=2(1-\lambda)$ these are modes with $h$ close to two. Indeed, the analysis of \cite{Maldacena:2016hyu} shows the presence of modes labeled by an integer $n$ which, when $x \gtrsim 1/\beta\mathcal{J}$, take the form
\beq
\delta_\epsilon g = \epsilon_n\, e^{-\i n y}\, \frac{2\pi \i}{\beta}\left[\frac{\sin \frac{n x}{2}}{\tan \frac{x}{2}} - n \cos \frac{ n x}{2}\right],
\eeq
with an amplitude $\epsilon_n$ (not to be confused with the deformation parameter $\epsilon$ which breaks supersymmetry). Note that the mode vanishes for $n=-1,0,1$. The eigenvalue is
\beq
\lambda = \frac{3|n|}{\beta \mathcal{J} } + \ldots.
\eeq
These are precisely the Schwarzian modes since their profile (away from $\tau_1=\tau_2$) can be written as
\beq
\delta_\epsilon G_{\psi\bar{\psi}} = [ \Delta \epsilon'(\tau_1) + \Delta \epsilon'(\tau_2) + \epsilon(\tau_1) \partial_{\tau_1} + \epsilon(\tau_2) \partial_{\tau_2}] \, G_{\psi\bar{\psi}},
\eeq
We introduce $\epsilon(\tau) = \sum_n \epsilon_n e^{\i 2\pi n \tau/\beta}$ which can therefore be identified with an infinitesimal time reparametrization. To quadratic order their action is
\bea
I &=& \frac{N}{4 q^2} \sum_n \underbrace{\frac{3 |n|}{\beta\mathcal{J}}}_{\rm eigenvalue} \underbrace{  \frac{4\pi^2 |n| (n^2-1)}{3}}_{\rm normalization} \overline{\epsilon}_n \epsilon_n= \frac{N\pi^2}{ q^2 \beta \mathcal{J}} \sum_n (n^4-n^2) \overline{\epsilon}_n \epsilon_n,\\
&=&-\frac{\alpha_S N}{\mathcal{J}} \int \d \tau \Big\{ \tan \frac{\pi f(\tau)}{\beta}, \tau\Big\}\Big|_{\rm quadratic~order},~~~\alpha_S = \frac{1}{2q^2}.
\ea
This is the quadratic expansion of the Schwarzian action under $f(\tau) = \tau+\epsilon(\tau)$, as indicated in the second line. It would be interesting, at least in large $q$, to derive the non-linear Schwarzian action. Instead we will limit ourselves to the quadratic approximation.

\medskip

Next we consider the symmetric sector which includes the $\U(1)$ phase mode which is in the same multiplet as the Schwarzian in the supersymmetric case. This was studied in the large $q$ limit in \cite{Davison:2016ngz,gu2020notes} and also, more relevant for the present article, in \cite{Peng:2020euz}. In this sector the quadratic action simplifies
\beq
I=I_{\rm classical}- \frac{N}{4q^2} \int \d \tau_1 \d \tau_2 \,\, \delta g (\tau_1,\tau_2)\, \Big\{ \frac{1}{2}\partial_{\tau_1}\partial_{\tau_2} \Big\} \delta g(\tau_1,\tau_2) .
\eeq
The interaction term does not contribute since its variation vanishes when $\delta g = - \delta \bar{g}$. The eigenvalue equation becomes 
\beq
\left[ n^2 + 4 \partial_x^2 - \frac{2 (-\lambda) v^2}{\sin^2 \frac{v x+ (1-v) \pi}{2}}\right] \psi_n (x) = 0
\eeq
A comparison with the four-point function ladder expansion shows that the eigenvalue of the quadratic action should be identified with $h(h-1) = -2 \lambda$. Therefore in the symmetric sector a light mode corresponds to a fluctuation with $h$ close to one. The eigenvalues and precise modes can be obtained from a solution of this equation supplemented by the coincident-time boundary condition. We first point out the form of the modes which have $h=1$ when $\beta \mathcal{J}=\infty$. They are proportional to
\beq
\delta g  \sim a_n e^{\i n y} \sin \frac{n x}{2}  ,
\eeq
and vanish when $n=0$. This fluctuation arises from a local $\U(1)$ transformation $G \to e^{\i a(\tau_1) -\i a(\tau_2) } G$. The quadratic action for the phase mode is 
\beq
I_{\U(1)} = K \int \d \tau (\partial_\tau a)^2,
\eeq
where $K$, proportional to $N/\mathcal{J}$, is the compressibility. We will not focus on deriving the numerical value of the coefficient here, since when considering $\mathcal{N}=2$ SYK it is constrained by supersymmetry and it is completely determined once the coefficient of the Schwarzian modes are known (this was explicitly verified in \cite{Peng:2020euz}). This coefficient can also be extracted from a classical analysis of the on-shell action as explained in Appendix D of \cite{Davison:2016ngz}.

\medskip

\subsubsection*{Large $q$ SYK at $\epsilon\neq0$}

Let us now turn on the $\U(1)$ breaking deformation $\epsilon\neq 0$. Before taking the large $q$ limit we first establish the free theory correlators around which we will expand. The solution is the same as the one spelled out in section \ref{sec:susybreakdef},
\beq
G_{\psi \bar{\psi},0}(\tau_1,\tau_2) = \frac{1}{1-\epsilon^2} \frac{{\rm sgn}(\tau_{12})}{2},~~~G_{\psi\psi,0} (\tau_1,\tau_2) = \frac{\i \epsilon}{1-\epsilon^2} \frac{{\rm sgn}(\tau_{12})}{2}.
\eeq
The correlator in the right is only non-vanishing when the $\U(1)$ symmetry is violated. The large $q$ limit is based on the following ansatz
\beq
G_{\psi\bar{\psi}}(\tau_1,\tau_2)=G_{\psi\bar{\psi},0} \left( 1 + \frac{1}{q} g_{\psi \bar{\psi}}(\tau_1,\tau_2)\right),~~~G_{\psi\bar{\psi}}(\tau_1,\tau_2)=G_{\psi\psi,0} \left( 1 + \frac{1}{q} g_{\psi \psi}(\tau_1,\tau_2)\right).
\eeq
At large $q$ the mean field action, after integrating out the self-energies, takes the simple form
\begin{equation}\label{eqeqcomplex_SYK_action}
\begin{split}
    I &=\frac{N}{2(1-\epsilon^2)q^2} \int \d\tau_1 \d\tau_2 \Big\{\frac{1+\epsilon^2}{4(1-\epsilon^2)}\partial_{\tau_1} g_{\psi \bar{\psi}}(\tau_1,\tau_2) \,\partial_{\tau_2} g_{\psi \bar{\psi}}(\tau_2,\tau_1)- \mathcal{J}^2 e^{\frac{1}{2} g_{\psi \bar{\psi}}(\tau_1,\tau_2)+\frac{1}{2} g_{\psi \bar{\psi}}(\tau_2,\tau_1)}\\
    &+\frac{\epsilon^2}{2(1-\epsilon^2)}\Big[\frac{1}{2}\partial_{\tau_1}g_{\psi\psi}(\tau_1,\tau_2)\partial_{\tau_2}g_{\bar\psi\bar\psi}(\tau_2,\tau_1)-\partial_{\tau_1}g_{\psi\bar\psi}(\tau_1,\tau_2)\partial_{\tau_2}(g_{\psi\psi}(\tau_2,\tau_1)+g_{\bar\psi\bar\psi}(\tau_2,\tau_1))\Big]\\
    &+\frac{\epsilon^4}{8(1-\epsilon^2)}\Big[\partial_{\tau_1}g_{\psi\psi}(\tau_1,\tau_2)\partial_{\tau_2}g_{\psi\psi}(\tau_2,\tau_1)+\partial_{\tau_1}g_{\bar\psi\bar\psi}(\tau_1,\tau_2)\partial_{\tau_2}g_{\bar\psi\bar\psi}(\tau_2,\tau_1)\Big]\Big\}\\
\end{split}
\end{equation}
The first line resembles the action of double-scaled CSYK, while the terms in the second and third lines represent the $\U(1)$ breaking deformation. In terms of the original variance in the couplings $J$, the large $q$ coupling is given by $\mathcal{J}^2 = q J^2 2^{1-q}/(1-\epsilon^2)^{q-1}$, so we require $\mathcal{J}$ to have a finite limit.

\medskip

To verify that this action is correct we can take the large $q$ limit of the Schwinger-Dyson equations directly and confirm they reproduce the equations of motion of our action. At finite $q$ we have the equations arising from varying the self-energies
\bea
G^{-1}_{\psi \bar{\psi}} &=& - \i \omega - \Sigma_{\psi \bar{\psi}} + \frac{\epsilon^2 \omega^2}{-\i \omega - \Sigma_{\psi \bar{\psi}}},\\
G^{-1}_{\psi \psi} &=& \epsilon \omega  + \frac{ (-\i \omega- \Sigma_{\psi \bar{\psi}})^2}{\epsilon \omega}
\ea
The equations that arise from varying the two-point function lead to
\beq
\Sigma_{\psi \bar{\psi} } = J^2 G_{\psi \bar{\psi} }^{q-1} =\frac{J^2 2^{1-q}}{(1-\epsilon^2)^{q-1}} {\rm sgn}(\tau) e^{g_{\psi \bar{\psi}}(\tau)} = \frac{1}{q} \mathcal{J}^2 {\rm sgn}(\tau) e^{g(\tau)}.
\eeq
and the first equation becomes the Liouville equation
\bea
\partial_\tau^2 [{\rm sgn}(\tau) g_{\psi \bar{\psi}}(\tau)] &=& 2\mathcal{J}^2 \frac{1+\epsilon^2}{1-\epsilon^2} \,{\rm sgn}(\tau) e^{g_{\psi \bar{\psi}}(\tau)}.
\ea
The second Schwinger-Dyson equation can be combined with the first to give, in the large $q$ limit, a relation that determines the $\U(1)$ breaking correlator, namely
\beq
\partial_\tau^2 [{\rm sgn}(\tau)g_{\psi \psi}(\tau)] =\frac{2}{1+\epsilon^2} \partial_\tau^2 [{\rm sgn}(\tau)g_{\psi \bar\psi}(\tau)].
\eeq
The solution to these equations is quite simple. The first equation has the same solution as large $q$ SYK with a rescaling of $\mathcal{J}$. The second equation is solved by 
\beq
g_{\psi \psi}(\tau) = \frac{2}{1+\epsilon^2} g_{\psi \bar{\psi}}(\tau).
\eeq
We verify in the Appendix \ref{App:LargeQ} that the equations of motion derived from \eqref{eqeqcomplex_SYK_action} match with the ones above. 

\medskip

Notice that in the large $q$ limit, $g_{\psi\psi}$ is proportional to $g_{\psi\bar{\psi}}$ meaning that in the conformal limit they are comparable. This might naively seem to be in contradiction with the observation in the previous section that the $\U(1)$ breaking UV terms do not affect the conformal solution. The resolution is that indeed $G_{\psi\psi}$ decays faster than $G_{\psi\bar\psi}$ in the IR limit, making it negligible, but only by an extra power of $1/q$. Therefore for any finite $q$, $G_{\psi\psi}$ becomes negligible in the IR, but this conclusion can fail if we first take the large $q$ limit instead. This subtlety will not seem to affect the fact that the leading correction to the Schwarzian action is local in time.

\medskip

The next step is to expand the action to quadratic order and evaluate it on the light modes that emerge in the low-energy limit. We will work perturbatively in $\epsilon$ and focus on finding the leading order correction to the $\U(1)$ phase mode. The quadratic expansion of the terms in the first line of \eqref{eqeqcomplex_SYK_action} are already known to reproduce the Schwarzian action coupled to the $\U(1)$ phase mode and therefore we will not repeat that analysis. At quadratic order the only $\U(1)$ breaking term we might add is
\beq
I = I_{\rm Sch} + I_{\rm \U(1)}+ \alpha_{\rm def}N \int \d \tau \,a(\tau)^2.
\eeq
The first two terms are the Schwarzian action and the $\U(1)$ phase mode action. The third term is the only symmetry-breaking deformation we can have at quadratic level. Any term mixing $a$ with $\varepsilon$ is forbidden since they arise from orthogonal sectors. 

Our goal here is to obtain in the large $q$ limit the value of $\alpha_{\rm def}$. In order to do this, it is simplest to work with a subset of modes that are time-independent. In this case the action to be reproduced is $I = \alpha_{\rm def} N \beta a^2$. This case is simpler than an arbitrary time-dependent phase mode since $\U(1)$ preserving mean fields, such as $g_{\psi\bar\psi}$, will not change and we can solely focus on $\delta g_{\psi\psi}$. Taking this into account, when evaluating the quadratic action we can just focus on the following term 
\beq
I \supset - \frac{\epsilon^2}{8}\frac{N}{q^2} \int \d\tau_1 \d\tau_2  \,\,  \delta g_{\psi \psi}(\tau_1,\tau_2) \partial_1\partial_2 \,\delta g_{\bar\psi\bar\psi}(\tau_2,\tau_1).
\eeq
The zero-mode which gets a non-zero action due to this term corresponds to $\delta g_{\psi \bar{\psi}} = 0$ and $\delta g_{\psi\psi} = 2 \i a g_{\psi\psi}$ (there is an extra constant piece which does not contribute to the action due to the derivatives). We can combine this with the fact that for small deformations $g_{\psi\psi} \approx 2 g_{\psi\bar{\psi}}$ and get
\beq
I = \frac{\epsilon^2}{2q^2}N \int \d \tau_1\d\tau_2 \,\, a^2\, g_{\psi\bar\psi} \partial_1\partial_2 g_{\psi\bar\psi}=\frac{\epsilon^2}{q^2}N a^2 \frac{\beta^2\mathcal{J}^2}{4\pi^2}\,\int \d \theta_1\d\theta_2 \,\,  g_{\psi\bar\psi}  e^{g_{\psi\bar\psi}} .
\eeq
In the last equality we used the equation of motion and changed variables to the dimensionless $\theta = 2\pi \tau/\beta$ (in this subsection we are not considering supersymmetry so $\theta$ should not be confused with a Grassman variable). The integral can be explicitly evaluated leading to
\beq
\int \d \theta_1\d\theta_2 \,\,  g_{\psi\bar\psi}  e^{g_{\psi\bar\psi}} =  \frac{4 \pi (\pi v (1+ \cos \pi v) - 2 \sin \pi v)}{v} \sim 8 \pi^2 (v-1) \sim 16 \pi^2 \frac{1}{\beta \mathcal{J}}.
\eeq
The integrand becomes localized at the coincidence point $\tau_1=\tau_2$ when $v\to1$, consistent with the expectation that the action is local. This leads to an action
\beq
I = \frac{\epsilon^2}{q^2} N a^2 \frac{\beta^2 \mathcal{J}^2}{4\pi^2 } 16 \pi^2 \frac{1}{\beta \mathcal{J}} =  \frac{4\epsilon^2}{q^2} N  \beta \mathcal{J} \, a^2,~~~~\Rightarrow~~~\alpha_{\rm def} =\epsilon^2\frac{4\mathcal{J}}{q^2}
\eeq
Plugging this back in the action for the Schwarzian and phase mode we obtain
\beq
I = I_{\rm Sch} + I_{\rm \U(1)}+ \epsilon^2\frac{4\mathcal{J}}{q^2}N \int \d \tau \,a(\tau)^2.
\eeq
In the large $q$ limit we can also evaluate the $\U(1)$ breaking action at small $\epsilon$ but for finite phase mode profiles $a(\tau)$. We do this calculation in Appendix \ref{app:U1effective} and obtain
\beq
I = I_{\rm Sch} + I_{\rm \U(1)}- \epsilon^2\frac{2\mathcal{J}}{q^2}N \int \d \tau \,[\cos(2 a(\tau))-1].
\eeq
This form is consistent with the periodicity of $a(\tau)$ as a phase mode, and might be important in attempting to give a gravity interpretation of this result.

We can estimate the value of $\epsilon$ for which this $\U(1)$ breaking term affects the Schwarzian regime of the model. The first two terms in the action will be of order $1/(\beta \mathcal{J})N/q^2$ while the third term is of order $\epsilon^2 \beta \mathcal{J} N/q^2$. Therefore the third term is negligible
\beq
\epsilon \lesssim \frac{1}{\beta \mathcal{J}}\ll 1.
\eeq
If we wish to focus on the regime of temperatures where the quantum effects arising from the Schwarzian mode are important, we need to take $\beta \mathcal{J} \sim N$ (in the large $q$ limit this is really $N/q^2$ but we work in the regime where $q^2/N \ll 1$). Therefore for the deformation to affect the quantum Schwarzian regime but \emph{not} the classical Schwarzian regime we need 
\beq
\epsilon \lesssim \frac{O(1)}{N}. 
\eeq
For deformations smaller than this value, its effect will be concentrated in the quantum region of the spectrum. This point will be important when we apply similar ideas to deformations of the $\mathcal{N}=2$ SYK model.

\subsection{$\mathcal{N}=2$ SYK: Large $\hat{q}$ action}

We will now repeat the previous analysis in the context of $\mathcal{N}=2$ SYK. Since the large $\hat{q}$ Liouville type action of $\mathcal{N}=2$ SYK has not been derived in the literature, we will fill this gap here. We will consider the $\U(1)$ breaking deformation of SYK in the next section. 

In the large $\hat{q}$ limit we consider the following ansatz for all correlators
\begin{equation}
\begin{split}
G_{\psi \bar{\psi} }(\tau_1,\tau_2) =& \frac{\text{sgn}(\tau_{12})}{2}\left[ 1 + \frac{g_{\psi \bar\psi}(\tau_1,\tau_2)}{\hat{q}} \right],\quad\Sigma_{\bar\psi \psi }(\tau_1,\tau_2) = \frac{\sigma_{\bar\psi\psi }(\tau_1,\tau_2)}{\hat{q}}
 \\
G_{b\bar{b}}(\tau_1,\tau_2) =& - \delta(\tau_{12}) + \frac{g_{b\bar{b}}(\tau_1,\tau_2)}{2\hat{q}}, \quad \qquad \, \, \Sigma_{\bar{b}b}(\tau_1,\tau_2) \,\, = \, \frac{\sigma_{\bar{b}b }(\tau_1,\tau_2)}{\hat{q}}\\
G_{\psi \bar{b}}(\tau_1,\tau_2) =& \frac{g_{\psi\bar{b}}(\tau_1,\tau_2)}{\hat{q}}, \,  \qquad \qquad \qquad \qquad \Sigma_{\bar{b}\psi}(\tau_1,\tau_2) = \frac{\sigma_{\bar{b}\psi}(\tau_1,\tau_2)}{\hat{q}}\\
G_{b\bar\psi}(\tau_1,\tau_2) =& \frac{g_{b\bar\psi }(\tau_1,\tau_2)}{\hat{q}}, \,  \qquad \qquad \qquad \qquad \Sigma_{\bar\psi b}(\tau_1,\tau_2) = \frac{\sigma_{\bar\psi b}(\tau_1,\tau_2)}{\hat{q}}.
\end{split}
\end{equation}
We next insert this ansatz in the mean field action and expand at large $\hat{q}$. This leads to an action that is quadratic in the rescaled self-energies $\sigma(\tau_1,\tau_2)$ and therefore will be integrated out. We present details on this calculation in Appendix \ref{App:LargeQ}. The action becomes
\begin{equation}
\begin{split}
I &= \frac{N}{8\hat{q}^2} \int \d\tau_1 \d\tau_2\, \Big\{-\partial_{\tau_1} g_{\psi \bar{\psi}}(\tau_1,\tau_2) \,\partial_{\tau_2} g_{\psi \bar{\psi}}(\tau_2,\tau_1) + g_{b \bar{b}}(\tau_1,\tau_2) \,g_{b \bar{b}}(\tau_2,\tau_1)\\
&+\partial_{\tau_1}g_{\psi \bar{b}}(\tau_1,\tau_2)  g_{b\bar{\psi}}(\tau_2,\tau_1) - g_{b\bar{\psi}}(\tau_1,\tau_2) \partial_{\tau_2} g_{\psi \bar{b}}(\tau_2,\tau_1)\\
&-\mathcal{J} (g_{b \bar{b}}+g_{b \bar{\psi}} g_{\psi \bar{b}}) \, \exp \, g_{\psi \bar{\psi}} -{\rm h.c.}\Big\}
\end{split}
\end{equation}
The arguments of the last term are all $(\tau_1,\tau_2)$. In Appendix \ref{App:CHECKSUSY} we verify the supersymmetry of this action under the transformation
\begin{equation}\label{ssksk}
\begin{split}
\delta_Q\bar\psi_i=\varepsilon_1\bar{b}_i,\quad\delta_Q b^i=\varepsilon_1\partial_\tau\psi^i,\quad\delta_Q\psi^i=\delta_Q\bar{b}_i=0,\\
\delta_{\bar{Q}}\psi^i=\varepsilon_2b^i,\quad\delta_{\bar{Q}}\bar{b}_i=\varepsilon_2\partial_\tau\bar\psi_i,\quad\delta_{\bar{Q}}\bar\psi_i=\delta_{\bar{Q}} b^i=0,
\end{split}
\end{equation}
where $\varepsilon_1$ and $\varepsilon_2$ are the supersymmetry transformation parameters. 

The large $\hat{q}$ action for $\mathcal{N}=2$ SYK resembles a complexification of the action of two-dimensional $\mathcal{N}=1$ Liouville theory where the right- and left-moving supersymmetries map to two supersymmetries of SYK. It would be interesting to study further this generalization of Liouville as a 2d CFT, but we leave this for future work. 

\bigskip

Having found the large $\hat{q}$ action, we can derive the equations of motion and find the classical solution for the mean fields. (We verify in an appendix these match the large $\hat{q}$ limit of the Schwinger-Dyson equations.) Since the field $g_{\psi \bar{b}}$ is fermionic, it necessarily has to vanish on-shell. The equation for $g_{\psi\bar\psi}$ and $g_{b\bar{b}}$, assuming vanishing charge and also time-translation invariance, are 
\beq
\partial_\tau^2 g_{\psi\bar{\psi}}(\tau) = \mathcal{J} g_{b\bar{b}}(\tau)\, e^{g_{\psi\bar\psi}(\tau)},~~~~g_{b\bar{b}}(\tau)=\mathcal{J}e^{g_{\psi\bar\psi}(\tau)}.
\eeq
These equations are valid for $\tau \neq 0$. Inserting the second equation in the RHS of the first equation leads to a Liouville problem  $\partial_\tau^2 g_{\psi\bar\psi} = \mathcal{J}^2 e^{2 g_{\psi\bar\psi}(\tau)}$ just like in complex SYK upon a rescaling $g_{\psi\bar{\psi}}\to g_{\psi\bar{\psi}}/2$. The solution to this equation with appropriate boundary conditions is
\beq
e^{g_{\psi\bar\psi}(\tau)} = \frac{\cos \frac{\pi v}{2}}{\cos \big( \frac{\pi v}{2} -\frac{\pi v \tau}{2\beta}\big)},~~~~\beta \mathcal{J} = \frac{\pi v}{2 \cos \frac{\pi v}{2}}.
\eeq
The solution, in a vanishing $R$-charge ensemble, is the same as $\mathcal{N}=1$ SYK found in \cite{Fu:2016vas}.

\bigskip

The next step is to expand the action to quadratic order and keep all the modes with an action of order $1/\beta \mathcal{J}$ which otherwise would be zero-modes in the low-energy limit. At the quadratic level this problem was solved in \cite{Peng:2020euz} using the approach of ladder diagrams (instead of using the large $\hat{q}$ action). We can simplify the discussion by exploiting supersymmetry. First, notice that one can easily integrate out $g_{b\bar{b}}$ and obtain an action
\beq
I \supset \frac{N}{8\hat{q}^2} \int \d\tau_1\d\tau_2\left\{ -\partial_{\tau_1} g_{\psi\bar{\psi}}(\tau_1,\tau_2)\partial_{\tau_2} g_{\psi\bar{\psi}}(\tau_2,\tau_1)+\mathcal{J}^2 e^{g_{\psi\bar{\psi}}(\tau_1,\tau_2)+g_{\psi\bar{\psi}}(\tau_2,\tau_1)} \right\},
\eeq
up to terms that involve fermionic correlators. We can compute the quadratic action of the Schwarzian modes using these two terms alone. Time reparametrizations will not source fermionic fluctuations given that they vanish on-shell and therefore the extra terms would not participate. The two terms above look identical, up to normalizations, to complex SYK so we can import that result. A determination of the coupling of the Schwarzian action, combined with supersymmetry, leads to
\beq
I = - \frac{\alpha_S N}{\mathcal{J}} \int \d\tau\, \frac{1}{2}\Big \{ (\partial_\tau^2 \epsilon)^2 - (\partial_\tau \epsilon)^2 + 4(\partial_\tau a)^2 + 2 (4 \eta \partial_\tau^3 \bar{\eta} + \eta \partial_\tau \bar{\eta}) \Big\}.
\eeq
where $\alpha_S = \frac{1}{8 \hat{q}^2}$. This is the linearized version of the $\mathcal{N}=2$ Schwarzian action. This was verified at the non-linear level by solving the theory in the double-scaled regime and comparing with the Schwarzian results in \cite{berkooz2020double}. The $\U(1)$ phase mode also arises from bosonic terms, while to obtain the action of fermionic modes parametrized by $\eta(\tau)$ we need to look at fermionic correlators. In the next section we will find the leading order correction to this action in the presence of a $\U(1)$ breaking deformation.

\subsection{$\mathcal{N}=2$ SYK: $\U(1)$ breaking deformation}

Finally we can incorporate the $\U(1)$ breaking deformation in the large $\hat{q}$ limit. We leave the details for Appendix \ref{App:LargeQ} and point out the main results. The large $\hat{q}$ mean fields $g_{XY}(\tau_1,\tau_2)$ associated to a correlator between fields $X$ and $Y$ are defined in the same way as in the cases above and the precise definitions are given in equation \eqref{large_q_G}. We can integrate out the self energies and obtain an action for these fields. The result has two parts $I=I_1+I_2$, a bosonic part $I_1$ and a fermionic part $I_2$, and are given explicitly by
\begin{equation}
\begin{split}
    I_1 &= \frac{N}{4\hat{q}^2}\int \d\tau_1 \d\tau_2 \Big\{ \mathcal{J}(1-\epsilon^2) g_{b \bar{b}}\exp \, g_{\psi \bar{\psi}}\\
    &+\frac{1+\epsilon^2}{(1-\epsilon^2)^2}\frac{1}{2} \Big(-\partial_{\tau_1} g_{\psi \bar{\psi}}(\tau_1,\tau_2) \,\partial_{\tau_2} g_{\psi \bar{\psi}}(\tau_2,\tau_1) + g_{b \bar{b}}(\tau_1,\tau_2) \,g_{b \bar{b}}(\tau_2,\tau_1)\Big) \\
    &+\frac{\epsilon^2}{(1-\epsilon^2)^2}\left(\frac{1}{2}g_{bb}(\tau_1,\tau_2)g_{\bar{b}\bar{b}}(\tau_2,\tau_1)-g_{b\bar{b}}(\tau_1,\tau_2)g_{bb}(\tau_2,\tau_1)-g_{b\bar{b}}(\tau_1,\tau_2)g_{\bar{b}\bar{b}}(\tau_2,\tau_1)\right.\\
    &-\frac{1}{2}\partial_{\tau_1}g_{\psi\psi}(\tau_1,\tau_2)\partial_{\tau_2}g_{\bar\psi\bar\psi}(\tau_2,\tau_1)+\partial_{\tau_1}g_{\psi\bar\psi}(\tau_1,\tau_2)\partial_{\tau_2}g_{\psi\psi}(\tau_2,\tau_1)+\partial_{\tau_1}g_{\psi\bar\psi}(\tau_1,\tau_2)\partial_{\tau_2}g_{\bar\psi\bar\psi}(\tau_2,\tau_1)\Big)\\
    &+\frac{\epsilon^4}{(1-\epsilon^2)^2}\left(\frac{1}{4}g_{bb}(\tau_1,\tau_2)g_{bb}(\tau_2,\tau_1)+g_{\bar{b}\bar{b}}(\tau_1,\tau_2)g_{\bar{b}\bar{b}}(\tau_2,\tau_1)\right.\\
    &\left.-\frac{1}{4}\partial_{\tau_1}g_{\psi\psi}(\tau_1,\tau_2)\partial_{\tau_2}g_{\psi\psi}(\tau_2,\tau_1)-\frac{1}{4}\partial_{\tau_1}g_{\bar\psi\bar\psi}(\tau_1,\tau_2)\partial_{\tau_2}g_{\bar\psi\bar\psi}(\tau_2,\tau_1)\right)\Big\}\\
\end{split}
\end{equation}
\begin{equation}
\begin{split}
    I_2&=\frac{N}{4\hat{q}^2}\int \d\tau_1 \d\tau_2\, \Big\{ \mathcal{J}(1-\epsilon^2)\ g_{\psi\bar{b}} g_{b \bar\psi} \ \exp \, g_{\psi \bar{\psi}}\\
    &+\frac{1+\epsilon^2}{(1-\epsilon^2)^2}\frac{1}{2} \Big( \partial_{\tau_1}g_{\psi \bar{b}}(\tau_1,\tau_2)  g_{b\bar{\psi}}(\tau_2,\tau_1) - g_{b\bar{\psi}}(\tau_1,\tau_2) \partial_{\tau_2} g_{\psi \bar{b}}(\tau_2,\tau_1)\Big)\\
    &+\frac{\epsilon^2}{(1-\epsilon^2)^2}\Big( \frac{1}{2}g_{b\psi }(\tau_1,\tau_2)\partial_{\tau_2}g_{\bar{b}\bar\psi}(\tau_1,\tau_2)-\frac{1}{2}\partial_{\tau_1}g_{b\psi}(\tau_2,\tau_1)g_{\bar{b}\bar\psi}(\tau_2,\tau_1)+g_{b\bar\psi}(\tau_1,\tau_2)\partial_{\tau_2}g_{b\psi}(\tau_1,\tau_2)\\
    &-\partial_{\tau_1}g_{\psi \bar{b}}(\tau_1,\tau_2)g_{b\psi}(\tau_2,\tau_1)+g_{b\bar\psi}(\tau_1,\tau_2)\partial_{\tau_2}g_{\bar{b}\bar\psi}(\tau_1,\tau_2)-\partial_{\tau_1}g_{\psi \bar{b}}(\tau_1,\tau_2)g_{\bar{b}\bar\psi}(\tau_2,\tau_1)\Big)\\
    &+\frac{\epsilon^4}{(1-\epsilon^2)^2}\left(\frac{1}{4}\partial_{\tau_1}g_{\psi b}(\tau_1,\tau_2)  g_{b\psi}(\tau_2,\tau_1) -\frac{1}{4} g_{b\psi}(\tau_1,\tau_2) \partial_{\tau_2} g_{\psi b}(\tau_2,\tau_1)\right.\\
    &\left.+\frac{1}{4}\partial_{\tau_1}g_{\bar\psi \bar{b}}(\tau_1,\tau_2)  g_{\bar{b}\bar{\psi}}(\tau_2,\tau_1) - \frac{1}{4}g_{\bar{b}\bar{\psi}}(\tau_1,\tau_2) \partial_{\tau_2} g_{\bar\psi \bar{b}}(\tau_2,\tau_1)\right)\Big\}\\
\end{split}
\end{equation}
The action is quite complicated but we see that as $\epsilon\to0$ we recover $\mathcal{N}=2$ SYK. After some rearrangement, the equations of motion become
\begin{equation}
\begin{split}
    \frac{(1-\epsilon^2)}{1+\epsilon^2}\partial_1\partial_2g_{\psi\bar\psi}(\tau_2,\tau_1)=&\mathcal{J}g_{b \bar{b}}\exp g_{\psi \bar{\psi}}\\
    \frac{(1-\epsilon^2)}{1+\epsilon^2}g_{b\bar{b}}(\tau_2,\tau_1)=&\mathcal{J}\exp g_{\psi \bar{\psi}}\\
\end{split}
\end{equation}
Up to a rescaling of $\mathcal{J}$ these are precisely the same equations as the undeformed $\mathcal{N}=2$ SYK. We saw a similar phenomenon in the solution of complex SYK as well. The solution to the Liouville equation is
\begin{equation}
    e^{2g_{\psi\bar\psi}}=\left(\frac{\cos\frac{\pi v}{2}}{\cos(\pi v(\frac{1}{2}-\frac{|\tau|}{\beta}))}\right)^2
\end{equation}
with 
\begin{equation}
    \beta\mathcal{J}\frac{1+\epsilon^2}{(1-\epsilon^2)^2}=\frac{\pi v}{\cos\frac{\pi v}{2}}.
\end{equation}
The $\U(1)$ breaking correlators such as $g_{\psi\psi}$ or $g_{bb}$ are simply determined by the $\mathcal{N}=2$ SYK solution as well
\begin{equation}
\begin{split}    g_{\psi\psi}(\tau_2,\tau_1)=&\frac{2}{1+\epsilon^2}g_{\psi\bar\psi}(\tau_2,\tau_1)\\
  g_{bb}(\tau_2,\tau_1)=&\frac{2}{1+\epsilon^2}g_{b\bar{b}}(\tau_2,\tau_1)\\
\end{split}
\end{equation}
These equations that we obtained from our large $\hat{q}$ action can be reproduced by taking the large $\hat{q}$ limit of the Schwinger-Dyson equations from section \ref{sectionthree}.

Finally we can compute the coefficient of the $\U(1)$ breaking term in the low energy effective action. For small values of the $\U(1)$ phase mode $a(\tau)$ the action will take the form
\beq
I = I_{\mathcal{N}=2 ~{\rm Sch}} + \alpha_{\rm def}N \int \d \tau \,a(\tau)^2.
\eeq
The Schwarzian modes have an action at order $\epsilon^0$ but the $\U(1)$ breaking term $\alpha_{\rm def}$ will get a contribution at order $\epsilon^2$. To compute it, we apply a $\U(1)$ transformation to all fields and evaluate the action in the small $\epsilon$ limit. The calculation is similar to the $\U(1)$ breaking deformation of CSYK and we present the details in Appendix \ref{App:LargeQ}. The result is 
\begin{align}
    \alpha_{\rm def} = \frac{16\epsilon^2 \mathcal{J}}{(\pi\hat{q}^2)} \, .
\end{align}
We also verified that in the large $\hat{q}$ limit the action becomes proportional to $\cos(2a)-1$, although its not clear whether this specific form generalizes to finite $\hat{q}$.

\section{Discussion}\label{sec:discussion}
In this paper, we have introduced the simplest supersymmetry-breaking deformation of $\mathcal{N}=2$ SYK that interpolates between a spectrum with a large ground state degeneracy and a large gap to a spectrum with no ground state degeneracy and exponentially small gaps. We performed numerical simulations of the deformed SYK model and we introduced and a numerical solution of the large $N$ Schwinger-Dyson equations. We extracted in the large $\hat{q}$ limit some aspects of the deformed $\mathcal{N}=2$ Schwarzian theory describing the low energy phase of the model, including which reparametrization modes are made massive by the deformation. Our analysis leaves several interesting open directions to investigate:

\paragraph{Supersymmetry-breaking deformations of 2d gravity:} We mostly focused on a quantum mechanical analysis in SYK-like models which are known to have an emergent (approximate) description in terms of AdS$_2$ supergravity. It would be interesting to extend this two-dimensional bulk gravity picture to include the infrared effects of the supersymmetry-breaking deformations introduced here. Our deformation leaves the IR phase unchanged but affects the UV behavior of the model. This suggests a natural proposal: one should study $\mathcal{N}=2$ Jackiw-Teitelboim gravity in $\AdS_2$ with boundary conditions that preserve one supercharge instead of two, and that furthermore break the $\U(1)$ symmetry of the model. This is because the matching conformal solutions imply both the deformed and undeformed theory have an AdS$_2$ vacuum with unbroken supersymmetry, and the symmetry breaking that occurs at finite temperature is usually due to the boundary modes. One possibility could be to construct a deformation of $\mathcal{N}=2$ Jackiw-Teitelboim gravity using the supersymmetric defects analyzed in \cite{Turiaci:2023jfa}. 

\paragraph{Exact solution and implications for higher-dimensional black holes:} In the context of near-extremal black holes described by Jackiw-Teitelboim gravity and the Schwarzian theory, having access to an exact solution is crucial. The reason is that the Schwarzian (or a certain gravity mode in higher dimensions) becomes strongly coupled at low temperatures and therefore to decide whether a gap is present and whether the zero-temperature entropy translates into a ground state degeneracy we need to exactly quantize the theory. This kind of analysis is possible in the JT limit because the Schwarzian theory is exactly solvable using localization, but it is unclear whether our deformed effective action is solvable in the same way. Such an analysis would make it possible to analytically understand how the BPS states get lifted by the deformation. It is therefore important to find solvable deformations of Jackiw-Teitelboim gravity in which $\U(1)_R$ is broken by boundary conditions, connecting with the previous paragraph, and identify those that describe the quantum corrections to higher dimensional black holes. 

\medskip 

As explained in the introduction, the kind of higher dimensional black hole that might be described by this kind of susy breaking model is given by a near-extremal charged black hole in $\AdS_4$ with a small cosmological constant. For completeness, we can explicitly compare the situation of our SYK model with this case. Firstly, both models have a conformal phase. Our model breaks the $\mathcal{N}=2$ Schwarzian theory down to $\mathcal{N}=1$, while the black hole setup would break the $\mathcal{N}=4$ Schwarzian to $\mathcal{N}=0$\footnote{Unfortunately, there is no known $\mathcal{N}=4$ SYK model with a Schwarzian phase to work with \cite{Heydeman:2020hhw}.}. As mentioned earlier, both $\mathcal{N}=0$ and $1$ Schwarzians display neither a gap nor a large ground state degeneracy so their spectra are qualitatively similar, though they begin to differ at sufficiently low energies. Secondly, in the black hole case, the presence of the cosmological constant can also affect the scaling dimensions of matter fields since it modifies the size of the $S^2$ compared with the radius of $\AdS_2$. Our setup is more rigid in the sense that it affects solely the Schwarzian sector. Finally in the case of black holes in AdS, the near-horizon $R$-symmetry corresponding to rotations is unbroken, while our models break $R$-symmetry. Most of these differences could be addressed by studying a deformation of $\mathcal{N}=2$ SYK model by a Hamiltonian which looks like complex SYK with $q=2\hat{q}$, since the $R$-symmetry would be unbroken, but the details of the IR solutions would also be affected by the deformation. Moreover, this deformation would break all supercharges. We leave a more thorough analysis of the connection with gravity for future work\footnote{We thank Finn Larsen and Siyul Lee for discussions on related ideas.}.

\paragraph{Correlators:} Besides the spectrum, interesting physics of the model is present for matter correlators of Jackiw-Teitelboim gravity \cite{Mertens:2017mtv,lin2023looking}. Even though classically the correlators do not depend on the amount of supersymmetry, quantum effects become enhanced at late times even when $\beta J \ll N$. These quantum effects do depend on the number of supercharges.  For example, matter correlators of the $\mathcal{N}=2$ Schwarzian theory become time-independent at late times (related to the presence of exact zero energy BPS states), while the $\mathcal{N}\leq 1$ Schwarzian predicts a late-time decay with universal power laws that depend on the number of supercharges. With a solvable model we could study how these late-time properties interpolate between each other.

\paragraph{Chaos and supersymmetry breaking:} The connection between quantum chaos and black holes was briefly reviewed in the introduction. Chaotic theories display a spectrum which shares statistical properties with random matrix ensembles. Such ensembles depend on the amount of symmetries the problem displays, related to the spectrum of charges of the black hole. Ensembles with unitary global symmetries predict that charge sectors transforming in different unitary representations are statistically independent. For the example related to supersymmetric black holes, the ensemble corresponding to $\mathcal{N}=2$ quantum mechanics was derived in \cite{Turiaci:2023jfa}. The main two features are that the different supermultiplets are statistically independent, and the wavefunctions of BPS states are random (a fact suggested earlier in \cite{lin2023looking}). It is worth investigating features of chaotic theories with approximate symmetries in general (see for example \cite{Baumgartner:2024orz}) and the relation to gravity. This kind of analysis applied when supersymmetry is an approximate symmetry might give a characterization of how the BPS black holes are organized in the bigger Hilbert space.

\section*{Acknowledgements}
%~~~~~~~~~~~~~~~~~~~~~~~~~~~~~~~~
It is a pleasure to thank Y. Chen, C. Johnson, R. Emparan, D. Jafferis, F. Larsen, S. Lee, J. Maldacena, M. Mezei, A. Shapere, D. Stanford for valuable discussions. MTH is supported by Harvard University and the Black Hole Initiative, funded in part by the Gordon and Betty Moore Foundation (Grant 8273.01) and the John Templeton Foundation (Grant 62286). MTH completed part of this work while at the Kavli Institute for Theoretical Physics (KITP), supported in part by grant NSF PHY-2309135. GJT was supported by the University of Washington and the DOE award DE-SC0024363. Part of GJTs work was performed at Aspen Center for Physics, which is supported by National Science Foundation grant PHY-2210452.

\newpage
\appendix

\section{Large $q$ actions}\label{App:LargeQ}
The goal of this section is to derive the large-$q$ action for the supersymmetric SYK model and the complex SYK model, both with the symmetry breaking deformations. 

\subsection{Deformed SUSY SYK model}
We expand the bilocal fields in the large $q$ limit to the order of $\mathcal{O}\left(\frac{1}{\hat{q}}\right)$, 
\begin{equation}\label{large_q_G}
\begin{split}
G_{\psi \bar{\psi} }(\tau_1,\tau_2) =& \frac{1}{1-\epsilon^2}\frac{\text{sgn}(\tau_{12})}{2}\left[ 1 + \frac{g_{\psi \bar\psi}(\tau_1,\tau_2)}{\hat{q}} \right],\ \Sigma_{\bar\psi\psi}(\tau_1,\tau_2) =\frac{\sigma_{\bar\psi\psi}(\tau_1,\tau_2)}{\hat{q}},\\
G_{b\bar{b}}(\tau_1,\tau_2) =& \frac{1}{1-\epsilon^2}\left[-\delta(\tau_{12}) + \frac{g_{b\bar{b}}(\tau_1,\tau_2)}{2q}\right],\ \Sigma_{b\bar{b}}(\tau_1,\tau_2) =\frac{\sigma_{b \bar{b}}(\tau_1,\tau_2)}{\hat{q}},\\
G_{\psi \psi }(\tau_1,\tau_2) =& \frac{\i\epsilon}{1-\epsilon^2}\frac{\text{sgn}(\tau_{12})}{2}\left[ 1 + \frac{g_{\psi \psi}(\tau_1,\tau_2)}{\hat{q}} \right],\ \Sigma_{\psi \psi}(\tau_1,\tau_2) = -\i\epsilon\frac{\sigma_{\psi \psi}(\tau_1,\tau_2)}{\hat{q}}, \\
G_{bb}(\tau_1,\tau_2) =& \frac{\i\epsilon}{1-\epsilon^2}\left[-\delta(\tau_{12}) + \frac{g_{bb}(\tau_1,\tau_2)}{2q}\right],\ \Sigma_{bb}(\tau_1,\tau_2) = -\i\epsilon\frac{\sigma_{b b}(\tau_1,\tau_2)}{\hat{q}}, \\
G_{\bar\psi \bar\psi }(\tau_1,\tau_2) =& \frac{-\i\epsilon}{1-\epsilon^2}\frac{\text{sgn}(\tau_{12})}{2}\left[ 1 + \frac{g_{\bar\psi \bar\psi }(\tau_1,\tau_2)}{\hat{q}} \right],\ \Sigma_{\bar\psi \bar\psi}(\tau_1,\tau_2) = \i\epsilon\frac{\sigma_{\bar\psi \bar\psi}(\tau_1,\tau_2)}{\hat{q}},\\
G_{\bar{b} \bar{b}}(\tau_1,\tau_2) =& \frac{-\i\epsilon}{1-\epsilon^2}\left[-\delta(\tau_{12}) + \frac{g_{\bar{b} \bar{b}}(\tau_1,\tau_2)}{2q}\right],\ \Sigma_{\bar{b}\bar{b}}(\tau_1,\tau_2) = \i\epsilon\frac{\sigma_{\bar{b}\bar{b}}(\tau_1,\tau_2)}{\hat{q}}, \\
G_{\psi \bar{b}}(\tau_1,\tau_2) =&\frac{g_{\psi\bar{b}}(\tau_1,\tau_2)}{\hat{q}},\ \Sigma_{\bar{b}\psi}(\tau_1,\tau_2) = \frac{\sigma_{\bar{b}\psi}(\tau_1,\tau_2)}{\hat{q}},\\
G_{\bar\psi b}(\tau_1,\tau_2) =&\frac{g_{\bar\psi b}(\tau_1,\tau_2)}{\hat{q}},\ \Sigma_{b\bar\psi}(\tau_1,\tau_2) = \frac{\sigma_{b\bar\psi}(\tau_1,\tau_2)}{\hat{q}},\\
G_{\psi b}(\tau_1,\tau_2) =&-\i\epsilon\frac{g_{\psi b}(\tau_1,\tau_2)}{\hat{q}},\ \Sigma_{b\psi}(\tau_1,\tau_2) = \i\epsilon\frac{\sigma_{b\psi}(\tau_1,\tau_2)}{\hat{q}},\\
G_{\bar\psi\bar{b}}(\tau_1,\tau_2) =&-\i\epsilon\frac{g_{\bar\psi b}(\tau_1,\tau_2)}{\hat{q}},\ \Sigma_{\bar{b}\bar\psi}(\tau_1,\tau_2) = \i\epsilon\frac{\sigma_{\bar{b}\bar\psi}(\tau_1,\tau_2)}{\hat{q}}.
\end{split}
\end{equation}
where $g$'s and $\sigma$'s are expected to be finite. On top of the action (\ref{Eqn:effective_action}), the following effective action of the deformed SUSY SYK model also incorporates the fermionic bilocal fields. We will expand this action in the large-$\hat{q}$ limit by expanding the bilocal fields above.
\begin{equation}
\begin{split}
    \frac{I}{N}=&\frac{K}{N}+\int d\tau_1 d\tau_2(-JG_{b\bar{b}}G^{\hat{q}-1}_{\psi\bar\psi}+G_{\psi\bar\psi}(\tau_2,\tau_1)\Sigma_{\bar\psi\psi}(\tau_1,\tau_2)+G_{b\bar{b}}(\tau_2,\tau_1)\Sigma_{\bar{b}b}(\tau_1,\tau_2)\\
    &+G_{\psi\psi}(\tau_2,\tau_1)\Sigma_{\bar\psi}(\tau_1,\tau_2)+G_{\bar\psi\bar\psi}(\tau_2,\tau_1)\Sigma_{\bar\psi\bar\psi}(\tau_1,\tau_2)+G_{bb}(\tau_2,\tau_1)\Sigma_{bb}(\tau_1,\tau_2)\\
    &+G_{\bar{b}\bar{b}}(\tau_2,\tau_1)\Sigma_{\bar{b}\bar{b}}(\tau_1,\tau_2)+G_{\psi\bar{b}}(\tau_2,\tau_1)\Sigma_{\bar{b}\psi}(\tau_1,\tau_2)+G_{\bar\psi b}(\tau_2,\tau_1)\Sigma_{b\bar\psi}(\tau_1,\tau_2))\\
    &+G_{\psi b}(\tau_2,\tau_1)\Sigma_{b\psi}(\tau_1,\tau_2)+G_{\bar\psi\bar{b}}(\tau_2,\tau_1)\Sigma_{\bar{b}\bar\psi}(\tau_1,\tau_2)).
\end{split}
\end{equation}
where kinetic terms $K$ in the effective action in terms of $g$ and $\sigma$ is
\begin{equation}
    \frac{K}{N}=
    \frac{1}{2}\log \text{Ber}
    \begin{pmatrix}
        -\frac{\i\epsilon}{2}\sigma_b+\frac{\i\epsilon}{\hat{q}}\sigma_{bb} & -\frac{1}{2}(\sigma_b+\frac{1}{\hat{q}}\sigma_{\bar{b}b}(\tau_2,\tau_1))& -\frac{\i\epsilon}{2\hat{q}}\sigma_{b\psi} &-\frac{1}{2\hat{q}}\sigma_{b\bar\psi}\\
        -\frac{1}{2}(\sigma_b+\frac{1}{\hat{q}}\sigma_{\bar{b}b})& \frac{\i\epsilon}{2}\sigma_b-\frac{\i\epsilon}{\hat{q}}\sigma_{\bar{b}\bar{b}} & -\frac{1}{2\hat{q}}\sigma_{\bar{b}\psi} & -\frac{\i\epsilon}{2\hat{q}}\sigma_{\bar{b}\bar\psi}\\
        -\frac{\i\epsilon}{2\hat{q}}\sigma_{b\psi}(\tau_2,\tau_1) & -\frac{1}{2\hat{q}}\sigma_{\bar{b}\psi}(\tau_2,\tau_1) & -\frac{\i\epsilon}{2}\sigma_f+\frac{\i\epsilon}{\hat{q}}\sigma_{\psi\psi}
        & -\frac{1}{2}(\sigma_f-\frac{1}{\hat{q}}\sigma_{\bar\psi\psi}(\tau_2,\tau_1)) \\
        -\frac{1}{2\hat{q}}\sigma_{b\bar\psi}(\tau_2,\tau_1) & -\frac{\i\epsilon}{2\hat{q}}\sigma_{\bar{b}\bar\psi}(\tau_2,\tau_1) & -\frac{1}{2}(\sigma_f+\frac{1}{\hat{q}}\sigma_{\bar\psi\psi}) & \frac{\i\epsilon}{2}\sigma_f-\frac{\i\epsilon}{\hat{q}}\sigma_{\bar\psi\bar\psi}
    \end{pmatrix}.
\end{equation}
In this action, except for those fields with flipped time variables explicitly written above, the rest of the $\sigma$'s have time variables in normal order $(\tau_1,\tau_2)$. By flipping the time variables, we relate some bilocal fields to their counterparts, e.g., $\sigma_{b\psi}$ and $\sigma_{\psi b}$. We expand the action in the large $\hat{q}$ limit by inserting the large-$\hat{q}$ expansion of the bilocal fields in (\ref{large_q_G}), leading to an action of $g$ and $\sigma$. From now on, we calculate the action in four parts separately: the action including bilocal fields of bosons (like $\sigma_{b\bar{b}}$), of fermions (like $\sigma_{\psi\bar\psi}$), the action including the mixed bilocal fields (like $\sigma_{b\bar\psi}$), and the interaction term. They are
\begin{equation}
\begin{split}
    \frac{I_b}{N}-\frac{I_{b(\hat{q}=\infty)}}{N}=&-\frac{1}{\hat{q}^2(1-\epsilon^2)^2}\text{Tr}(\frac{1}{2}(1+\epsilon^2)\sigma_b^{-1}*\sigma_{b\bar{b}}*\sigma_b^{-1}*\sigma_{b\bar{b}}\\
    &+\epsilon^4(\sigma_b^{-1}*\sigma_{bb}*\sigma_b^{-1}*\sigma_{bb}+\sigma_b^{-1}*\sigma_{\bar{b}\bar{b}}*\sigma_b^{-1}*\sigma_{\bar{b}\bar{b}})\\
    &+2\epsilon^2\sigma_b^{-1}*\sigma_{b\bar{b}}*\sigma_b^{-1}*(\sigma_{\bar{b}\bar{b}}+\sigma_{bb})\\
    &+2\epsilon^2\sigma_b^{-1}*\sigma_{bb}*\sigma_b^{-1}*\sigma_{\bar{b}\bar{b}})\\
    &+\frac{1}{2\hat{q}^2(1-\epsilon^2)}\text{Tr}(g_{b\bar{b}}*\sigma_{b\bar{b}}+\epsilon^2g_{\bar{b}\bar{b}}*\sigma_{\bar{b}\bar{b}}+\epsilon^2g_{bb}*\sigma_{bb}),
\end{split}
\end{equation}
\begin{equation}\label{I_f}
\begin{split}
    \frac{I_f}{N}-\frac{I_{f(\hat{q}=\infty)}}{N}=&\frac{1}{\hat{q}^2(1-\epsilon^2)^2}\text{Tr}(\frac{1}{2}(1+\epsilon^2)\sigma_f^{-1}*\sigma_{\psi\bar\psi}*\sigma_f^{-1}*\sigma_{\psi\bar\psi}\\
    &+\epsilon^4(\sigma_f^{-1}*\sigma_{\psi\psi}*\sigma_f^{-1}*\sigma_{\psi\psi}+\sigma_f^{-1}*\sigma_{\bar\psi\bar\psi}*\sigma_f^{-1}*\sigma_{\bar\psi\bar\psi})\\
    &+2\epsilon^2\sigma_f^{-1}*\sigma_{\psi\bar\psi}*\sigma_f^{-1}*(\sigma_{\bar\psi\bar\psi}+\sigma_{\psi\psi})\\
    &+2\epsilon^2\sigma_f^{-1}*\sigma_{\psi\psi}*\sigma_f^{-1}*\sigma_{\bar\psi\bar\psi})\\
    &+\frac{1}{2\hat{q}^2(1-\epsilon^2)}\text{Tr}(g_{\psi\bar\psi}*\sigma_{\psi\bar\psi}+\epsilon^2g_{\bar\psi\bar\psi}*\sigma_{\bar\psi\bar\psi}+\epsilon^2g_{\psi\psi}*\sigma_{\psi\psi}),
\end{split}
\end{equation}
\begin{equation}
\begin{split}
    \frac{I_m}{N}-\frac{I_{m(\hat{q}=\infty)}}{N}=&\frac{1}{2q^2(1-\epsilon^2)^2}\text{Tr}(\sigma_b^{-1}*\sigma_{b\bar\psi}*\sigma_f^{-1}*\sigma_{\psi\bar{b}}+\sigma_b^{-1}*\sigma_{\bar{b}\psi}*\sigma_f^{-1}*\sigma_{\bar\psi b}\\
    &+\epsilon^4(\sigma_b^{-1}*\sigma_{b\psi}*\sigma_f^{-1}*\sigma_{\psi b}+\sigma_b^{-1}*\sigma_{\bar{b}\bar\psi}*\sigma_f^{-1}*\sigma_{\bar\psi\bar{b}})\\
    &+\epsilon^2\sigma_b^{-1}*\sigma_{\bar{b}\bar\psi}*\sigma_f^{-1}*(-\sigma_{\psi b}+\sigma_{\psi \bar{b}}+\sigma_{\bar\psi b})\\
    &+\epsilon^2\sigma_b^{-1}*\sigma_{\bar{b}\psi}*\sigma_f^{-1}*(-\sigma_{\psi b}+\sigma_{\psi \bar{b}}+\sigma_{\bar\psi \bar{b}})\\
    &+\epsilon^2\sigma_b^{-1}*\sigma_{b\bar\psi}*\sigma_f^{-1}*(-\sigma_{\psi b}+\sigma_{\bar\psi b}+\sigma_{\bar\psi \bar{b}})\\
    &-\epsilon^2\sigma_b^{-1}*\sigma_{b\psi}*\sigma_f^{-1}*(\sigma_{\psi \bar{b}}+\sigma_{\bar\psi b}+\sigma_{\bar\psi \bar{b}})\\
    &+\frac{1}{\hat{q}^2}\text{Tr}(g_{\psi\bar{b}}*\sigma_{\bar{b}\psi}+g_{\bar\psi b}*\sigma_{b\bar\psi}+g_{\psi b}*\sigma_{b\psi}+g_{\bar\psi\bar{b}}*\sigma_{\bar{b}\bar\psi}),
\end{split}
\end{equation}
\begin{equation}
\frac{I_I}{N}-\frac{I_{I(\hat{q}=\infty)}}{N}=-\frac{J}{\hat{q}}\frac{1}{2^{\hat{q}-2}(1-\epsilon^2)^{\hat{q}}} \int d\tau_1 d\tau_2 \ g_{b \bar{b}}\exp \, g_{\psi \bar{\psi}}.
\end{equation}
The next step is to integrate out $\sigma$'s and to obtain an action of $g$'s only. We first solve for $\sigma$'s from their equations of motion. Each $\sigma$ only exists in one of the actions above, so we can do the following procedures seperately for each part of the action. Taking the bilocal fields of bosons, $\sigma_{b\bar{b}}$, $\sigma_{bb}$ and $\sigma_{\bar{b}\bar{b}}$ as an example, their solutions are
\begin{equation}
\begin{split}
    \sigma_b^{-1}*\sigma_{bb}*\sigma_b^{-1}=&-\frac{1}{2(1-\epsilon^2)}(2g_{b\bar{b}}-g_{\bar{b}\bar{b}}-\epsilon^2g_{bb}),\\
    \sigma_b^{-1}*\sigma_{\bar{b}\bar{b}}*\sigma_b^{-1}=&-\frac{1}{2(1-\epsilon^2)}(2g_{b\bar{b}}-g_{bb}-\epsilon^2g_{\bar{b}\bar{b}}),\\
    \sigma_b^{-1}*\sigma_{b\bar{b}}*\sigma_b^{-1}=&-\frac{1}{(1-\epsilon^2)}((1+\epsilon^2)g_{b\bar{b}}-\epsilon^2(g_{bb}+g_{\bar{b}\bar{b}})).\\
\end{split}
\end{equation}
We apply the following manipulations to insert them back into the action
\begin{equation}\label{eom_sigma}
\begin{split}
    \text{Tr}[g_{b\bar{b}}*\sigma_{b\bar{b}}]=&\text{Tr}[\sigma_b*g_{b\bar{b}}*\sigma_b*(\sigma_b^{-1}*\sigma_{b\bar{b}}*\sigma_b^{-1})],\\
    \text{Tr}[\sigma_b^{-1}*\sigma_{bb}*\sigma_b^{-1}*\sigma_{bb}]=&\text{Tr}[\sigma_b*(\sigma_b^{-1}*\sigma_{bb}*\sigma_b^{-1})*\sigma_b*(\sigma_b^{-1}*\sigma_{bb}*\sigma_b^{-1})].
\end{split}
\end{equation}
where each combination of $\sigma_b^{-1}*\sigma*\sigma_b^{-1}$ ca be replaced by the $g$'s in Eqn.(\ref{eom_sigma}). By inserting them back into the action, we obtain an action of $g$'s. Then, we do the same derivation for the action including bilocal fields of fermions (like $\sigma_{\psi\bar\psi}$) and the action including the mixed bilocal fields (like $\sigma_{b\bar\psi}$). Combining them with the interaction term, We write the resulting action in two parts, the bosonic part $I_1$ and the fermionic part $I_2$. 
\begin{equation}\label{large_q_action_bosonic}
\begin{split}
    \frac{I_1}{N} &=-\frac{J}{\hat{q}}\frac{1}{2^{\hat{q}}(1-\epsilon^2)^{\hat{q}}} \int d\tau_1 d\tau_2 \ (g_{b \bar{b}}\exp \, g_{\psi \bar{\psi}}+g_{ \bar{b}b}\exp \, g_{ \bar{\psi}\psi})\\
    &-\frac{1+\epsilon^2}{(1-\epsilon^2)^2}\frac{1}{8\hat{q}^2} \int d\tau_1 d\tau_2\, (-\partial_{\tau_1} g_{\psi \bar{\psi}}(\tau_1,\tau_2) \,\partial_{\tau_2} g_{\psi \bar{\psi}}(\tau_2,\tau_1) + g_{b \bar{b}}(\tau_1,\tau_2) \,g_{b \bar{b}}(\tau_2,\tau_1) \\
    &-\frac{\epsilon^2}{(1-\epsilon^2)^2}\frac{1}{4\hat{q}^2}\int d\tau_1 d\tau_2\left(\frac{1}{2}g_{bb}(\tau_1,\tau_2)g_{\bar{b}\bar{b}}(\tau_2,\tau_1)-g_{b\bar{b}}(\tau_1,\tau_2)g_{bb}(\tau_2,\tau_1)-g_{b\bar{b}}(\tau_1,\tau_2)g_{\bar{b}\bar{b}}(\tau_2,\tau_1)\right.\\
    &-\frac{1}{2}\partial_{\tau_1}g_{\psi\psi}(\tau_1,\tau_2)\partial_{\tau_2}g_{\bar\psi\bar\psi}(\tau_2,\tau_1)+\partial_{\tau_1}g_{\psi\bar\psi}(\tau_1,\tau_2)\partial_{\tau_2}g_{\psi\psi}(\tau_2,\tau_1)+\partial_{\tau_1}g_{\psi\bar\psi}(\tau_1,\tau_2)\partial_{\tau_2}g_{\bar\psi\bar\psi}(\tau_2,\tau_1)\\
    &-\frac{\epsilon^4}{(1-\epsilon^2)^2}\frac{1}{4\hat{q}^2}\int d\tau_1 d\tau_2\left(\frac{1}{4}g_{bb}(\tau_1,\tau_2)g_{bb}(\tau_2,\tau_1)+g_{\bar{b}\bar{b}}(\tau_1,\tau_2)g_{\bar{b}\bar{b}}(\tau_2,\tau_1)\right.\\
    &\left.-\frac{1}{4}\partial_{\tau_1}g_{\psi\psi}(\tau_1,\tau_2)\partial_{\tau_2}g_{\psi\psi}(\tau_2,\tau_1)-\frac{1}{4}\partial_{\tau_1}g_{\bar\psi\bar\psi}(\tau_1,\tau_2)\partial_{\tau_2}g_{\bar\psi\bar\psi}(\tau_2,\tau_1)\right),
\end{split}
\end{equation}
\begin{equation}\label{large_q_action_fermionic}
\begin{split}
    \frac{I_2}{N} &=\frac{J}{\hat{q}}\frac{1}{2^{\hat{q}}(1-\epsilon^2)^{\hat{q}}} \int d\tau_1 d\tau_2\, \ (g_{\psi\bar{b}} g_{b \bar\psi} \ \exp \, g_{\psi \bar{\psi}}+g_{\bar\psi b} g_{\bar{b} \psi} \ \exp \, g_{ \bar{\psi}\psi})\\
    &-\frac{1+\epsilon^2}{(1-\epsilon^2)^2}\frac{1}{8\hat{q}^2} \int d\tau_1 d\tau_2 \partial_{\tau_1}g_{\psi \bar{b}}(\tau_1,\tau_2)  g_{b\bar{\psi}}(\tau_2,\tau_1) - g_{b\bar{\psi}}(\tau_1,\tau_2) \partial_{\tau_2} g_{\psi \bar{b}}(\tau_2,\tau_1)\\
    &-\frac{\epsilon^2}{(1-\epsilon^2)^2}\frac{1}{4\hat{q}^2}\int d\tau_1 d\tau_2 \frac{1}{2}g_{b\psi }(\tau_1,\tau_2)\partial_{\tau_2}g_{\bar{b}\bar\psi}(\tau_1,\tau_2)-\frac{1}{2}\partial_{\tau_1}g_{b\psi}(\tau_2,\tau_1)g_{\bar{b}\bar\psi}(\tau_2,\tau_1)\\
    &+g_{b\bar\psi}(\tau_1,\tau_2)\partial_{\tau_2}g_{b\psi}(\tau_1,\tau_2)-\partial_{\tau_1}g_{\psi \bar{b}}(\tau_1,\tau_2)g_{b\psi}(\tau_2,\tau_1)\\
    &\left.+g_{b\bar\psi}(\tau_1,\tau_2)\partial_{\tau_2}g_{\bar{b}\bar\psi}(\tau_1,\tau_2)-\partial_{\tau_1}g_{\psi \bar{b}}(\tau_1,\tau_2)g_{\bar{b}\bar\psi}(\tau_2,\tau_1)\right)\\
    &-\frac{\epsilon^4}{(1-\epsilon^2)^2}\frac{1}{4\hat{q}^2}\int d\tau_1 d\tau_2\left(\frac{1}{4}\partial_{\tau_1}g_{\psi b}(\tau_1,\tau_2)  g_{b\psi}(\tau_2,\tau_1) -\frac{1}{4} g_{b\psi}(\tau_1,\tau_2) \partial_{\tau_2} g_{\psi b}(\tau_2,\tau_1)\right.\\
    &\left.+\frac{1}{4}\partial_{\tau_1}g_{\bar\psi \bar{b}}(\tau_1,\tau_2)  g_{\bar{b}\bar{\psi}}(\tau_2,\tau_1) - \frac{1}{4}g_{\bar{b}\bar{\psi}}(\tau_1,\tau_2) \partial_{\tau_2} g_{\bar\psi \bar{b}}(\tau_2,\tau_1)\right).
\end{split}
\end{equation}
By applying the supersymmetry transformations, Eq.(\ref{susyq=1}), one can check that the action preserves $\mathcal{N}=1$ supersymmetry. To proceed, we define a useful rescaled coupling 
\begin{equation}
\mathcal{J}=\frac{J\hat{q}}{(2(1-\epsilon^2))^{\hat{q}-2}}.
\end{equation}

We solve the equations of motion for the $g$'s. The equations of motion for fermionic $g$'s (e.g., $g_{b\psi}$) have zero solutions, so we focus on the equations of motion for bosonic $g$'s (e.g., $g_{\psi\bar\psi}$). They are
\begin{equation}
\begin{split}
    -(1+\epsilon^2)\partial_1\partial_2g_{\psi\bar\psi}(\tau_2,\tau_1)+\epsilon^2\partial_1\partial_2 (g_{\psi\psi}(\tau_2,\tau_1)+g_{\bar\psi\bar\psi}(\tau_2,\tau_1))=&\mathcal{J}g_{b \bar{b}}\exp g_{\psi \bar{\psi}},\\
    \partial_1\partial_2(g_{\bar\psi\bar\psi}(\tau_2,\tau_1)+\epsilon^2g_{\psi\psi}(\tau_2,\tau_1))=&2\partial_1\partial_2g_{\psi\bar\psi}(\tau_2,\tau_1),\\
    \partial_1\partial_2(g_{\psi\psi}(\tau_2,\tau_1)+\epsilon^2g_{\bar\psi\bar\psi}(\tau_2,\tau_1))=&2\partial_1\partial_2g_{\psi\bar\psi}(\tau_2,\tau_1),\\
    (1+\epsilon^2)g_{b\bar{b}}(\tau_2,\tau_1)-\epsilon^2(g_{bb}(\tau_2,\tau_1)+g_{\bar{b}\bar{b}}(\tau_2,\tau_1))=&\mathcal{J}\exp g_{\psi \bar{\psi}},\\
    g_{\bar{b}\bar{b}}(\tau_2,\tau_1)+\epsilon^2g_{bb}(\tau_2,\tau_1)=&2g_{b\bar{b}}(\tau_2,\tau_1),\\
    g_{bb}(\tau_2,\tau_1)+\epsilon^2g_{\bar{b}\bar{b}}(\tau_2,\tau_1)=&2g_{b\bar{b}}(\tau_2,\tau_1).
\end{split}
\end{equation}
After some rearrangement, they become
\begin{equation}
\begin{split}
    -\frac{(1-\epsilon^2)^2}{1+\epsilon^2}\partial_1\partial_2g_{\psi\bar\psi}(\tau_2,\tau_1)=&\mathcal{J}g_{b \bar{b}}\exp g_{\psi \bar{\psi}},\\
    (1+\epsilon^2)\partial_1\partial_2g_{\psi\psi}(\tau_2,\tau_1)=&2\partial_1\partial_2g_{\psi\bar\psi}(\tau_2,\tau_1),\\
    (1+\epsilon^2)\partial_1\partial_2g_{\bar\psi\bar\psi}(\tau_2,\tau_1)=&2\partial_1\partial_2g_{\psi\bar\psi}(\tau_2,\tau_1),\\
    \frac{(1-\epsilon^2)^2}{1+\epsilon^2}g_{b\bar{b}}(\tau_2,\tau_1)=&\mathcal{J}\exp g_{\psi \bar{\psi}},\\
    (1+\epsilon^2)g_{bb}(\tau_2,\tau_1)=&2g_{b\bar{b}}(\tau_2,\tau_1),\\
    (1+\epsilon^2)g_{\bar{b}\bar{b}}(\tau_2,\tau_1)=&2g_{b\bar{b}}(\tau_2,\tau_1).
\end{split}
\end{equation}
These equations can also be obtained by expanding the SD equations of $G$ and $\Sigma$ (\ref{SDequations1})-(\ref{SDequations7}) in large $\hat{q}$ limit. After more cancellations, we found that $g_{\psi\bar\psi}$ can be solved by the following Liouville equation, and all the other fields can be determined by the solution of $g_{\psi\bar\psi}$,
\begin{equation}
    \partial^2_\tau g_{\psi\bar\psi}(\tau)=\frac{(1+\epsilon^2)^2}{(1-\epsilon^2)^4}\mathcal{J}^2\exp (2g_{\psi\bar\psi}) \, .
\end{equation}
All other bosonic correlators are determined by $g_{\psi\bar\psi}$. The solution to the Liouville equation is
\begin{equation}
    e^{2g_{\psi\bar\psi}}=\left(\frac{\cos\frac{\pi v}{2}}{\cos(\pi v(\frac{1}{2}-\frac{|\tau|}{\beta}))}\right)^2,
\end{equation}
with 
\begin{equation}
    \beta\mathcal{J}\frac{1+\epsilon^2}{(1-\epsilon^2)^2}=\frac{\pi v}{\cos\frac{\pi v}{2}}.
\end{equation}

\subsection{Deformed complex SYK model}
The effective action in terms of $G$ and $\Sigma$ for the deformed complex SYK model is
\begin{equation}
\begin{split}
    \frac{I}{N}=&-\log\text{Pf}\begin{bmatrix}-\frac{\i\epsilon}{2}\sigma_f-\Sigma_{\psi\psi}&-\frac{1}{2}\sigma_f-\frac{1}{2}\Sigma_{\psi\bar\psi}\\-\frac{1}{2}\sigma_f-\frac{1}{2}\Sigma_{\psi\bar\psi}&\frac{\i\epsilon}{2}\sigma_f-\Sigma_{\bar\psi\bar\psi}\end{bmatrix}+\int d\tau_1 d\tau_2[G_{\psi\bar\psi}(\tau_1,\tau_2)\Sigma_{\psi\bar\psi}(\tau_2,\tau_1)\\
    &+G_{\psi\psi}(\tau_1,\tau_2)\Sigma_{\psi\psi}(\tau_2,\tau_1)+G_{\bar\psi\bar\psi}(\tau_1,\tau_2)\Sigma_{\bar\psi\bar\psi}(\tau_2,\tau_1)-\frac{J^2}{q}G^q(\tau_1,\tau_2)]
\end{split}
\end{equation}
First, we expand the bilocal fields $G$'s and $\Sigma$'s in large $q$ limit. The action can be separated into two parts: the interaction term and all the other terms. The second part is the same as Eqn. (\ref{I_f}), and the interaction term in large $q$ limit is,
\begin{equation}
\frac{I_I}{N}-\frac{I_{I(q=\infty)}}{N}=\frac{J^2}{q}\frac{1}{2^{q}(1-\epsilon^2)^q} \int d\tau_1 d\tau_2 \ \exp \, g_{\psi \bar{\psi}}\\
\end{equation}
Since the second part is the same as in SUSY model, when we solve for $\sigma$'s and insert them back into the action, the resultant action has the same terms with bilocal fields of fermions. The action is
\begin{equation}\label{complex_SYK_action}
\begin{split}
    \frac{I}{N} &=-\frac{J^2}{q}\frac{1}{2^{q}(1-\epsilon^2)^q} \int d\tau_1 d\tau_2 \ \exp \, g_{\psi \bar{\psi}}\\
    &-\frac{1+\epsilon^2}{(1-\epsilon^2)^2}\frac{1}{8q^2} \int d\tau_1 d\tau_2\, -\partial_{\tau_1} g_{\psi \bar{\psi}}(\tau_1,\tau_2) \,\partial_{\tau_2} g_{\psi \bar{\psi}}(\tau_2,\tau_1) \\
    &-\frac{\epsilon^2}{(1-\epsilon^2)^2}\frac{1}{4q^2}\int d\tau_1 d\tau_2\left(-\frac{1}{2}\partial_{\tau_1}g_{\psi\psi}(\tau_1,\tau_2)\partial_{\tau_2}g_{\bar\psi\bar\psi}(\tau_2,\tau_1)\right.\\
    &\left.+\partial_{\tau_1}g_{\psi\bar\psi}(\tau_1,\tau_2)\partial_{\tau_2}g_{\psi\psi}(\tau_2,\tau_1)+\partial_{\tau_1}g_{\psi\bar\psi}(\tau_1,\tau_2)\partial_{\tau_2}g_{\bar\psi\bar\psi}(\tau_2,\tau_1)\right)\\
    &-\frac{\epsilon^4}{(1-\epsilon^2)^2}\frac{1}{4q^2}\int d\tau_1 d\tau_2\left(-\frac{1}{4}\partial_{\tau_1}g_{\psi\psi}(\tau_1,\tau_2)\partial_{\tau_2}g_{\psi\psi}(\tau_2,\tau_1)\right.\\
    &\left.-\frac{1}{4}\partial_{\tau_1}g_{\bar\psi\bar\psi}(\tau_1,\tau_2)\partial_{\tau_2}g_{\bar\psi\bar\psi}(\tau_2,\tau_1)\right)\\
\end{split}
\end{equation}
We get two Liouville equations 
\bea
\partial_\tau^2 g_{\psi \bar{\psi}} &=& \mathcal{J}^2 \frac{(1+\epsilon^2)}{(1-\epsilon^2)^2} e^{g_{\psi \bar{\psi}}(\tau)}\\
\partial_\tau^2 g_{\psi \psi} &=& \mathcal{J}^2\frac{2}{(1-\epsilon^2)^2} e^{g_{\psi \bar{\psi}}(\tau)} = \frac{2}{1+\epsilon^2} \partial_\tau^2 g_{\psi \overline{\psi}}
\ea
where $\mathcal{J}^2=\frac{J^2q}{(2(1-\epsilon^2))^{q-2}}$. The solution is quite simple. The first equation has the same solution as large $q$ SYK with a rescaling of $\mathcal{J}$. The second equation is solved by 
\beq
g_{\psi \psi}(\tau) = \frac{2}{1+\epsilon^2} g_{\psi \bar{\psi}}(\tau).
\eeq
\section{The nonlinear $\U(1)$ mode}
\label{app:U1effective}
\subsection{Deformed SUSY SYK model}
\begin{figure}[h]
    \centering
    \includegraphics[width=0.35\textwidth]{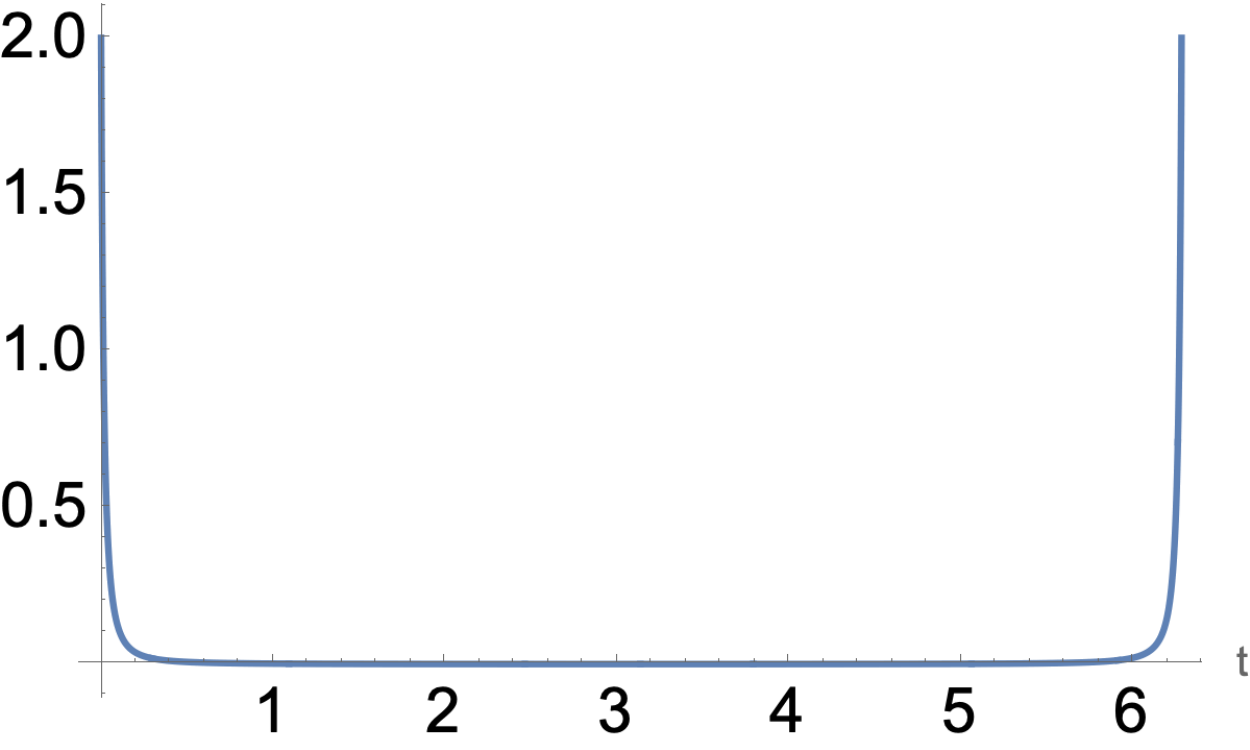}
    \caption{$\left(\frac{\cos\frac{\pi v}{2}}{\cos\left(\frac{\pi v}{2}-\frac{ v |\tau_1-\tau_2|}{2}\right)}\right)^2$. }
    \label{integrand}
\end{figure}

In this section, we derive the action of a phase mode fluctuation $a(\tau)$. We apply a $\U(1)$ transformation on the on-shell solutions of $g$'s:
\begin{equation}
\begin{split}
    g_{\psi\bar\psi}(\tau_1,\tau_2)\rightarrow e^{\frac{\i}{\hat{q}}(a(\tau_1)-a(\tau_2))}g_{\psi\bar\psi}(\tau_1,\tau_2)\\
    g_{\psi\psi}(\tau_1,\tau_2)\rightarrow e^{\frac{\i}{\hat{q}}(a(\tau_1)+a(\tau_2))}g_{\psi\psi}(\tau_1,\tau_2)\\
    g_{\bar\psi\bar\psi}(\tau_1,\tau_2)\rightarrow e^{-\frac{\i}{\hat{q}}(a(\tau_1)+a(\tau_2))}g_{\bar\psi\bar\psi}(\tau_1,\tau_2)\\
    g_{b\bar{b}}(\tau_1,\tau_2)\rightarrow e^{\i\frac{1-\hat{q}}{\hat{q}}(a(\tau_1)-a(\tau_2))}g_{b\bar{b}}(\tau_1,\tau_2)\\
    g_{bb}(\tau_1,\tau_2)\rightarrow e^{\i\frac{1-\hat{q}}{\hat{q}}(a(\tau_1)+a(\tau_2))}g_{bb}(\tau_1,\tau_2)\\
    g_{\bar{b}\bar{b}}(\tau_1,\tau_2)\rightarrow e^{-\i\frac{1-\hat{q}}{\hat{q}}(a(\tau_1)+a(\tau_2))}g_{\bar{b}\bar{b}}(\tau_1,\tau_2)\\
\end{split}
\end{equation}
where $a(\tau)$ are finite. Note that $\psi$ carries $\U(1)_R$ charge of $1/\hat{q}$. The change of the action will be the effective action of $a$
\begin{equation}
    \frac{I[a]}{N}=\frac{I[g'[a,g]]}{N}-\frac{I[g]}{N} \, .
\end{equation}
Now, we look at every bosonic term in the action and derive the change of the term when applying the $\U(1)$ transformation.
\begin{equation}
    \delta\left(\int d\tau_1 d\tau_2\, (\partial_{\tau_1} g_{\psi \bar{\psi}}(\tau_1,\tau_2) \,\partial_{\tau_2} g_{\psi \bar{\psi}}(\tau_2,\tau_1))\right)=\int d\tau_1 d\tau_2\, \left(\mathcal{O}\left(\frac{1}{\hat{q}}\right)+\mathcal{O}\left(\frac{1}{\hat{q}^2}\right)\right)
\end{equation}
The change in the kinetic term of $g_{\psi\bar\psi}$ is of order $\mathcal{O}(\frac{1}{\hat{q}})$ in large $\hat{q}$ limit. Next is the mass term of $g_{b\bar{b}}$,
\begin{equation}
    \delta\left(\int d\tau_1 d\tau_2\, g_{b \bar{b}}(\tau_1,\tau_2) \,g_{b \bar{b}}(\tau_2,\tau_1)\right)=0
\end{equation}
Note that the time variable in the second $g$ switches order, so the extra phases cancel. The following are the three deformation terms with bosons:
\begin{equation}
    \delta\left(\int d\tau_1 d\tau_2(g_{bb}(\tau_1,\tau_2)g_{\bar{b}\bar{b}}(\tau_2,\tau_1))\right)=0
\end{equation}
\begin{equation}
    \delta\left(\int d\tau_1 d\tau_2(g_{b\bar{b}}(\tau_1,\tau_2)g_{bb}(\tau_2,\tau_1))\right)=\int d\tau_1 d\tau_2(e^{-\i 2a(\tau_1)}-1)g_{b\bar{b}}(\tau_1,\tau_2)g_{bb}(\tau_2,\tau_1)+\mathcal{O}\left(\frac{1}{\hat{q}}\right)
\end{equation}
\begin{equation}
    \delta\left(\int \d\tau_1 \d\tau_2(g_{b\bar{b}}(\tau_1,\tau_2)g_{\bar{b}\bar{b}}(\tau_2,\tau_1))\right)=\int d\tau_1 d\tau_2(e^{\i 2a(\tau_1)}-1)g_{b\bar{b}}(\tau_1,\tau_2)g_{\bar{b}\bar{b}}(\tau_2,\tau_1)+\mathcal{O}\left(\frac{1}{\hat{q}}\right)
\end{equation}
The last two terms are of order 1 in the large $\hat{q}$ limit. Combining them and taking large-$\hat{q}$ limit, we obtain
\begin{equation}\label{mass_term}
\begin{split}
    &\delta\left(\int \d\tau_1 \d\tau_2\left(g_{b\bar{b}}(\tau_1,\tau_2)g_{bb}(\tau_2,\tau_1)+g_{b\bar{b}}(\tau_1,\tau_2)g_{\bar{b}\bar{b}}(\tau_2,\tau_1)\right)\right)\\
    =&2\int d\tau_1 d\tau_2(\cos(2a(\tau_1))-1)g_{b\bar{b}}(\tau_1,\tau_2)g_{bb}(\tau_2,\tau_1)))\\
\end{split}
\end{equation}
The following are the three deformation terms with fermions, 
\begin{equation}
    \delta\left(\int d\tau_1 d\tau_2(\partial_{\tau_1}g_{\psi\psi}(\tau_1,\tau_2)\partial_{\tau_2}g_{\bar\psi\bar\psi}(\tau_2,\tau_1))\right)=\int d\tau_1 d\tau_2\,\mathcal{O}\left(\frac{1}{\hat{q}}\right)
\end{equation}
\begin{equation}
    \delta\left(\int d\tau_1 d\tau_2(\partial_{\tau_1}g_{\psi\bar\psi}(\tau_1,\tau_2)\partial_{\tau_2}g_{\psi\psi}(\tau_2,\tau_1)\right)=\int d\tau_1 d\tau_2\,\mathcal{O}\left(\frac{1}{\hat{q}}\right)
\end{equation}
\begin{equation}
    \delta\left(\int d\tau_1 d\tau_2(\partial_{\tau_1}g_{\psi\bar\psi}(\tau_1,\tau_2)\partial_{\tau_2}g_{\bar\psi\bar\psi}(\tau_2,\tau_1)\right)=\int d\tau_1 d\tau_2\,\mathcal{O}\left(\frac{1}{\hat{q}}\right)
\end{equation}
They are all of order $\mathcal{O}(\frac{1}{\hat{q}})$ in large $\hat{q}$ limit. The last term is the interaction,
\begin{equation}
\begin{split}
    &\delta\left(\int d\tau_1 d\tau_2\, (g_{b \bar{b}} \exp g_{\psi \bar{\psi}}+g_{ \bar{b}b}\exp g_{\bar{\psi}\psi })\right)\\
    =&\int d\tau_1 d\tau_2(e^{-\i(a(\tau_1)-a(\tau_2))}-1)g_{b\bar{b}}\exp g_{\psi \bar{\psi}}+(e^{\i(a(\tau_1)-a(\tau_2))}-1)g_{\bar{b}b}\exp g_{\bar{\psi}\psi }+\mathcal{O}\left(\frac{1}{\hat{q}}\right)
\end{split}
\end{equation}
The on-shell solution of $g_{b\bar{b}}$ and $g_{\bar{b}b}$ coincides, so does $g_{\psi\bar\psi}$ and $g_{\bar\psi\psi}$, so the two terms combine,
\begin{equation}
    \delta\left(\int d\tau_1 d\tau_2\, (g_{b \bar{b}} \exp g_{\psi \bar{\psi}}+g_{ \bar{b}b}\exp g_{\bar{\psi}\psi })\right)
    =2\int d\tau_1 d\tau_2(\cos(a(\tau_1)-a(\tau_2))-1)g_{b\bar{b}}\exp g_{\psi \bar{\psi}} \, ,
\end{equation}
since $g_{b\bar{b}}\exp g_{\psi \bar{\psi}}$ acts like a Delta function. This term gives the kinetic term of the $\U(1)$ mode, while the Eqn. (\ref{mass_term}) gives a mass term as below.
\begin{equation}
\begin{split}
    -\frac{2\epsilon^2}{(1-\epsilon^2)^2}\frac{1}{\hat{q}^2}\int d\tau_1 d\tau_2(\cos(2a(\tau_1))-1)g_{b\bar{b}}(\tau_1,\tau_2)g_{bb}(\tau_2,\tau_1)
\end{split}
\end{equation}
Inserting the on-shell values of $g$'s,
\begin{equation}
\begin{split}
    g_{bb}=g_{\bar{b}\bar{b}}=\frac{2}{1+\epsilon^2}g_{b\bar{b}}=\frac{2}{(1-\epsilon^2)^2}\mathcal{J}\exp g_{\psi\bar\psi}=\frac{2}{(1-\epsilon^2)^2}\mathcal{J}\left|\frac{\cos\frac{\pi v}{2}}{\cos\left(\frac{\pi v}{2}-\frac{ v |t|}{2}\right)}\right|
\end{split}
\end{equation}
The action becomes
\begin{equation}
\begin{split}
    -2\epsilon^2\frac{1+\epsilon^2}{(1-\epsilon^2)^2}\frac{4\mathcal{J}^2}{\hat{q}^2}\left(\int d\tau_1(\cos(2a)-1)\int d\tau\left(\frac{\cos\frac{\pi v}{2}}{\cos(\pi v(\frac{1}{2}-\frac{|\tau|}{\beta}))}\right)^2\right)
\end{split}
\end{equation}
Besides the function of $a$, the integrand is shown in Figure \ref{integrand}. It approaches zero when $t$ moves away from zero, so it acts like a delta function, restraining $\tau_1-\tau_2$ to small values. 
The integrals over $\tau$ are
\begin{equation}
\begin{split}
    \int d\tau\left(\frac{\cos\frac{\pi v}{2}}{\cos(\pi v(\frac{1}{2}-\frac{|\tau|}{\beta}))}\right)^2=\frac{1}{2\pi}\int_0^{2\pi} \beta d\theta\left(\frac{\cos\frac{\pi v}{2}}{\cos(\pi v(\frac{1}{2}-\frac{|\theta|}{2\pi}))}\right)^2=\frac{2}{\mathcal{\pi J}}\frac{(1-\epsilon^2)^2}{1+\epsilon^2}
\end{split}
\end{equation}
where we made the change of variable $\tau \to \theta = 2\pi \tau /\beta$. The first integral is of order $\mathcal{O}(1)$ when $v$ approaches 1. The whole action is
\begin{equation}
\begin{split}
    -\frac{(1-\epsilon^2)^2}{1+\epsilon^2}\frac{8\beta\mathcal{J}}{\pi^2\hat{q}^2}\epsilon^2\int d\theta_1(\cos(2a)-1)
\end{split}
\end{equation}

\subsection{Deformed complex SYK model}
For complex SYK model, we can compute the action of a phase mode fluctuation $a(\tau)$ that is time independent $\dot{a}=0$. The U(1) transformation are 
\begin{equation}
\begin{split}
    g_{\psi\psi}(\tau_1,\tau_2)\rightarrow e^{\i2a}g_{\psi\psi}(\tau_1,\tau_2)\\
    g_{\bar\psi\bar\psi}(\tau_1,\tau_2)\rightarrow e^{-\i2a}g_{\bar\psi\bar\psi}(\tau_1,\tau_2)
\end{split}
\end{equation}
There are two terms in the action (\ref{complex_SYK_action}) that contributes to the mass term of the $\U(1)$ mode, 
\begin{equation}
    \delta\left(\partial_{\tau_1}g_{\psi\bar\psi}\partial_{\tau_2}g_{\psi\psi}+\partial_{\tau_1}g_{\psi\bar\psi}\partial_{\tau_2}g_{\bar\psi\bar\psi}\right)=2(\cos(2a)-1)\partial_{\tau_1}g_{\psi\bar\psi}\partial_{\tau_2}g_{\bar\psi\bar\psi}
\end{equation}
The prefactor measures the mass of the $\U(1)$ mode. Other than numerical prefactors, it is important to make sure we understand the powers of $\beta$. To determine this, we can proceed in the following way. First, notice that the derivatives and integral measure are invariant under changing $\tau \to \theta = 2\pi \tau /\beta$. We can then use the equation of motion that the mean field satisfies which produces an overall factor of $\beta^2 \mathcal{J}^2$. Then we are left with an integral
\beq
\int \d \theta_1 \d \theta_2 g_{\psi \bar{\psi}} e^{g_{\psi \bar\psi}}= \frac{4 \pi (\pi v (1+ \cos \pi v) - 2 \sin \pi v)}{v} \sim 8 \pi^2 (v-1) \sim 16 \pi^2 \frac{1}{\beta \mathcal{J}}
\eeq
Then the action is 
\beq
I = \# \frac{N}{q^2} \epsilon^2 \beta \mathcal{J} \, (1-\cos(2a))
\eeq
The fact that its proportional to beta is consistent with this term coming from $\int \d \tau \, a(\tau)^2$ as expected from symmetry principles. Besides this term, all other modes of the Schwarzian or phase mode already have an action to zeroth order in $\epsilon$.

\section{Supersymmetry of the large $\hat{q}$ action}\label{App:CHECKSUSY}
In this section, we will check that the large-$\hat{q}$ action $I_1+I_2$ for the deformed SUSY model has global $\mathcal{N}=1$ supersymmetry. Applying the supersymmetry transformations (\ref{susyq=1}) on bilocal fields,
\begin{equation}
\begin{split}
&\delta_\mathcal{Q}g_{\psi\bar\psi}=g_{b\bar\psi}-g_{\psi\bar{b}},\ \delta_\mathcal{Q}g_{\psi\psi}=g_{b\psi}-g_{\psi b},\
\delta_\mathcal{Q}g_{\bar\psi\bar\psi}=g_{\bar{b}\bar\psi}-g_{\bar\psi\bar{b}}\\
&\delta_\mathcal{Q}g_{b\bar{b}}=\partial_{\tau_1}g_{\psi\bar{b}}+\partial_{\tau_2}g_{b\bar\psi},\ 
\delta_\mathcal{Q}g_{bb}=\partial_{\tau_1}g_{\psi b}+\partial_{\tau_2}g_{b\psi},\ 
\delta_\mathcal{Q}g_{\bar{b}\bar{b}}=\partial_{\tau_1}g_{\bar\psi \bar{b}}+\partial_{\tau_2}g_{\bar{b}\bar\psi}\\
&\delta_\mathcal{Q}g_{b\bar\psi}=\partial_{\tau_1}g_{\psi\bar\psi}+g_{b\bar{b}},\ 
\delta_\mathcal{Q}g_{\psi\bar{b}}=g_{\bar{b}\bar{b}}-\partial_{\tau_2}g_{\psi\bar\psi},\ 
\delta_\mathcal{Q}g_{\bar\psi\bar{b}}=g_{\bar{b}\bar{b}}-\partial_{\tau_2}g_{\bar\psi\bar\psi}\\
&\delta_\mathcal{Q}g_{\bar{b}\bar\psi}=\partial_{\tau_1}g_{\bar\psi\bar\psi}+g_{\bar{b}\bar{b}},\ 
\delta_\mathcal{Q}g_{\psi b}=g_{bb}-\partial_{\tau_2}g_{\psi\psi},\ 
\delta_\mathcal{Q}g_{\bar\psi\bar{b}}=\partial_{\tau_1}g_{\psi\psi}+g_{bb}\\
\end{split}
\end{equation}
Inserting these into the large-$\hat{q}$ action (\ref{large_q_action_bosonic}) and (\ref{large_q_action_fermionic}), every term in (\ref{large_q_action_bosonic}) has a counterpart in (\ref{large_q_action_fermionic}); together, they are invariant under supersymmetry. For example, for the interaction term,
\begin{equation}
\begin{split}
   &\delta_{\mathcal{Q}}\int d\tau_1 d\tau_2(g_{b \bar{b}}\exp \, g_{\psi \bar{\psi}}- g_{\psi\bar{b}} g_{b \bar\psi} \ \exp \, g_{\psi \bar{\psi}})\\
   =&\int d\tau_1 d\tau_2(\partial_{\tau_1}g_{\psi\bar{b}}+\partial_{\tau_2}g_{b\bar\psi})\exp \, g_{\psi \bar{\psi}}+g_{b\bar\psi}\partial_{\tau_2}(\exp g_{\psi\bar\psi})+g_{\psi\bar{b}}\partial_{\tau_1}(\exp g_{\psi\bar\psi})\\
\end{split}
\end{equation}
They combine into total derivatives, so the interaction terms have supersymmetry. The next two terms are the kinetic term of $g_{\psi\bar\psi}$ and mass term of $g_{b\bar{b}}$, together with their fermionic counterparts, 
\begin{equation}
\begin{split}
\delta_{\mathcal{Q}}\int d\tau_1 d\tau_2\, (-\partial_{\tau_1} g_{\psi \bar{\psi}}\,\partial_{\tau_2} g_{\psi \bar{\psi}} + g_{b \bar{b}}\,g_{b \bar{b}}+\partial_{\tau_1}g_{\psi \bar{b}} g_{b\bar{\psi}}- g_{b\bar{\psi}}\partial_{\tau_2} g_{\psi \bar{b}})=0
\end{split}
\end{equation}
Following the same strategy, one can prove all the other terms and their fermionic counterparts are invariant under the supersymmetry transformation. 
\bibliographystyle{utphys2}
{\small \bibliography{Biblio}{}}

\end{document}